\documentclass[11pt]{article} 
\usepackage[utf8]{inputenc}
\usepackage{multirow}
\usepackage{float}
\usepackage{amsfonts}
\usepackage{amsmath}
\usepackage{amssymb}
\usepackage{amsthm}
\usepackage{mathtools}
\usepackage{dsdshorthand}
\usepackage{tikz}
\usepackage{makecell}
\usepackage{ulem}
\usepackage{tensor}
\usetikzlibrary{decorations}
\usepackage{multirow}
\usetikzlibrary{trees}
\usetikzlibrary{decorations.pathmorphing}
\usetikzlibrary{decorations.markings}
\usetikzlibrary{external}
\usetikzlibrary{intersections}
\usetikzlibrary{shapes,arrows}
\usetikzlibrary{arrows.meta}
\usetikzlibrary{calc}
\usetikzlibrary{shapes.misc}
\usetikzlibrary{decorations.text}
\usetikzlibrary{backgrounds}
\usetikzlibrary{fadings}
\usetikzlibrary{tikzmark,calc,arrows,shapes,decorations.pathreplacing}
\tikzset{
	cross/.style={cross out, draw=black, minimum size=2*(#1-\pgflinewidth), inner sep=0pt, outer sep=0pt},
	branchCut/.style={postaction={decorate},
		snake=zigzag,
		decoration = {snake=zigzag,segment length = 2mm, amplitude = 2mm}	
}}
\usepackage{caption}
\usepackage{color}
\definecolor{darkgreen}{rgb}{0,0.5,0}
\definecolor{darkblue}{rgb}{0,0,0.6}
\definecolor{purple}{rgb}{0.4,.2,0.7}
\usepackage[margin = 2.5cm]{geometry}
\usepackage{graphicx}
\usepackage[hyperindex,breaklinks,hyperfootnotes = false, colorlinks = true, linkcolor = darkblue, citecolor = purple]{hyperref}
\usepackage{subcaption}
\usepackage{ytableau}
\usepackage[english]{babel}
\usepackage[autostyle, english = american]{csquotes}
\MakeOuterQuote{"}

\renewcommand{\tilde}{\widetilde}

\renewcommand{\hat}{\widehat}

\usepackage{comment}

	\newcommand{\ee}{\end{equation}}
\newcommand{\bea}{\begin{eqnarray}}
	\newcommand{\eea}{\end{eqnarray}}

\def\nref#1{(\ref{#1})}

\def\nref#1{(\ref{#1})}
\def\n{\nabla}
\def\tll{\tilde{\lambda}_{l}}

\def\mcD{\mathcal{D}}

\def\deltaco{\delta_{\text{co}}}
\def\lds{\ell_{\text{dS}}}

\usepackage{tocloft}
\begin{document}
	
	\thispagestyle{empty}
	\begin{center}
		~\vspace{5mm}
		
		\vskip 2cm 
		
		{\LARGE \bf 
		  One-loop aspects of de Sitter axion wormholes
		}

		\vspace{0.5in}

        Victor Ivo$^1$ and Haifeng Tang$^2$
		
		\vspace{0.5in}

		$^1$
		{\it  Jadwin Hall, Princeton University,  Princeton, NJ 08540, USA }
        \\
		~
		\\
        $^2$
        {\it Leinweber Institute for Theoretical Physics, Stanford University, Stanford, CA 94305, USA}

	\end{center}
	
	\vspace{0.5in}
	
	\begin{abstract}

    We discuss aspects of the Euclidean path integral around axion-supported de Sitter wormholes, at one-loop order. We numerically compute the phase of the path integral around these solutions, as well as for a certain "multiple wormholes" generalization, and interpret this phase in different regimes. When the geometry is well approximated by a sphere with a small handle, the wormhole admits an effective description as a sphere with two local operator insertions, whose positions fluctuate around the antipodal configuration. The antipodal configuration is an extremum of the position integral for the operators, but we show that it is an unstable one. Accordingly, the phase of the wormhole solution can be viewed as the Polchinski phase in the sphere, multiplied by an additional phase from the integral over positions of the effective local operators. Using our expressions for the one-loop determinant, we also estimate the EFT coefficients of the dual bilocal operators in odd spacetime dimensions, to one-loop order. Lastly, we also discuss "maximal flux" solutions, which have $S^{1}\times S^{D-1}$ geometry. Their Lorentzian continuations are Einstein static universes, so we call them "Einstein wormholes". In this limit, we determine the spectrum of fluctuations analytically and show that the phase of the path integral around this solution is entirely accounted for by the well-known instability of the Einstein static universe.  
        
    \end{abstract}
	
	\vspace{1in}
	
	\pagebreak
	
	\setcounter{tocdepth}{3}
	{\hypersetup{linkcolor=black}\tableofcontents}

\section{Introduction}

Understanding aspects of quantum gravity in de Sitter space is clearly very important. However, in the absence of a top-down construction of de Sitter quantum gravity, it is reasonable to try to learn as much about de Sitter as we can from semiclassical gravity. Indeed, developments in recent years have emphasized that the gravitational path integral is a valuable tool in understanding quantum aspects of gravity \cite{Almheiri:2019qdq, Penington:2019kki, Iliesiu:2020qvm, Heydeman:2020hhw}.

A natural question then is what is the role of spacetime wormholes in de Sitter quantum gravity \cite{Myers:1988sp, Halliwell:1989pu, Gutperle:2002km, Chen:2020tes, Aguilar-Gutierrez:2023ril, Aguilar-Gutierrez:2023hls, Fumagalli:2024msi, Yang:2025lme}. An interesting on-shell wormhole was found by Myers in \cite{Myers:1988sp}, and further studied in \cite{Halliwell:1989pu, Gutperle:2002km, Aguilar-Gutierrez:2023ril, Aguilar-Gutierrez:2023hls}. This solution is a generalization of the Giddings Strominger wormhole \cite{Giddings:1987cg} to spacetimes with a positive cosmological constant. These wormholes are $SO(D)$ symmetric solutions in $D$ spacetime dimensions with Euclidean line element
\begin{equation}
ds^{2}=d\tau^{2}+a^{2}(\tau)d\Omega_{D-1}^{2}    
\end{equation}
and where $d\Omega_{D-1}^{2}$ is the line element of $S^{D-1}$. The wormhole solution is supported by the flux of a $D-2$ form field, parameterized by a number $Q$, along the $S^{D-1}$ sphere. The solution exists for any value of $Q$ such that $0<Q<Q_{c}$, but $Q$ itself has discrete values due to flux quantization. $Q_{c}$ is the maximum value of $Q$, which is fixed by the cosmological constant and Newton's constant. The wormholes with flux parameter $Q_{c}$ have a constant value for $a(\tau)$, so they have the geometry of a cylinder. We call the wormholes with this extremal charge "Einstein wormholes"\footnote{The reason is that these solutions are Einstein static universes in Lorentzian signature.}, or just extremal wormholes. 

\begin{figure}[h!]
    \centering
    \includegraphics[width=1\linewidth]{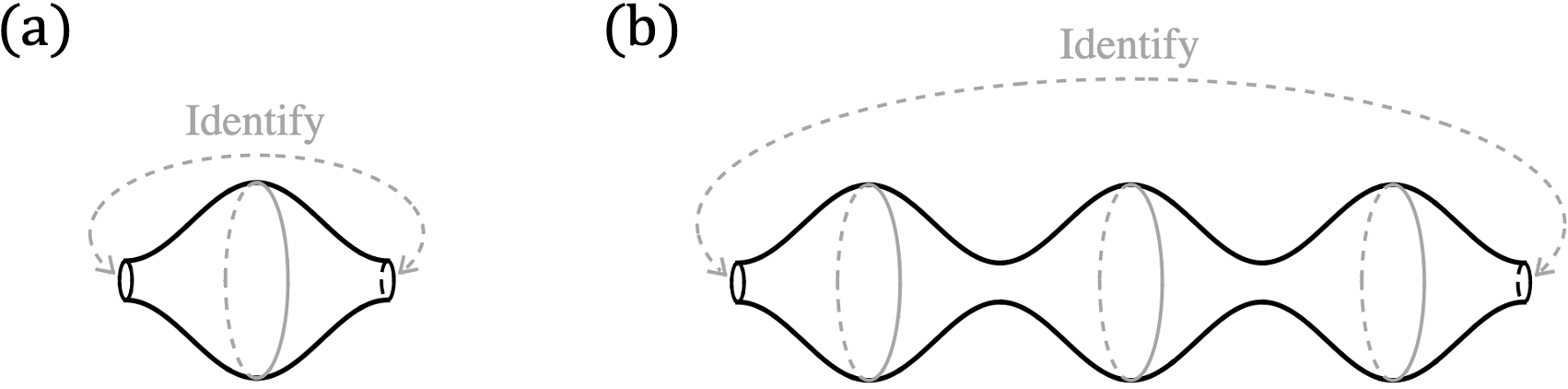}
    \caption{Illustration of the geometry of the Euclidean axion wormhole solution in de Sitter. The periodically identified direction is the $\tau$-direction, where in \textbf{(a)} the solution contains one fundamental period, and while in \textbf{(b)} it contains more than one fundamental period.}
    \label{wormholefig}
\end{figure}

Depending on how one analytically continues the wormhole solutions to Lorentzian signature, they could be interpreted as connected contributions to the density matrix of the universe, or equivalently, contributions to the wavefunction of two disconnected universes. While these Lorenzian continuations of the wormhole are interesting, here, we will be interested in the solutions with $S^{1} \times S^{D-1}$ topology. These solutions are obtained by taking the $\tau$ direction to be periodic, as illustrated in figure \ref{wormholefig}. 

This is possible because for $Q<Q_{c}$ the classical solutions are periodic in Euclidean time, $\tau$. This implies that one can choose the period of $\tau$ to be any integer multiple of a fundamental period, which is fixed by the classical equations of motion. The $Q=Q_{c}$ version of the solution is a bit more special. Since its geometry is that of a cylinder, the length of the $\tau$ direction can be any continuous value. 

In this paper, we study the one-loop determinant around these axion wormhole solutions, in both the $0<Q<Q_{c}$ and $Q=Q_{c}$ regimes. The spectrum of quadratic fluctuations around these wormholes is somewhat complicated, so to solve for it, we have to rely on numerics, which limits the amount of analytic results we can derive. However, this will be enough for the points we want to investigate in this paper. 

Our paper has two main points: \textbf{First}, we compute the phase of the Euclidean path integral, $Z_{\text{wormhole}}$, around this wormhole solution numerically\footnote{Note that the perturbative stability of this wormhole was studied in the original paper \cite{Aguilar-Gutierrez:2023ril}. However, the phase and this criterion of stability are two different properties of a saddle, and one should not mix the two up. The main difference is perhaps that, in \cite{Aguilar-Gutierrez:2023ril}, some components of the fluctuation field are integrated along an imaginary contour as Lagrangian multipliers. In our approach, the defining contour of all fields is along the real axis.}. To compute the phase, we use the prescription of \cite{Polchinski:1988ua, Maldacena:2024spf, Ivo:2025yek}, with a variation of a regularization proposed in \cite{Chen:2025jqm}. For other interesting papers that discuss the phase of the Euclidean path integral, see \cite{Anninos:2020hfj, Anninos:2021ene, Law:2025yec, Shi:2025amq, Law:2025ktz, Ivo:2025fwe, Ivo:2025xek, Giombi:2026sqa}. The expression for the phase of the wormhole is surprisingly simple. In a convention where every negative mode contributes to the phase with a factor of $(-i)$, the results divide as follows:
\begin{equation}
\label{whphase}
Z_{\text{wormhole}}\propto \begin{cases*}
    i^{2N}~~,~~\text{if $0<Q<Q_{c}$ and the wormhole has $N$ cycles}\\
    i^{1+2\lfloor N\rfloor},\text{if $Q=Q_{c}$ and $\tau \sim \tau+\frac{2\pi N}{\sqrt{2(D-1)}}$}
\end{cases*}
\end{equation}
where we use $\propto$ here to mean "same phase as". As discussed before, the $Q=Q_{c}$ wormhole has a continuum length parameter $N$ for the $\tau$ circle. It is a free modulus of this solution, which we need to fix by hand, or leave the integration over it to the end of the calculation.

\begin{figure}[h!]
    \centering
    \includegraphics[width=0.7\linewidth]{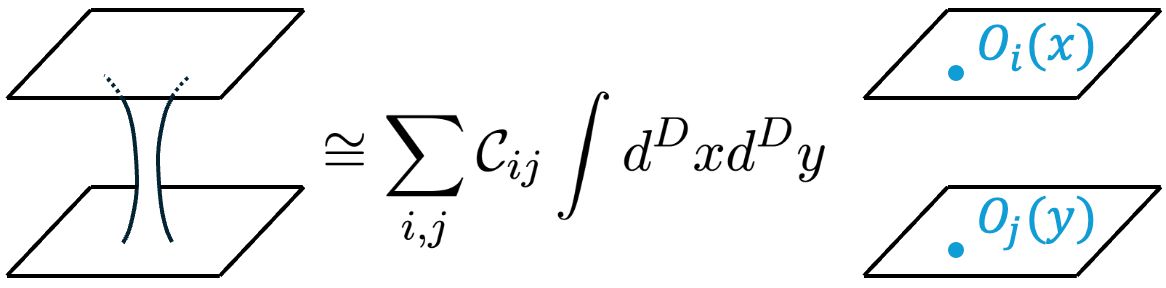}
    \caption{An illustration of the sum over bi-local operator insertions that reproduces the effect of the wormhole at large distances. The sum runs over all operators $O_{i,j}$ consistent with the symmetries of the problem, and the $\mathcal{C}_{ij}$ are EFT coefficients. The wormhole drawn on the left-hand side is supposed to stand for the path integral over the small wormhole and all its moduli. }
    \label{fig:whbilocal}
\end{figure}

Furthermore, we also provide a physical interpretation for the phase of the wormhole in \nref{whphase} in two different limits. For the first limit, when $Q$ is very small, the wormhole with $N=1$ cycles is a sphere with a small handle connecting the pode and antipode. We then propose that, using the ideas from \cite{Coleman:1988cy, Giddings:1988cx, Klebanov:1988eh, Witten:2026twr}, we can replace the effect of the wormhole at large distances by a sum over integrated local operator insertions in the spacetimes connected by the wormhole (see Figure \nref{fig:whbilocal}). 

The idea is that these local operator insertions mimic the effect of many states propagating through the wormhole. While the sum over insertions is infinite, it truncates quickly because higher-dimensional operators should be suppressed by powers of the scale of the wormhole. The most relevant contribution for us is therefore that of the bilocal operator with the lowest dimension. For the wormhole with $n$ units of magnetic flux, this bilocal operator is the product of two dimensionless\footnote{We will refer to these operators as "dimensionless" throughout the paper in the sense that their two-point functions are numbers, with no units. This is not a statement about the short-distance scaling behaviour of their two-point function.} magnetic operators $K_{\pm n}$ that create and destroy $n$ units of $D-1$ form flux, respectively (see \cite{Witten:2026twr}).

We would consider the full integral over the $x$ and $y$ positions of $K_{n}(x)K_{-n}(y)$ if we were interested in the full contribution of off-shell wormholes where the $x$ and $y$ positions are unfixed moduli. However, the solution we are interested in has definite positions for the wormhole mouths, which would be the pode and antipodes of $S^{D}$. Therefore, we propose that what actually corresponds to the on-shell wormhole we discuss is
\begin{equation}
\label{zwhop}
Z_{\text{wormhole}}\approx Z_{\text{GR}}(S^{D}) \times Z_{D-2}(S^{D})\times \mathcal{C}_{n,-n}\int d^{D}x\,d^{D}y\,\langle K_{n}(x)K_{-n}(y)\rangle|_{\text{antipodal saddle}}
\end{equation}
where $Z_{\text{GR}}(S^{D})$ and $Z_{D-2}(S^{D})$ are the one-loop path integrals of the pure gravity theory and the $D-2$ form theory in the sphere, respectively. The $\langle \rangle$ symbol stands for the expectation value of an operator in the sphere. $\mathcal{C}_{n,-n}$ is what we named the effective field theory (EFT) coefficient of this bilocal operator, which is fixed by matching the leading order terms of \nref{zwhop}. 

As indicated in \nref{zwhop}, what reproduces $Z_{\text{wormhole}}$ is not the full $x$, $y$ integral in \nref{zwhop}, but instead its contribution from near the extremum of the two-point function of $K_{\pm n}$ where $x$ and $y$ are antipodal points in the sphere. This is a saddle for the two-point function, because the operators are the furthest away they could be in $S^{D}$.

However, by expanding $x$ and $y$ away from the antipodal position, we find that the two-point function increases instead of decreasing, which means that this configuration is an unstable saddle. More specifically, fixing $x$ and displacing $y$, we find that the two-point function behaves as a wrong-sign Gaussian in the $D$ independent displacements of $y$. Since moving $x$ and $y$ together leaves the two-point function invariant, the $x$,$y$ integral has $D$ negative modes and the phase for the wormhole is
\begin{equation}
\label{whint}
Z_{\text{wormhole}}\sim i^{D+2}\times 1 \times (-i)^{D}=i^{2}
\end{equation}
where we used that the sphere partition function of pure gravity has a phase of $Z_{\text{GR}}(S^{D})\sim i^{D+2}$ \cite{Polchinski:1988ua, Maldacena:2024spf}, that the form path integral has no phase \cite{Ball:2024xhf}, and we assumed $\mathcal{C}_{n,-n}>0$. Therefore, we can understand the phase of the wormhole at small charge as coming from the fact that the dual leading operator's positions are fluctuating around an unstable extremum of their two-point function. In particular, perhaps the most interesting aspect of this exercise is that it tells us what the interpretation of the on-shell wormhole is, in terms of the effective operator description. 

The negative modes in the operator description seem, at least naively, incompatible with the proposal of \cite{Aguilar-Gutierrez:2023ril} that these wormholes are stable contributions to the path integral. However, we should mention that the same could be said for the saddle that prepares the maximal entropy state for an observer in de Sitter \cite{Witten:2023xze}. The reason is that, in this state, the trajectory for the observer in Euclidean signature is a great circle, with negative modes associated with sliding the geodesics from the maximal length curve \cite{Maldacena:2024spf}. Nevertheless, due to its interpretation, we expect the sphere with a great circle geodesic to be a well-defined contribution to the path integral. 

Another comment, before we move on, is that the wormhole with $N$ cycles has a similar explanation for its phase. Namely, it has the phase of $N$ spheres and $2N$ operator insertions, pairwise in antipodal positions in each sphere, which gives $(i^{D+2}(-i)^{D})^{N}=i^{2N}$, as we expected from \nref{whphase}.

The phase of the wormholes we discussed then continues the same as we increase the flux $Q$ because it has no reason to change, until the flux $Q$ reaches the maximum value $Q_{c}$. For $Q=Q_{c}$, the wormhole solution is qualitatively different, and we need a new analysis. For the extremal solution wormholes, with $Q=Q_{c}$, we can connect their phase with the classical physical instabilities of the solution in Lorentzian signature, following the idea in \cite{Ivo:2025yek}. In summary, perturbing the size of the $S^{D-1}$ in this Einstein static universe is an unstable perturbation that will increase exponentially in time \cite{Eddington:1930zz}. When treated as a linearized perturbation of this wormhole, this mode is a tachyon which will lead to a phase in the Euclidean signature computation\footnote{The situation is a bit more subtle, because this mode also has a wrong sign kinetic term. The effect of this for the phase is that this unstable mode will contribute with factors of $i$ instead of $(-i)$.}, precisely of the form in \nref{whphase}.

The \textbf{second} point of this paper is to properly set up the calculation of the one-loop determinant around the wormhole, and from this discuss many interesting features of its one-loop spectrum. While we do not attempt to directly evaluate the one-loop determinant using techniques such as the heat-kernel regularization, we leave the final answers in terms of many functional determinants and zero-mode contributions. The spectra of these functional determinants can be solved explicitly by numerics from the eigenvalue equations that we discuss in the paper, and we also discuss the appropriate zero-mode contributions.

In particular, we use this setup to estimate the EFT coefficient $\mathcal{C}_{n,-n}$ in \nref{zwhop} in terms of the parameters of the wormhole solution, in odd spacetime dimensions. To solve for $\mathcal{C}_{n,-n}$, we need to discuss the ratio of the wormhole path integral to the original sphere answer, which we discuss explicitly in section \nref{smwhopis}. 

To evaluate this ratio based only on the results we discussed in the paper, we need to propose a rule in subsection \nref{lochand} for evaluating functional determinants in spacetimes with small handles, based only on the low-lying spectrum of the functional. This is a reasonable rule for odd $D$, but the Weyl anomaly makes the analysis more complicated in even $D$. With this rule as an assumption, we are able to find, for odd $D$
\begin{equation}
\label{lbdint}
\mathcal{C}_{n,-n}\sim \bigg(\frac{n \kappa_{\text{phy}}}{f_{\theta,\text{phy}}}\bigg)^{-\frac{2D}{(D-2)}}\bigg(\frac{n}{\kappa_{\text{phy}} f_{\theta,\text{phy}}}\bigg)^{\frac{\frac{D(D-1)}{2}+2D}{2}}(n \kappa_{\text{phy}} f_{\theta,\text{phy}})^{-\frac{1}{2}}\exp\bigg(-\frac{2 \pi^{2} n}{\kappa _{\text{phy}}f_{\theta,\text{phy}}}\sqrt{\frac{D-1}{D-2}}-\Delta S_{\text{ct}}\bigg)
\end{equation}
with $\kappa_{\text{phy}}^{2}=8 \pi G_{N}$, $f_{\theta,\text{phy}}$ the coupling constant of the $D-2$ form theory, and $n$ the integer that parameterizes the quantized flux $Q$ of the wormhole. To see these definitions more cleanly, see our section \nref{prelim}. 

Note that we also included a term $\Delta S_{\text{ct}}$ in \nref{lbdint} because, in treating GR as an effective field theory, we should allow for a series of higher derivative corrections to the Einstein-Hilbert action. We call $\Delta S_{\text{ct}}$ the change of these corrections, between the wormhole spacetime and the sphere. Higher derivative terms are suppressed, but in higher dimensions, they can be more relevant in the $G_{N}$ expansion than the functional determinant contribution we computed. Also, as a matter of principle, what should be well-defined is the combination of the functional determinant we discussed and the action with appropriate counterterms. The structure of these counterterms, however, is strongly dimension-dependent, and they have to be discussed separately for whatever $D$ one is interested in.

While these terms are extra unknowns, we should note that the one-loop contribution in \nref{lbdint} seems to be intuitively reasonable. Let us briefly discuss where all the terms come from: The exponential term is the usual action of this wormhole, which is the main suppression of the EFT coefficients. The first term in \nref{lbdint} is the radius of the neck of the wormhole, to the power of $-2D$. One can view this term as being there for dimensional reasons, since $-2D$ is the length dimension of the EFT coefficient, and the size of the wormhole is its natural length scale.

Furthermore, there is a one-loop enhancement to $\mathcal{C}_{n,-n}$, which is the action of the wormhole to the power of half $2D+\frac{D(D-1)}{2}$. This enhancement looks like the usual contribution of zero-mode factors for instantons \cite{Coleman:1977py}. Indeed, this number is the number of moduli of the wormhole in flat space. To be more specific, the wormhole in flat space has $D+D$ moduli that correspond to moving the wormhole mouths in the ambient spacetime. The $\frac{D(D-1)}{2}$ is the dimension of $SO(D)$, and the moduli associated with it are a relative $SO(D)$ twist between the two wormhole mouths. The last term in \nref{lbdint} is a bit more mysterious, and comes from the ghost determinant of the $(D-2)$ form theory in the wormhole spacetime. This term would likely not be present for wormholes stabilized by scalar fields.

We should comment that in even $D$, we expect the answer to have a further correction because of the conformal anomaly. However, \nref{lbdint} still seems interesting to us as a proof of principle. The reason is that even though the mouths of axion wormholes, in the form description, are known to be dual to the magnetic operator $K_{\pm n}$ \cite{Coleman:1989zu, Witten:2026twr}, as far as we are aware the contribution from the one-loop determinant to the EFT coefficient in \nref{lbdint} was not computed before. It seems to us, in particular, that \nref{lbdint} provides the first example of such a wormhole EFT coefficient being computed to one-loop\footnote{There was an attempt to do a matching of this sort in the past by the authors of \cite{Coleman:1989zu}, where they studied wormholes stabilized by massive matter fields. There, they attempted to extract a normalization factor like the one we discussed in $D=4$ purely from the anomaly of the effective action $\Gamma(\phi)$ for gravity coupled to a scalar, computed in \cite{tHooft:1974toh}. We believe that while this anomaly is an important term in $D=4$, the zero and light modes that we discussed in the paper give a non-trivial contribution to the path integral as well. It is unclear if ignoring their contribution should lead to the correct answer in this specific problem.}. Of course, at our level of analysis, we were not able to obtain the $O(1)$ numerical prefactor in \nref{lbdint}, and that would be an interesting future direction. In particular, one would like to obtain \nref{lbdint} from a more honest heat kernel regularization computation, and understand how $\Delta S_{\text{ct}}$ might modify it for various values of $D$.

The paper is structured as follows: In section \nref{prelim} we review the classical solution for the wormhole, in particular what happens when the charge is at its critical value $Q=Q_{c}$. In section \nref{fluc} we discuss the general strategy to compute the one-loop determinant, in particular the important decomposition of all relevant fluctuations into $SO(D)$ harmonics. We also discuss the numerical result for the phase. In section \nref{lightzero} we discuss novel modes for the linearized fluctuations which exist in the wormhole spacetime and have no sphere counterpart. In particular, one of these modes has to do with moving with wormhole mouths with respect to each other, and is enough to explain the answer for the phase we obtained in \nref{whphase} for $Q<Q_{c}$. In section \nref{smwhopis}, we discuss the computation of the coefficient $\mathcal{C}_{n,-n}$ in \nref{lbdint} for the leading operators dual to the wormhole mouths. In section \nref{einwh} we discuss the one-loop spectrum of the Einstein wormhole more explicitly. In particular, we are able to find its spectrum analytically. We also discuss how to obtain the phase of this wormhole from the physical instabilities picture of \cite{Ivo:2025yek}. In section \nref{disc}, we end with some final remarks and interesting future directions.

\section{Classical solution}
\label{prelim}

We are going to consider the theory to be Einstein gravity with a positive cosmological constant $\Lambda$, and a minimally coupled $D-2$ form field $B_{a_{1}...a_{D-2}}$. The overall Euclidean action of the theory is
\begin{equation}
\label{action}
I_{E}=-\frac{1}{2\kappa^{2}}\int (R-2\Lambda)+\frac{1}{2(D-1)!}\int F_{a_{1}..a_{D-1}}F^{a_{1}..a_{D-2}}
\end{equation}
with $\kappa^{2}=8 \pi G_{N}$, $G_{N}$ the Newton's constant and $F=dB$. Also, we used the following notation
\begin{equation}
\int =\int d^{D}x\sqrt{g}
\end{equation}
which we will use consistently throughout the rest of the paper. 

Note that we also picked a convention where the action of the $D-2$-form is canonically normalized. In this convention, the flux quantization condition for $F$ will read
\begin{equation}
\label{fluxq}
\int F_{a_{1}...a_{D-1}}dx^{a_{1}}..dx^{a_{D-1}}=\frac{2 \pi n}{f_{\theta}}
\end{equation}
where $f_{\theta}$\footnote{The $\theta$ in the subscript is because the $(D-2)$ form theory is dual to a theory of 0-forms, which we call $\theta$. We will discuss this point more later.} is the coupling of the $D-2$-form theory. The cosmological constant $\Lambda$ is given by
\begin{equation}
\Lambda=\frac{(D-1)(D-2)}{2\lds^{2}}
\end{equation}
with $\lds$ the de Sitter length of the pure gravity solution of the theory. For convenience, from now on we will pick a convention where $\lds=1$. Note that this is achievable by starting from the original action \nref{action} with dimensionful parameters, and then rescaling the coordinates appropriately by $\lds$. 

Doing so, we obtain an action of the form \nref{action}, but with all fields and couplings dimensionless. The dimensionless couplings $\kappa$ and $f_{\theta}$ that we work with are then related to their dimensionful counterparts $\kappa_{\text{phy}}$ and $f_{\theta,\text{phy}}$ via
\begin{equation}
\label{coupscale}
\kappa^{2}=\frac{\kappa_{\text{phy}}^{2}}{\lds^{D-2}}~~,~~ f_{\theta}^{2}=f_{\theta,\text{phy}}^{2}\lds^{D-2}
\end{equation}
such that restoring the $\lds$ dependence later will be simple \footnote{One should note, however, that this rescaling can affect the one-loop determinant calculation later. This is because, for even $D$, the log divergences of the one-loop determinant are sensitive to this rescaling in a non-trivial way.}.

The equations of motion of the theory are
\begin{equation}
\begin{gathered}
R_{ab}=(D-1)g_{ab}+\frac{\kappa^{2}}{(D-2)!}F_{a a_{2}..a_{D-1}}\tensor{F}{_b^{a_{2}..a_{D-1}}}-\frac{\kappa^{2}}{(D-1)!}g_{ab}F_{c_{1}..c_{D-1}}F^{c_{1}..c_{D-1}}\\
\n^{a}F_{a a_{2}...a_{D-1}}=0
\end{gathered}
\end{equation}

The solutions we will be interested in are $SO(D)$ symmetric, with line element
\begin{equation}
ds^{2}=d\tau^{2}+a^{2}(\tau)d\Omega_{D-1}^{2}
\end{equation}
where $d\Omega_{D-1}^{2}$ is the line element of $S^{D-1}$. To stabilize this solution, we take the background form to have a non-trivial magnetic flux over the $S^{D-1}$ direction, with
\begin{equation}
F_{a_{1}...a_{D-1}}=\frac{Q}{a^{D-1}}\bar{\epsilon}_{a_{1}...a_{D-1}}
\end{equation}
where $\bar{\epsilon}$ is the volume form of $S^{D-1}$ normalized via $\bar{\epsilon}_{a_{1}..a_{D-1}}\bar{\epsilon}^{a_{1}..a_{D-1}}=(D-1)!$. The flux quantization condition \nref{fluxq} will then restrict $Q$ to respect
\begin{equation}
Q=\frac{2 \pi n}{\text{Vol}(S^{D-1})f_{\theta}}
\label{eq: quantization of Q}
\end{equation}

The classical background is then completely fixed by the Friedmann equation
\begin{equation}
\label{fried}
a'^{2}=1-a^{2}-\frac{\kappa^{2}Q^{2}}{(D-1)(D-2)a^{2(D-2)}}
\end{equation}
where we denote $'=\partial_{\tau}$. Equation \nref{fried} makes manifest that, up to an overall scaling with $\lds$, the wormhole solution is a function solely of the dimensionless quantity
\begin{equation}
\kappa Q=\frac{2 \pi n}{\text{Vol}(S^{D-1})}\frac{\kappa}{f_{\theta}}=\frac{2 \pi n}{\text{Vol}(S^{D-1})}\frac{\kappa_{\text{phy}}}{f_{\theta,\text{phy}}\lds^{D-2}}
\end{equation}

Equation \nref{fried} admits minimum and maximum values for the radius $a_{\pm}$ which are solutions to
\begin{equation}
\label{aminmax}
1-a_{\pm}^{2}=\frac{\kappa^{2}Q^{2}}{(D-1)(D-2)a_{\pm}^{2(D-2)}}
\end{equation}

The solutions to this equation are such that $a_{-} \rightarrow0$ and $a_{+}\rightarrow 1$ as $Q \rightarrow 0$. A useful way to think about the solutions is that if $Q=0$, the solution to \nref{fried} is $a(\tau)=\sin \big(\tau\big)$. This describes the metric of a round sphere, with $\tau=\frac{\pi}{2}$ the maximum of $a$, and $\tau=0$ and $\tau=\pi$ its two zeros. The zeros correspond to the pode and the antipode of the sphere.

The idea is then that turning on the magnetic flux will modify the Friedman equation \nref{fried} strongly near $a=0$, and will result in $a$ approaching $a_{-}$ instead of $a=0$ near the pode and antipodes. In this paper, we are going to study the classical compact manifolds obtained by identifying the endpoints of the $\tau$ interval along a few cycles of $a(\tau)$.

Since the solution is periodic, one can identify the solution after one oscillation cycle of $a(\tau)$ or after multiple cycles. The "single cycle" solution is the one that, as $Q \rightarrow 0$, has the geometry approaching that of a round sphere with one small handle. The solution with $N$ cycles, similarly, approaches that of $N$ round spheres, with neighbouring spheres connected via a small handle. 

To get more intuition, it is useful to discuss the metric of the handle region more explicitly when $\kappa Q$ is very small. At these small values of the charge, the scale factor of the metric $a(\tau)$ has a minimum size $a(\tau)=a_{-}=r_{0}$ fixed by equation \nref{aminmax}, which we can solve for approximately at small $\kappa Q$ as
\begin{equation}
\begin{gathered}
\label{rodef}
r_{0}^{(D-2)} \approx \frac{\kappa Q}{\sqrt{(D-1)(D-2)}}=\frac{2\pi n \kappa}{\sqrt{(D-1)(D-2)}\text{Vol}(S^{D-1})f_{\theta}}\\ \rightarrow r_{0,\text{phy}}^{(D-2)}\approx \frac{2\pi n \kappa_{\text{phy}}}{\sqrt{(D-1)(D-2)}\text{Vol}(S^{D-1})f_{\theta,\text{phy}}}
\end{gathered}
\end{equation}
where we also wrote down the physical length of the throat $r_{0,\text{phy}}=r_{0}\lds$. As expected, it does not depend on $\lds$, only on the gravitational and magnetic constants. 

In this regime, we can approximate the metric in the handle region as
\begin{equation}
\label{smwhreg}
ds^{2}\approx r_{0}^{2}\bigg[dx^{2}+\varrho^{2}(x)d\Omega_{D-1}^{2}\bigg]
\end{equation}
where $\varrho(x)$ satisfies
\begin{equation}
\bigg(\frac{d\varrho}{dx}\bigg)^{2}=1-\varrho^{-2(D-2)}
\end{equation}
and where $\varrho$ goes from $1$ to infinity. The useful point about the metric \nref{smwhreg} is that its only scale is $r_{0}$, and more importantly, it is the metric of a wormhole in flat space (see \cite{Witten:2026twr}). 

Therefore, the small charge solution consisting of $N$ cycles is composed of $N$ spheres connected cyclically by regions \nref{smwhreg}, which we can think of as being flat space wormholes. More specifically, they are wormholes that connect the pode and antipodes of the relevant spheres, with the spheres acting as the asymptotic regions of the wormholes.

However, we should also mention that the analytical solution to \nref{fried} for general $\kappa Q$ is only known (see \cite{Aguilar-Gutierrez:2023ril}) for $D=3$, so we have to solve for the background numerically at generic values of $D$\footnote{In the $D\rightarrow+\infty$ limit however, one can find a simple analytical solution to \eqref{fried}, for general $Q$. These geometries are round spheres cut and then patched together. See appendix \nref{app: large D} for more details.}. One can, however, note general properties of the solution without knowing its analytic form. For example, equation \nref{fried} does not have real periodic solutions if $Q$ is bigger than a critical value. On a technical level, the reason is that for $Q$ too large, one is no longer able to find real solutions to equation \nref{aminmax}. Physically, this is because there is a maximal amount of charge that a positive cosmological constant spacetime can hold. This fact is also true for charged black hole solutions. For AdS or flat space, $Q$ can be infinitely large. 

Therefore, there is an extremal solution where $Q$ is equal to its maximal, or critical, value, $Q_{c}$, and $a(\tau)$ is equal to a constant, with
\begin{equation}
Q^{2}=Q_{c}^{2}=\frac{(D-2)}{\kappa^{2}}\bigg(\frac{D-2}{D-1}\bigg)^{D-2}~,~\text{with} ~~ a^{2}(\tau)=a_{c}^{2}=\frac{(D-2)}{(D-1)}
\label{eq: Qc}
\end{equation}

We will refer to this solution as "Einstein wormhole", since in Lorentzian signature, it is an Einstein static universe. Sometimes we also call it just "extremal wormhole". This is an interesting solution due to its simplicity; that is, the geometry is just that of a cylinder. We also have analytic control of $a(\tau)$ for the near-extremal wormhole with $Q^{2}= Q_{c}^{2}(1-\epsilon)$ and small $\epsilon$, which can be solved for perturbatively in $\epsilon$. For more details, see Appendix \nref{eincl}.

Another feature of the Einstein wormhole is that, unlike the solutions with non-critical flux, the length of the $\tau$ direction does not need to be an integer multiple of a fundamental length. The reason is that the length of the wormhole is an unfixed modulus, which can take any value. Therefore, when we refer to the Einstein wormhole, we will generally mean the $Q=Q_{c}$ solution where we fixed the length of the $\tau$ circle, $L$, by hand. To obtain the overall contribution from these wormholes, one would have to integrate over the length $L$ appropriately later.

A useful way of thinking about all the classical wormhole solutions that we discussed is by using Hodge duality to write the $D-1$ field strength in terms of a 1-form. To do so, it is convenient to first fix the relation between the volume form of the manifold, which we denote $\epsilon_{a_{1}..a_{D}}$, and the volume form of $S^{D-1}$ we introduced before, as
\begin{equation}
\epsilon_{a_{1}...a_{D}}=D\,\bar{\epsilon}_{[a_{1}...a_{D-1}}(d\tau)_{a_{D}]}
\end{equation}

With this normalization, $\epsilon$ satisfies $\epsilon_{a_{1}..a_{D}}\epsilon^{a_{1}...a_{D}}=D!$. Using that, we define a dual field strength $\tilde{F}_{a}$ as
\begin{equation}
F_{a_{1}..a_{D-1}}=\epsilon_{a_{1}..a_{D}}\tilde{F}^{a_{D}}~~,~~ \tilde{F}_{a}=\frac{Q}{a^{D-1}}(d\tau)_{a}
\end{equation}

In terms of this dual variable, the stress tensor is instead
\begin{equation}
T_{ab}=\frac{1}{(D-2)!}F_{a a_{2}..a_{D-1}}\tensor{F}{_b^{a_{2}..a_{D-1}}}-\frac{1}{2(D-1)!}g_{ab}F_{c_{1}..c_{D-1}}F^{c_{1}..c_{D-1}}=-\tilde{F}_{a}\tilde{F}_{b}+\frac{1}{2}g_{ab}\tilde{F}^{c}\tilde{F}_{c}
\end{equation}

In the literature \cite{Hawking:1995ap, Aguilar-Gutierrez:2023ril}, it is often a common trick to think of this stress tensor as being minus the stress tensor of a massless scalar $\theta$, with $\tilde{F}_{a}=\n_{a}\theta$. Here we will not attempt to apply this trick off-shell, but it will often be less cumbersome for us to write formulas in terms of the background $\tilde{F}_{a}$ instead of the background $D-1$ field strength. 

In addition to this, throughout the paper, we will often think of the background $D-2$ form $B_{a_{1}..a_{D-2}}$ using instead a background two-form $\tilde{B}^{a_{D-1}a_{D}}$ defined via
\begin{equation}
\label{Btilde}
B_{a_{1}...a_{D-2}}=\frac{(-1)^{D-1}}{2}\epsilon_{a_{1}...a_{D-1}a_{D}}\tilde{B}^{a_{D-1}a_{D}}~~,~~ \tilde{F}_{a}=\n^{b}\tilde{B}_{ab}
\end{equation}

Last but not least, we should mention that the $D-2$ form theory we are studying in a given geometry is dual to a periodic scalar theory \cite{Donnelly:2016mlc, Witten:2026twr} via
\begin{equation}
I=\frac{1}{2(D-1)!}\int F_{a_{1}..a_{D-1}}F^{a_{1}..a_{D-1}} \longleftrightarrow \frac{1}{2}\int \n_{a}\theta\, \n^{a}\theta
\end{equation}
with $\theta$ a period scalar. Under the duality, the flux quantization condition \nref{fluxq} for $F$ maps into the fact that the scalar $\theta$ is periodic as $\theta \sim \theta+f_{\theta}$. We did not try to construct wormhole solutions from the dual scalar formulation because there are no wormhole saddles along the real axis of $\theta$. Perhaps one could bypass this issue by studying saddles with imaginary $\theta$, but to avoid possible drama, we restrict ourselves to the $(D-2)$ form formulation instead.  

\subsection{On-shell action}
\label{onactsec}

It will be useful for us to discuss the one-shell action of the wormhole solution. The action was discussed at some level in \cite{Aguilar-Gutierrez:2023ril}, where the authors studied it numerically and gave an approximate format for it in a general number of dimensions. They also found its analytic form in $D=3$. 

Here, we just wish to fill the gaps by finding the exact leading action perturbatively around $Q=0$ and $Q=Q_{c}$. We will evaluate the action for the $N=1$ solutions, and the solutions with $N$ cycles will have $N$ times that action. The expression for the on-shell action can be conveniently written as
\begin{equation}
\label{onact}
I_{E}=\text{Vol}(S^{D-1})\int d\tau\, a^{D-1}\bigg(-\frac{(D-1)}{\kappa^{2}}+\frac{Q^{2}}{a^{2(D-1)}}\bigg)
\end{equation}

The first term in \nref{onact} is proportional to the volume of the manifold, and the second term comes from the magnetic flux. We now discuss how to evaluate \nref{onact} for the relevant regimes of $Q$. 

\textbf{Small $Q$:} If $\kappa Q \approx 0$, the deviation of the action from the sphere answer is dominated by the $Q^{2}$ flux term in \nref{onact}. This term will be dominated by a small region, where the metric of the wormhole is described by \nref{smwhreg}. The overall scaling of this contribution with $Q$ will therefore be $\sim \frac{Q^{2}}{r_{0}^{D-2}}\sim \frac{Q}{\kappa }$. The correction to the volume term in \nref{onact} will be subleading to that, because in the bulk of the sphere, the corrections are of order $Q^{2}$, and the contribution from the handle region is of order $\kappa^{-2}r_{0}^{D}\sim \kappa^{-2}(\kappa Q)^{\frac{D}{(D-2)}}$, which is always a faster decay than linear for finite $D$. 

We can therefore compute the leading $Q$ correction to the action by computing the contribution from the flux term in \nref{onact}, on the region \nref{smwhreg}. Doing so, we find
\begin{equation}
\label{onactflat}
I_{E}\approx I_{S^{D}}+\pi \text{Vol}(S^{D-1})\sqrt{\frac{D-1}{D-2}}\frac{Q}{\kappa}=I_{S^{D}}+\frac{2 \pi^{2}n}{\kappa f_{\theta}}\sqrt{\frac{D-1}{D-2}}
\end{equation}
with $I_{S^{D}}$ the action of the pure gravity sphere saddle. Note that this matches the action in \cite{Aguilar-Gutierrez:2023ril} for $D=3$. In particular, since $\kappa f_\theta=\kappa_{\text{phy}}f_{\theta,\text{phy}}$, the correction to the action does not depend on $\lds$, which is what we expect from a wormhole in flat space. We also checked that \nref{onactflat} is a good numerical approximation to \nref{onact} at small $\kappa Q$.

\textbf{Near extremal $Q$:} If $Q=Q_{c}$, the action of the wormhole is zero, so it is interesting to discuss how the action of the wormhole approaches zero as $Q \rightarrow Q_{c}$. To be more specific, for $Q^{2}=Q_{c}^{2}(1-\epsilon)$ we can find a solution for $a(\tau)$ perturbatively in $\sqrt{\epsilon}$, as we did in Appendix \nref{eincl}. On this perturbative solution, $a$ is constant to leading order, and has oscillations with amplitude $O(\sqrt{\epsilon})$. 

It is then straightforward to derive the action at $O(\epsilon)$ by using the equations of motion to rewrite \nref{onact} as
\begin{equation}
\label{actein}
I_{E}=-(D-1)(D-2)\text{Vol}(S^{D-1})\int d\tau\, a^{D-3}a'^{2}\approx -\frac{\pi(D-1)\epsilon}{\sqrt{2(D-1)}\kappa^{2}}\bigg(\frac{D-2}{D-1}\bigg)^{\frac{(D-1)}{2}}\text{Vol}(S^{D-1})
\end{equation}
where in the second line we used equation \nref{neareina} for the approximate solution of $a(\tau)$ near $Q=Q_{c}$. The action \nref{actein} matches the $D=3$ action in \cite{Aguilar-Gutierrez:2023ril} after we relate $\epsilon$ to $Q$ appropriately.

\section{One-loop generalities and numerics}
\label{fluc}

\subsection{Generalities and setup}
\label{genset}

We would like to understand the one-loop determinant around the solutions we described in section \nref{prelim}. To do so, we need to set up the formalism we are going to use to compute the one-loop spectrum. Denoting the field fluctuations collectively by $\Phi$, and the gauge transformations collectively as $\alpha$, the one-loop determinant we want to compute is a path integral of the following form
\begin{equation}
\label{z1loop}
Z_{1-\text{loop}}=\frac{\int D\Phi\,e^{-\delta_{2}I_{E}}}{\int D\alpha}
\end{equation}
with $\delta_{2}I_{E}$ the quadratic action for the fluctuations around the saddle, and $D\Phi$ the measure over fluctuations. We divide by the path integral over the group of gauge symmetries, $\int D\alpha$, to be discussed more later. 

The strategy to compute the one-loop determinant is then as follows: First, one picks a local norm $(\cdot,\cdot)$ on the space of fluctuations. The local norm will then fix a unique local measure $\int D\Phi$ for the path integral, up to an overall factor that can be absorbed into local counterterms.

However, it is convenient to fix a specific normalization for the measure, which we do by imposing the following relation
\begin{equation}
\label{normnor}
\int D\Phi\,e^{-\frac{1}{2}(\Phi,\Phi)}=1
\end{equation}

The next step is then to compute the quadratic action of fluctuations around the classical background, $\delta_{2}I_{E}$, and to write it in terms of a hermitian fluctuation operator $M_{0}$ as
\begin{equation}
\delta_{2}I_{E}=\frac{1}{2}(\Phi,M_{0}\Phi)
\end{equation}

We denote the fluctuation operator by $M_{0}$ because the theory we are considering has some gauge redundancies that need to be dealt with before we can use the action to define a usual fluctuation operator. To do so, we have to first discuss more explicitly the specific problem we want to solve.

For the problem we are interested in, the linearized fluctuations are those of the metric $\delta g_{ab}$ and the $D-2$ form $\delta B_{a_{1}....a_{D-2}}$. For convenience, we will parameterize these fluctuations as
\begin{equation}
\delta g_{ab}=2\kappa h_{ab}~~,~~ \delta B_{a_{1}...a_{D-2}}=\frac{\sqrt{2}}{2}(-1)^{D-1}\epsilon_{a_{1}...a_{D-2}a_{D-1}a_{D}}b^{a_{D-1}a_{D}}
\end{equation}
where we introduced the 2-form $b_{ab}$ because it is generally simpler to think in terms of 2-forms instead of $D-2$ forms. Furthermore, we take the local norm for the fluctuations to be given by\footnote{We also discuss more general local norms in Appendix \nref{eigensec}.}
\begin{equation}
\label{locnorm}
(\Phi,\Phi)=\frac{1}{4\kappa^{2}}\int\delta g_{ab}\delta g^{ab}+\frac{1}{(D-2)!}\int \delta B_{a_{1}..a_{D-2}}\delta B^{a_{1}...a_{D-2}}=\int (h_{ab}h^{ab}+b_{ab}b^{ab})
\end{equation}

Then, we need to find the quadratic action for the fluctuations. Using the background solution to simplify some terms, the quadratic action around the classical solution, $\delta_{2}I_{E}$, can be written as
\begin{equation}
\begin{gathered}
\label{qdact}
\delta_{2}I_{E}=\frac{1}{2}\int \tilde{h}^{ab}\big(-\n^{2}h_{ab}+2\n_{(a}\n^{c}\tilde{h}_{b)c}-2\tensor{R}{_a^c_b^d}h_{cd}+2\kappa^{2}h_{c(a}\tensor{q}{_{b)}^c}-2\kappa^{2}q_{ab}h)\\
+\sqrt{2}\kappa\int \tilde{h}^{ab}(\tilde{F}_{a}f_{b}+\tilde{F}_{b}f_{a})+\int \n^{b}b_{ab}\n_{c}b^{ac}
\end{gathered}
\end{equation}
where we defined the tensors
\begin{equation}
q_{ab}=\tilde{F}_{a}\tilde{F}_{b}~~,~~ q=g^{ab}q_{ab}~~,~~ f_{a}=\n^{b}b_{ab}\text{, and}~~ h=g^{ab}h_{ab}~~, ~~
\tilde{h}_{ab}=h_{ab}-\frac{1}{2}g_{ab}h
\end{equation}

From here, one would like to write this action as the expectation value of a hermitian operator $M$ in the inner product associated with the norm \nref{locnorm}. However, as we discussed before, this operator would have infinitely many zero modes because of the gauge symmetry of the theory.

To be more specific, since we are working with gravity coupled to a $D-2$ form, the theory should be invariant under coordinate transformations and gauge transformations of the $D-2$ form. The diffeomorphisms $\xi^{a}$ act on the fluctuations as
\begin{equation}
\label{diffact}
h_{ab}=\frac{1}{\sqrt{2}}(\n_{a}\xi_{b}+\n_{b}\xi_{a})~~,~~ b^{ab}=\kappa(\n_{c}(\xi^{c}\tilde{B}^{ab})-(\n_{c}\xi^{a})\tilde{B}^{cb}-(\n_{c}\xi^{b})\tilde{B}^{ac})
\end{equation}
where to find the $b_{ab}$ transformation, we took the Hodge dual of the coordinate transformation of $B_{a_{1}...a_{D-2}}$, and we used the background field $\tilde{B}^{a_{1}a_{2}}$ defined in \nref{Btilde}. 

We express the $D-2$ form gauge transformation in terms of the original fluctuation $\delta B_{a_{1}..a_{D-2}}$, and a $D-3$ form $C_{a_{1}a_{2}...a_{D-3}}$ as
\begin{equation}
\label{formact}
\delta B_{a_{1}...a_{D-2}}=(dC)_{a_{1}..a_{D-2}}=(D-2)\n_{[a_{1}}C_{a_{2}..a_{D-2}]}
\end{equation}

The motivation for doing so is that this version of the transformation is more convenient for the ghost determinant calculation, discussed in Appendix \nref{ghostdet} and \nref{formdet}. Of course, one can analogously see \nref{formact} as a gauge transformation acting on $b_{ab}$, parameterized by a three-form which is the Hodge dual of $C_{a_{1}..a_{D-3}}$. 

We should also mention that there could be discrete gauge symmetries associated with both \nref{diffact} and \nref{formact}. Here, we will focus on gauge symmetries connected to the identity, assuming that the discrete part is taken into account later\footnote{The discrete part would be relevant for the wormhole with $N$ cycles we discussed in section \nref{prelim}, since this wormhole has a cyclic permutation symmetry.}.

To fix the gauge redundancy, we choose gauge fixing conditions, and we get rid of almost all of the gauge redundancies by inserting an appropriate ghost determinant in the path integral, and associated gauge fixing terms in the action. The relevant gauge conditions are
\begin{equation}
\label{gfcond}
P_{a}=-\sqrt{2}\n^{b}\tilde{h}_{ab} ~~\text{, and }~~ L_{a_{1}..a_{D-3}}=-\n^{c}\delta B_{c a_{1}..a_{D-3}}~~ 
\end{equation}

Having established the gauge fixing conditions $P_{a}$ and $L_{a_{1}..a_{D-3}}$, we can introduce gauge fixing terms in the action \nref{qdact} to make the fluctuation operator well defined. To be more specific, we add
\begin{equation}
\label{gfterm}
I_{gf}=\frac{1}{2}\int P_{a}P^{a}+\frac{1}{2(D-3)!}\int L_{a_{1}..a_{D-3}}L^{a_{1}..a_{D-3}}=\int \n^{b}\tilde{h}_{ab}\n_{c}\tilde{h}^{ac}+\frac{1}{6}\int (db)_{abc}(db)^{abc}
\end{equation}
with
\begin{equation}
(db)_{abc}=3\n_{[a}b_{bc]}
\end{equation}

For more details on the gauge fixing procedure and the associated ghost determinant, see Appendix \nref{ghostdet}. Also, note that while we use the specific gauge fixing term \nref{gfterm} in the main text, we discuss more general gauge fixing terms in Appendix \nref{altgf}, which we will use to check gauge fixing invariance of many results.

The overall action with the gauge fixing term is
\begin{equation}
\begin{gathered}
I_{E}^{(2)}=\delta_{2}I_{E}+I_{gf}=\frac{1}{2}\int \tilde{h}^{ab}\big(-\n^{2}h_{ab}-2\tensor{R}{_a^c_b^d}h_{cd}+2\kappa^{2}h_{c(a}\tensor{q}{_{b)}^c}-2\kappa^{2}q_{ab}h)\\
+\sqrt{2}\kappa\int \tilde{h}^{ab}(\tilde{F}_{a}f_{b}+\tilde{F}_{b}f_{a})+\frac{1}{2}\int b^{ab}\Delta_{2} b_{ab}
\end{gathered}
\end{equation}
with $\Delta_{2}$ the Hodge Laplacian of 2-forms in the manifold, given by
\begin{equation}
\Delta_{2} b_{ab}=-\n^{2}b_{ab}+\tensor{R}{_a^c}b_{cb}+\tensor{R}{_b^c}b_{ac}-\tensor{R}{_a_b^c^d}b_{cd}
\end{equation}

Using this gauge fixed action, we can write a well-defined eigenvalue problem. That is, taking the action of the form
\begin{equation}
\label{actm}
I_{E}^{(2)}=\frac{1}{2}(\Phi,M\Phi)~~\text{,with}~~ \Phi=\begin{pmatrix}
    h_{ab}\\b_{ab}
\end{pmatrix}
\end{equation}
the eigenvalue equation $M\Phi=\lambda \Phi$ is then equivalent to
\begin{equation}
\begin{gathered}
\label{eigeneq}
\lambda\bigg(h_{ab}-\frac{1}{D-2}g_{ab}h\bigg)=(-\n^{2}h_{ab}-2\tensor{R}{_a^c_b^d}h_{cd}+2\kappa^{2}h_{c(a}\tensor{q}{_{b)}^c}-2\kappa^{2}q_{ab}h)+\sqrt{2}\kappa(\tilde{F}_{a}f_{b}+\tilde{F}_{b}f_{a})\\
\lambda b_{ab}=\Delta b_{ab}+\sqrt{2}\kappa(\n_{a}(\tilde{h}_{bc}\tilde{F}^{c})-\n_{b}(\tilde{h}_{ac}\tilde{F}^{c}))
\end{gathered}
\end{equation}

Having considered these points, the one-loop determinant will be given by
\begin{equation}
\label{1lpsum}
Z_{1-\text{loop}}=Z_{\text{ghost}}\times Z_{M,0}\times \, \text{Det}' M^{-\frac{1}{2}}
\end{equation}
where the $\text{Det}'$ notation stands for the functional determinant of an operator with zero modes removed. The $Z_{M,0}$ term is the contribution from residual zero modes of $M$, which needs to be discussed separately from the functional determinant. We discuss these zero modes in more detail in section \nref{zeromod}. The $Z_{\text{ghost}}$ term is the contribution from the gauge fixing ghosts, which we discuss in more detail in Appendix \nref{ghostdet}.

It is useful to summarize the structure of all the terms in \nref{1lpsum}. First, we argue in Appendix \nref{decgag} that the ghost determinant $Z_{\text{ghost}}$ decouples into the product of two terms: The first term is a ghost determinant for the gauge condition $L_{a_{1}..a_{D-3}}$ in \nref{gfcond} with respect to the gauge transformation in \nref{formact}, which we discuss in Appendix \nref{formdet}. This ghost determinant is built out of the functional determinant of many Hodge Laplacians $\Delta_{p}$, and their associated zero mode contributions (see \cite{Donnelly:2016mlc}). The second term is a ghost determinant for the de Donder gauge condition $P^{a}$ in \nref{gfcond} with respect to the gauge transformation \nref{diffact}, which we discuss in Appendix \nref{ddgh}. Its overall contribution is a functional determinant of the variation of $P^{a}$ with respect to the $\xi^{a}$ transformation, and a term that comes from the division by the $SO(D)$ isometries of the background, which we call $Z_{\text{iso}}$. These can be seen as a contribution from ghost zero modes.

The residual zero modes of $M$, in $Z_{M,0}$, consist of two types of modes: The first set of possible zero modes corresponds to harmonic $D-2$ forms (or equivalently $2$ forms) of the manifold. These contributions only exist for $D=3$, and the path integral over them is compact. However, we can show that when they exist, they always cancel another extra contribution from the ghost determinant (see \nref{ghostL}). The second set of zero modes corresponds to an integral over possible twists on the wormhole. That is, there are other wormhole solutions connected to the one we introduced, where one identifies the wormhole coordinates over a period as $(\tau,x)\sim (\tau+L, U_{\text{iso}} \cdot x)$. $L$ is the length of the wormhole's $\tau$ direction, $x$ an $S^{D-1}$ coordinate and $U_{\text{iso}}$ an $SO(D)$ group element. The wormhole we discussed in \nref{prelim} has a trivial twist $U_\text{iso}$, but in principle, one should integrate over all possible twists. We call the contribution from this integral $Z_{\text{twist}}$.

Putting all these pieces together, we can write a more explicit formula for the one-loop determinant of the wormhole as
\begin{equation}
\label{1lpend}
Z_{1-\text{loop}}=Z_{\text{twist}}Z_{\text{iso}}|\text{Det}'(\delta_{\xi}P)|\frac{\text{Det}'(M^{-\frac{1}{2}})\text{Det}'(\Delta_{0}^{-\frac{1}{2}})}{\text{Det}'(\Delta_{D-2}^{-\frac{1}{2}})}{}\text{det}^{\frac{1}{2}}\bigg(\frac{f_{\theta}^{2}}{2 \pi}\Gamma_{0}\bigg){}\text{det}^{-\frac{1}{2}}\bigg(\frac{f_{\theta}^{2}}{2 \pi}\Gamma_{1}\bigg)e^{S_{\text{ano},\text{em}}}
\end{equation}
with
\begin{equation}
\tensor{(\delta_{\xi}P)}{^b_a}=-\delta_{a}^{b}\n^{2}-\tensor{R}{^b_a}
\end{equation}
such that $|\text{Det}'(\delta_{\xi}P)|$ is the functional determinant from the de Donder gauge condition. The absolute value is because it comes from a ghost determinant\cite{Polchinski:1988ua}. We did not consider the effect of discrete gauge symmetries, and they should be included later in the calculation. The new $e^{S_{\text{ano},\text{em}}}$ term is a necessary anomaly contribution, because to compute the ghost determinant for the form gauge condition in Appendix \nref{formdet}, and thus derive \nref{1lpend}, we used electromagnetic duality. The duality has a relative anomaly if $D$ is even, but $S_{\text{ano},\text{em}}$ is zero otherwise (see Appendix \nref{formdet} and \cite{Donnelly:2016mlc} for details on $S_{\text{ano},\text{em}}$). 

Let us further discuss the contributions in \nref{1lpend}. We use $\Delta_{p}$ to stand for the Hodge Laplacian of a $p$-form. The $\Gamma_{p}$ stands for an inner product matrix between Harmonic $p$-forms of the manifold, and $\text{det}$ is their determinant.\footnote{We use $\text{det}$ for regular determinant, and $\text{Det}$ for functional determinant.}. These contributions come from the ghost determinant for forms that we discussed in Appendix \nref{formdet}. The $\Gamma$'s are generally present to give the contribution of zero modes in the path integral over forms. The $\Gamma$'s in \nref{1lpend} are non-trivial in our problem, because every manifold has one and only one harmonic zero-form, which corresponds to constant functions. These enter $\Gamma_{0}$. Furthermore, the wormhole we study has a harmonic 1-form associated with the $S^{1}$ direction, so $\Gamma_{1}$ is non-trivial. In particular, both $\Gamma_{0}$ and $\Gamma_{1}$ are one-dimensional in our context, and therefore contribute in a simple way.

While equation \nref{1lpend} is useful to keep in mind, we will not actually need it for the rest of this section\footnote{Equation \nref{1lpend} will be essential for section \nref{smwhopis}, however.}. Throughout the rest of this section, we will instead focus on discussing how one would compute the $\text{Det}'M^{-\frac{1}{2}}$ piece of \nref{1lpend}. Since this is the functional determinant of $M$, we have to discuss how to solve the eigenvalue problem \nref{eigeneq}, which is what we will do for the rest of this section. However, $M$ has negative modes, and one should discuss how to deal with them. We will discuss this in more detail in the next subsection.

\subsection{Negative modes and prescription for phase}

Since we are dealing with Euclidean gravity, there are additional technical problems, because the fluctuation operator $M$ in \nref{eigeneq} will have infinitely many negative modes \cite{Gibbons:1978ac}. We will deal with the negative modes using the prescription from \cite{Polchinski:1988ua, Maldacena:2024spf, Ivo:2025yek}, where we treat each negative mode as a wrong sign gaussian, such as
\begin{equation}
\label{wrongs}
\int dx\, e^{(1-i\epsilon)x^{2}}
\end{equation}
that we can regulate by rotating the contour of integration of $x$ while avoiding the direction of maximum increase, via $x \rightarrow (-i)x$. 

Note that for this prescription to be well defined, we need the wrong sign Gaussians to be slightly complex, as in the integral \nref{wrongs}. We can implement this in an otherwise real one-loop determinant calculation, by taking $\hbar$ to be slightly complex as
\begin{equation}
\frac{1}{\hbar} \rightarrow \frac{1}{\hbar}(1-i\epsilon)
\end{equation}
which is what we assume was done throughout the rest of the paper.

Under these assumptions, each negative mode of the one-loop spectrum will contribute to the one-loop determinant with a factor of $(-i)$. The one-loop determinant will therefore have an overall phase which is $(-i)^{n_{E}}$, where $n_{E}$ is the number of negative modes in the spectrum of $M$. However, as we discussed, $n_{E}$ will formally be infinite.

To regularize this phase, a key idea is to note that the infinity in $n_{E}$ goes roughly as "one mode per point in spacetime"\footnote{The reasoning is that at small wavelengths, the pure trace part of the metric $h_{ab}\sim g_{ab}\psi$ decouple from the other modes, and becomes effectively a massless field with negative sign kinect term. Therefore, at small wavelengths, these modes will contribute one negative mode for each eigenvalue of $-\n^{2}$.}. Therefore, if we formally denote the "number of points in spacetime" by $n_{\text{sp}}$, the difference $n_{-}=n_{E}-n_{\text{sp}}$ is a candidate for a better behaved quantity. In fact, we can compute $n_{-}$ in a lattice regularized version of the problem where both $n_{E}$ and $n_{\text{sp}}$ are large but finite\footnote{See \cite{Chen:2025jqm} for a concrete realization of this idea.}.

To leverage this idea, we note that since $(-i)^{n_{\text{sp}}}$ is a factor of $(-i)$ per point in spacetime, we can think of it as a local contribution that can be absorbed into counterterms of the action. Assuming we can do that, we can compute the finite part of $(-i)^{n_{E}}$ by rewriting it as
\begin{equation}
\label{phasestep}
(-i)^{n_{E}}=(-i)^{n_{E}-n_{\text{sp}}}(-i)^{n_{\text{sp}}}\sim (-i)^{n_{E}-n_{\text{sp}}}
\end{equation}
with the symbol $\sim$ meaning ``has the same phase as'', and where we assumed that the factor of $(-i)^{n_{\text{sp}}}$ was appropriately dealt with by being absorbed into local counterterms \cite{Polchinski:1988ua}. Therefore, to compute the phase we can equivalently compute $(-i)^{n_{-}}$, with $n_{-}$ defined as
\begin{equation}
\label{nminus}
n_{-}=(\text{number of negative modes})-\text{(number of points in spacetime)}
\end{equation}

As we mentioned before, we can compute $n_{-}$ in a lattice regularization of the background spacetime as in \cite{Chen:2025jqm}. However, such a position-based regularization is challenging in higher-dimensional manifolds. 

To avoid doing that, here we will use a slightly different strategy to compute $n_{-}$. To understand the strategy, note that $n_{\text{sp}}$ can be more generally thought of as being the number of basis elements of a scalar mode $\phi(x)$ expanded in the regularized spacetime. The position basis is a specific instance of such a basis.

We could then, instead, think of $n_{\text{sp}}$ as the number of basis eigenmodes of a properly chosen scalar differential operator $G$. If the differential operator preserves the $SO(D)$ invariance of the background, its space of eigenmodes will be a direct sum over different angular momentum harmonics of $S^{D-1}$. 

Therefore, the number of eigenmodes of the differential operator $G$ will be a sum over its number of eigenmodes, $n_{l}$, over each angular momentum sector $l$ as
\begin{equation}
\label{nsp}
n_{\text{sp}}=\sum_{l}n_{l}
\end{equation}

At this level, this is just a formal relation since both $n_{\text{sp}}$ and $n_{l}$ are individually infinite. The key point, however, is that we can see the differential operator $G$ restricted to each angular momentum sector, as a radial differential operator $G_{l}$ along the $\tau$ direction.

Therefore, in a finite regularization of the problem where we define the $\tau$ direction using a discrete grid, the number $n_{l}$ will be finite. More specifically, $n_{l}$ will be the degeneracy of the $l$ sector scalar spherical harmonic, times the number of grid points of the $\tau$ direction.

This is a useful fact for us, because this is precisely the discretization that we will use to solve the eigenvalue equation \nref{eigeneq}, and the subtraction of negative modes by $n_{\text{sp}}$ will be important there. In particular, since $n_{\text{sp}}$ has the same degrees of freedom as one scalar, the $n_{-}$ as defined in \nref{nminus} tell us that the contribution to $n_{-}$ coming from scalar negative modes should come with the subtraction by \nref{nsp}. Therefore, the contribution from non-scalars should not come with the subtraction, since we only want to subtract by $n_{\text{sp}}$ once.

Before we discuss more specifics, we will first simplify the eigenvalue problem \nref{eigeneq} further by decomposing the fluctuation fields into many different representations of the background $SO(D)$ symmetry. This will allow us to, in particular, identify the relevant scalar fluctuations.

\subsection{Decomposition into representations of $SO(D)$}
\label{repsod}

To simplify \nref{eigeneq} we use the fact that the background is $SO(D)$ invariant. This implies, in particular, that fluctuation fields that transform as different representations of $SO(D)$ do not mix. 

Before we decompose the fluctuation fields, it is convenient to define the following decomposition of the metric
\begin{equation}
\label{decomp}
g_{ab}=n_{a}n_{b}+\gamma_{ab}~~\text{, with}~~ n_{a}=(d\tau)_{a}~~,~~ \gamma_{ab}=a^{2}(\tau)\hat{\gamma}_{ab}
\end{equation}
with $\hat{\gamma}_{ab}$ the metric of $S^{D-1}$. Using this decomposition, we can define a notion of tangent vectors. 

To be more specific, we call a vector field $v_{a}$ tangent if it is orthogonal to $n^{a}$ everywhere, e,g, $n^{a}v_{a}=0$. Analogously, a tensor $\tensor{v}{_{a_{1}..a_{p}}^{b_{1}..b_{q}}}$ is tangent if every possible contraction of it with $n^{a}$ is zero.

Furthermore, it is also convenient to define a notion of tangent covariant derivative for scalars and tangent tensors. To do so, take the covariant derivative $\n_{a}$ of the background. Then, we define the tangent covariant derivative, $\mcD_{a}$, of a tangent tensor $\tensor{v}{_{a_{1}..a_{p}}^{b_{1}..b_{q}}}$ as follows
\begin{equation}
\mcD_{a}\tensor{v}{_{a_{1}..a_{p}}^{b_{1}..b_{q}}}=\tensor{\gamma}{_{a}^{c}}\tensor{\gamma}{_{a_{1}}^{c_{1}}}...\tensor{\gamma}{_{a_{p}}^{c_{p}}}\tensor{\gamma}{_{d_{1}}^{b_{1}}}...\tensor{\gamma}{_{d_{q}}^{b_{q}}}\n_{c}\tensor{v}{_{c_{1}..c_{p}}^{d_{1}..d_{q}}}
\end{equation}

One should think of $\mcD_{a}$ as the local covariant derivative of the $S^{D-1}$ factor, e.g, of a $(D-1)$ dimensional sphere with local radius $a(\tau)$. Using the tangent covariant derivative, we can define many different tensor structures that behave differently with respect to $SO(D)$. 

For example, one can define a notion of transverse tangent vectors and tensors. We will call a tensor $v_{a_{1}..a_{p}}$ tangent and transverse if, on top of being tangent, it satisfies $\mcD^{a}v_{a_{1}a_{2}..a..a_{p}}=0$ for every possible contraction of $\mcD$ with $v$. If $v$ is completely symmetric or antisymmetric, requiring this quantity to be zero for a given contraction implies it is zero for every possible contraction.

We also use the decomposition of the metric \nref{decomp} to define the following tensors
\begin{equation}
P_{ab}=n_{a}n_{b}-\frac{1}{(D-1)}\gamma_{ab}~~,~~\mcD^{2}=\gamma^{cd}\mcD_{c}\mcD_{d}~~,~~ \mcD_{ab}=\mcD_{a}\mcD_{b}-\frac{1}{(D-1)}\gamma_{ab}\mcD^{2}
\end{equation}

With all these points into consideration, we decompose the metric and two-form fluctuations $h_{ab}$ and $b_{ab}$ as
\begin{equation}
\begin{gathered}
\label{decompf}
h_{ab}=g_{ab}\psi+P_{ab}A+2n_{(a}\mcD_{b)}f+\mcD_{ab}\eta+2n_{(a}v_{b)}+2\mcD_{(a}x_{b)}+\phi_{ab}\\
b_{ab}=2n_{[a}\mcD_{b]}\chi+2n_{[a}u_{b]}+2\mcD_{[a}\omega_{b]}+j_{ab}
\end{gathered}
\end{equation}
where $\{\psi,A,f,\eta,\chi\}$ are scalars, $\{v_{a},x_{a},u_{a},\omega_{a}\}$ are tangent tranverse vectors, $\phi_{ab}$ is a tangent traceless transverse symmetric two tensor, and $j_{ab}$ is a tangent transverse two-form. These 4 different groups of fields come into different representations of $SO(D)$, and therefore have eigenvalue problems decoupled from each other. Another property of \nref{decompf} is that all the fields in the decomposition are orthogonal to each other under the ultralocal norm \nref{locnorm} (and its generalization in \nref{eq: De Witt parameter}).

Now, we discuss some important details of the decomposition. First, we note that we can further decompose all the modes into appropriate $SO(D)$ spherical harmonics. $SO(D)$ invariance will then imply that different harmonics do not mix, and we will solve the eigenvalue equation harmonic by harmonic. The spherical harmonics for the 4 types of field structures that we are interested in are as follows.

\textbf{Scalars:} The scalar spherical harmonics, which we refer to by $Y_{l}$, are the eigenmodes of the local $S^{D-1}$ laplacian with eigenvalues
\begin{equation}
\label{scharm}
-\mcD^{2}Y_{l}=\frac{l(l+D-2)}{a(\tau)^{2}}Y_{l}
\end{equation}
with $l$ the angular momentum number, which is a non-negative integer. The factor of $a^{-2}$ in the eigenvalue equation comes from the local size of the $S^{D-1}$ factor. Also, note that these spherical harmonics should also be labelled by appropriate magnetic quantum numbers. However, the magnetic quantum numbers will not affect the eigenvalue problem we are interested in, so we will suppress them throughout the paper.

Having considered these points, we can expand the five scalars $\{\psi, A, f, \eta, \chi\}$ as
\begin{equation}
\begin{gathered}
\label{scexp}
\psi=\sum_{l=0}^{\infty}\psi_{l}(\tau)Y_{l} ~~,~~ A=\sum_{l=0}^{\infty}A_{l}(\tau)Y_{l}~~,~~
f=\sum_{l=1}^{\infty}f_{l}(\tau)Y_{l}\\ \eta=\sum_{l=2}^{\infty}\eta_{l}(\tau)Y_{l}~~,~~
\chi=\sum_{l=1}^{\infty}\chi_{l}(\tau)Y_{l}
\end{gathered}
\end{equation}
with the sum over magnetic quantum numbers implicit. In terms of this decomposition, the eigenvalue problem becomes a one-dimensional problem for the scalar fields with a given value of $l$. Note also that for some fields in \nref{scexp} the sum over $l$ does not run over all non-negative integers. The reason is that the fields $f$ and $\chi$ only generate non-trivial fluctuations if $l>0$, since the operator $\mcD_{a}$ annihilates constant functions. Analogously, the field $\eta$ only generates non-trivial fluctuations if $l>1$, since $\mcD_{ab}$ annihilates constant functions, and spherical harmonics with $l=1$\footnote{The reason is that the gradient of an $l=1$ scalar spherical harmonic is a conformal killing vector in the sphere.}. Therefore, for $l=0$ the scalar fluctuations are only $\psi_{l}$ and $A_{l}$, for $l=1$ the scalar fluctuations are $\psi_{l}$,$A_{l}$,$f_{l}$ and $\chi_{l}$, and for $l \geq 2$ the scalar fluctuations are all the five scalar fields.

\textbf{Tangent transverse vectors:} The tangent transverse vector spherical harmonics, which we call $Y_{l,a}$, are vectors tangent to the $S^{D-1}$ which respect
\begin{equation}
\label{tvharm}
-\mcD^{2}Y_{l,a}=\frac{(l(l+D-2)-1)}{a(\tau)^{2}}Y_{l,a}~~,~~ \mcD^{a}Y_{l,a}=0
\end{equation}
with $l\geq 1$. Note the $\mcD^{b}Y_{l,b}=0$ equation is the transverse condition, and the first equation gives the $-\mcD^{2}$ eigenvalue of $Y_{l, a}$\footnote{One should not confuse the scale factor $a(\tau)$ with the vector index $a$.}. However, taking $Y_{l, a}$ as a vector field defined in the entirety of the wormhole manifold, we have the freedom to multiply $Y_{l, a}$ by a function of $\tau$, and it will still respect \nref{tvharm}.

A natural thing to do would be to require $Y_{l, a}$ to have zero coordinate derivative with respect to $\tau$, in which case $Y_{l, a}$ would be, in coordinates, the usual $S^{D-1}$ spherical harmonic. However, it is slightly more convenient for us to require $Y_{l, a}$ to have zero covariant derivative along the normal direction. We therefore impose
\begin{equation}
n^{b}\n_{b}Y_{l,a}=0
\end{equation}
which fixes, with a given solution to \nref{tvharm}, a unique $Y_{l,a}$ up to a constant. 

Using this $Y_{l,a}$, we can expand the tangent transverse vector fields $\{v_{a},x_{a},u_{a},\omega_{a}\}$ as
\begin{equation}
\begin{gathered}
\label{tvexp}
v_{a}=\sum_{l=1}^{\infty}v_{l}(\tau)Y_{l,a} ~~,~~ x_{a}=\sum_{l=2}^{\infty}x_{l}(\tau)Y_{l,a}\\
u_{a}=\sum_{l=1}^{\infty}u_{l}(\tau)Y_{l,a}~~,~~\omega_{a}=\sum_{l=1}^{\infty}\omega_{l}(\tau) Y_{l,a}
\end{gathered}
\end{equation} 

Note that we restricted the sum in the $x$ expansion to $l \geq 2$, because for $l=1$, the operator $\mcD_{(a}Y_{b)}$ is zero, since these $Y_{l,b}$ are isometries of $S^{D-1}$. This implies that at $l=1$, the transverse vector fluctuations are just $v_{b},u_{b}$ and $\omega_{b}$. For $l>1$, all 4 transverse vector fluctuations exist.

\textbf{Tangent transverse two-forms:} The tangent transverse two-form spherical harmonics, $Y_{l,[ab]}$, are two-forms along the $S^{D-1}$ direction which respect 
\begin{equation}
\label{t2fharm}
-\mcD^{2}Y_{l,[ab]}=\frac{(l(l+D-2)-2)}{a(\tau)^{2}}Y_{l,[ab]}~~,~~ \mcD^{a}Y_{l,[ab]}=0
\end{equation}
with $l \geq 1$. Like in the tangent transverse vector case, we use the residual freedom of equation \nref{t2fharm} to set
\begin{equation}
n^{c}\n_{c}Y_{l,[ab]}=0
\end{equation}

We can then expand $j_{ab}$ as
\begin{equation}
\label{t2fexp}
j_{ab}=\sum_{l=1}^{\infty}j_{l}(\tau) Y_{l,[ab]}
\end{equation}

\textbf{Tangent transverse traceless symmetric tensors:} The tangent transverse traceless symmetric spherical harmonics, $Y_{l,(ab)}$, are symmetric tensors along the $S^{D-1}$ direction which respect 
\begin{equation}
\label{t2tharm}
-\mcD^{2}Y_{l,(ab)}=\frac{(l(l+D-2)-2)}{a(\tau)^{2}}Y_{l,(ab)}~~,~~ \mcD^{a}Y_{l,(ab)}=0~~,~~\gamma^{ab}Y_{l,(ab)}=0
\end{equation}
with $l \geq 2$. Like in the tangent transverse vector case, we use the residual freedom of equation \nref{t2tharm} to set
\begin{equation}
n^{c}\n_{c}Y_{l,(ab)}=0
\end{equation}

We can then expand $\phi_{ab}$ as
\begin{equation}
\label{t2texp}
\phi_{ab}=\sum_{l=2}^{\infty}\phi_{l}(\tau) Y_{l,(ab)}
\end{equation}

Having discussed these points, we are ready to discuss the eigenvalue problem for the 4 different types of fields. Since some of the eigenvalue equations are quite cumbersome, we leave the discussion of their explicit form to Appendix \nref{eigensec}.

\subsection{Features of the spectrum and phase}
\label{featspec}

Now we discuss the eigenvalue spectrum for all different representations of $SO(D)$. After the decomposition of the fields into spherical harmonics that we discussed in section \nref{repsod}, the eigenvalue equations for the modes in a given spherical harmonic are differential equations in $\tau$. We can then solve these eigenvalue equations numerically by discretizing the $\tau$ direction along an appropriate grid and solving the counterpart discretized problem. For more details, see Appendix \nref{numdeta}. 

In this section, we will only discuss aspects of the eigenvalue spectrum that we obtained via this numerical method. These observations will allow us, in particular, to compute the phase of the one-loop determinant, following a variant of the grid method in \cite{Chen:2025jqm}. Before we discuss the features of the spectrum for the different representations of $SO(D)$, let us make some general preliminary remarks. 

First, here we will mostly focus on the properties of the $Q<Q_{c}$ spectrum, since we discuss the analytical spectrum of the $Q=Q_{c}$ wormhole later in section \nref{einwh}. However, we will discuss the phase of both $Q<Q_{c}$ and $Q=Q_{c}$ solutions. Secondly, a general feature that we observed is that for small values of $\kappa Q$, the spectrum of all fields is approximately the same as their $S^{D}$ counterpart, with the exception of the presence of a finite number of new modes that we discuss explicitly. 

As a last point, we should also mention that the result for the phase we report below seems robust against changes in the gauge fixing term \nref{gfterm} and the chosen local norm \nref{locnorm}. We discuss the changes to the gauge fixing term and to the format of the local norm that we considered in Appendix \nref{eigensec} and \nref{numdeta}.

\textbf{Scalars:} The eigenvalue spectrum for the scalars consists of multiple modes with both positive and negative values for $\lambda$. This is indeed what one expects for the spectrum of these scalars. The reason is that at small distances, the eigenvalue spectrum is dominated by the kinetic term of these fields, so they all decouple. The field $\psi$ has a wrong sign kinect term, while all other scalars have the usual sign for their kinect term; therefore, there should be one tower of negative modes coming from high eigenvalue modes of $\psi$, one for each short wavelength eigenvalue of $-\n^{2}$.

However, at each different value of angular momentum, we can define a renormalized number of scalar negative modes $n_{-,s}(l)$ depending on the number of $\tau$ grid points we used to solve the eigenvalue equation. More specifically,
\begin{equation}
n_{-,\text{sc}}(l)=d(l)[(\text{number of scalar negative modes})-(\text{number of $\tau$ grid points})]
\end{equation}
where $d(l)$ is the degeneracy of the angular momentum $l$ scalar spherical harmonic of $S^{D-1}$. The subtraction by the number of grid points comes from the "$n_{\text{sp}}$" term that we discussed in section \nref{genset}. As we discussed, $n_{\text{sp}}$ can be thought of as the number of basis modes of a scalar field, so we include the subtraction by $n_{\text{sp}}$ here since it combines naturally with the tower of scalar negative modes from $\psi$.

In our regularization, we find $n_{-,\text{sc}}(l)$ to be finite, and only non-zero for $l=0$. The results of $n_{-,\text{sc}}(l)$ for $l=0$ will depend on the charge of the solution in a very simple way. If $0<Q<Q_{c}$ one finds that $n_{-,\text{sc}}(0)=-2N$, with $N$ the number of cycles of the wormhole solution.  

If $Q=Q_{c}$, then $n_{-,\text{sc}}$ will depend on the length $L$ of the $\tau$ direction of the solution via
\begin{equation}
\label{nminusein}
n_{-,\text{sc}}(0)=-1-2\Big\lfloor\frac{L}{2\pi}\sqrt{2(D-1)}\Big\rfloor 
\end{equation}
In section \nref{einwh}, we will be able to find this $Q=Q_{c}$ result analytically. 

Another relevant point to mention is that the scalar spectrum has no zero eigenvalues for $0<Q<Q_{c}$. Near $Q=0$, there are $2 \times N$ sets of new scalar light modes at $l=1$, whose eigenvalue size scale as $\sim \kappa Q$. We call these modes "new" because they have no $S^{D}$ counterpart. $N$ of these sets of modes is positive, and the other $N$ is negative. We discuss them in detail in section \nref{nearslight}. An interesting point is that since the degeneracy of the $l=1$ harmonic is $D$, we can track the fact that we found $n_{-,\text{sc}}$ to be zero at $l=1$ to the existence of these $D\times N$ new negative modes. Namely, $n_{-,\text{sc}}(1)$ would be $-D\times N$ in the analogous computation for a product of $N$ spheres \cite{Polchinski:1988ua}, but it is offset by these new light modes.

The spectrum also has 3 light modes near $Q=Q_{c}$ at the $l=0$ sector, which become zero modes at $Q=Q_{c}$. One of them corresponds to fluctuations of the length of the wormhole, and the other two are perturbations of the scale factor of the metric $a(\tau)$ with a specific frequency. We discuss them more in section \nref{neareinlight}.

\textbf{Tangent transverse vectors:} The eigenvalue spectrum of the tangent transverse vectors consists of positive modes and one set of zero modes. The zero modes are at the $l=1$ transverse vector representation and exist for any $Q$. They are locally a coordinate transformation, but not globally. They are therefore physical, and do not cancel against the volume of coordinate transformations. They are the manifestation of twisted wormhole solutions connected to the one we introduced in section \nref{prelim}, as we briefly discussed in section \nref{genset}. Therefore, we take these zero modes into consideration by performing an appropriate integral over twisted wormholes. We will discuss this contribution in more detail in section \nref{zeromod}.

\textbf{Tangent transverse two-forms:} The eigenvalue spectrum of the tangent transverse two-forms is non-negative for any $D$, and only has positive modes for $D>3$. In particular, no mode becomes light as $Q \rightarrow 0$ or $Q \rightarrow Q_{c}$. The two-forms have a zero mode for $D=3$, and any $Q$, because in $D=3$ there are harmonic two-forms in the manifold. The reason is that the topology of the wormhole is $S^{1} \times S^{2}$. These modes are exclusive to the wormhole, since they have no $S^{D}=S^{3}$ counterpart. Moreover, the integral over this mode is compact, and we will comment about it in more detail in section \nref{lightzero}. 

\textbf{Symmetric tangent transverse traceless tensors:} The eigenvalue spectrum of the symmetric tangent transverse traceless two tensors is positive, and has no zero or light modes.

\textbf{Summary:} Now, let us briefly summarize the most important points (see also Table \nref{tab: summary of physical spectrum}). First, note that all the negative modes from the wormhole come solely from the scalar fields $\{\psi, A, f, \eta, \chi\}$. Therefore, the phase from the wormhole comes only from the $n_{-,\text{sc}}$ contribution from the scalars, which is
\begin{equation}
\label{whphasesm}
n_{-}= \begin{cases*}
    -2N~~,~~\text{if $0<Q<Q_{c}$ and the wormhole has $N$ cycles}\\
    -(1+2\lfloor N\rfloor),\text{if $Q=Q_{c}$ and $\tau \sim \tau+\frac{2\pi N}{\sqrt{2(D-1)}}$}
\end{cases*}
\end{equation}
and the phase for the wormhole is $(-i)^{n_{-}}$, as advertised in the introduction. Another relevant aspect of the spectrum is that, while at small $\kappa Q$ most of the eigenvalues we found were small deformations of eigenvalues in $S^{D}$, the wormhole has a few new modes with no sphere counterpart. They are as follows: There are $2N$ sets of new light modes on the scalar sector at $l=1$. $N$ of these sets have negative eigenvalues, while the other $N$ have positive ones. These extra negative modes explain the difference between the phase we see from \nref{whphasesm} and the associated one for $N$ spheres. The other new modes are all zero modes. One of them is in the tangent transverse vector sector at $l=1$ and is always present. There is one extra zero mode in the tangent transverse two-form sector if $D=3$, and none otherwise. We discuss all of these new modes, and more, in section \nref{lightzero}.

\begin{table}[t]
\centering
\renewcommand{\arraystretch}{1.3} 
\setlength{\tabcolsep}{5pt}       

\begin{tabular}{|c|c|c|c|c|c|}
\hline
 Sectors& $l$ & $n_-$ & $n_\text{zero}$ & \makecell[c]{$n_\text{light},$\\$Q\rightarrow0$}  &  \makecell[c]{$n_\text{light},$\\$Q\rightarrow Q_c$}\\ \hline

\multirow{3}{*}{Scalars} 
  & $l=0$ & $-2N$  &  &  & $3$ \\ \cline{2-6}
  & $l=1$ &  &  & $2N\times D$ &  \\ \cline{2-6}
  & $l\geq2$ &  &  &  &  \\ \hline

\multirow{2}{*}{\makecell[c]{Tangent transverse\\ vectors}} 
  & $l=1$ &  & $1\times \frac{D(D-1)}{2}$ &  &  \\ \cline{2-6}
  & $l\geq2$ &  &  &  &  \\ \hline

\multirow{2}{*}{\makecell[c]{Tangent transverse\\ two-forms}} 
  & $l=1$ &  & $1$, iff $D=3$ &  &  \\ \cline{2-6}
  & $l\geq2$ &  &  &  &  \\ \hline

\makecell[c]{Symmetric tangent\\transverse traceless\\tensors} & $l\geq2$  &  &  &  &  \\ \hline
\end{tabular}
\caption{A summary of several important features of physical spectrum for general charge $0<Q<Q
_c$ and general dimension $D\geq3$, highlighting on the number of ``regularized'' negative modes $n_-$, zero modes $n_\text{zero}$ and light modes $n_\text{light}$ in two limits. To make the visualization clearer, we leave the cell blank if there are no such modes.}
\label{tab: summary of physical spectrum}
\end{table}

\section{More on zero and light modes}
\label{lightzero}

In this section, we discuss the zero modes of the fluctuation operator $M$ in \nref{eigeneq}, and the modes that become light as either $Q \rightarrow 0$ or as $Q \rightarrow Q_{c}$. An interesting aspect of these modes is that they are all modes exclusive to the wormhole spacetime, which have no sphere counterpart. 

We discuss the zero modes in section \nref{zeromod}, the near sphere light modes in section \nref{nearslight}, and the near Einstein light modes in \nref{neareinlight}. Importantly, in section \nref{nearslight} we will discuss further the connection between the wormhole at small $\kappa Q$ and the sphere with operator insertions. This will allow us to reinterpret the phase of the $N=1$ wormhole we derived in section \nref{featspec}, as being the usual Polchinski phase for the sphere, times a phase coming from an integral over operator insertions. The phase for the wormhole with $N$ cycles has a similar interpretation.

\subsection{Zero modes}
\label{zeromod}

As briefly discussed in section \nref{featspec}, for $0<Q<Q_{c}$ the eigenvalue spectrum has zero modes for the tangent transverse vector sector for general $D$, and zero modes for the tangent transverse two-form sector at $D=3$. We will now discuss them in a bit more detail.

\textbf{Tangent transverse vectors:} The tangent transverse vector sector has a set of zero modes, which are locally a coordinate transformation generated by
\begin{equation}
\label{twimod}
\xi_{a}=\sqrt{2}\bigg(\int^{\tau} \frac{d\tau' }{a(\tau')^{D+1}}\bigg)c_{a}
\end{equation}
and an appropriate form gauge transformation. The $c^{a}$ is an $SO(D)$ isometry of the background, e.g, $\n_{(a}c_{b)}=0$. Note that since the integral $\int \frac{d\tau}{a^{D+1}}$ is not periodic in $\tau$, $\xi_{a}$ is not invariant under monodromy and is therefore not a well-defined coordinate transformation globally. Therefore, this mode is not pure gauge.

To be more specific, the modes generated by these coordinate transformations are
\begin{equation}
h_{ab}=\frac{2c_{(a}n_{b)}}{a^{D+1}}
\end{equation}
and an appropriate $b_{ab}$ that respects the gauge fixing condition $(db)_{abc}=0$. The interpretation of these modes is a bit clearer once we note that the monodromy of \nref{twimod} around one period of the wormhole is proportional to an isometry of the background. 

This implies that starting from one $S^{D-1}$ slice and going around the $\tau$ circle via one period, we obtain another $S^{D-1}$ that differs from it by a relative isometry. Since the isometry maps $S^{D-1}$ to itself, this defines a compatible gluing of the two slices. We can therefore interpret the integral over these modes as an integral over twisted wormholes\footnote{The integral over these twists is also important to impose conservation of angular momentum through the wormhole, see \cite{Hawking:1987mz}.}. That is, if $x$ is an $S^{D-1}$ coordinate and $L$ is the length of the $\tau$ circle of the wormhole, we identify coordinates after a period as
\begin{equation}
\label{twigl}
(\tau,x)\sim (\tau+L,U_{\text{iso}} \cdot x)
\end{equation}
with $U_{\text{iso}}$ an element of $SO(D)$. This is similar to the "relative twist" mode one integrates over in JT gravity \cite{Mertens:2022irh, Saad:2019lba}, when gluing two trumpets. The relative isometry is equal to the monodromy of $\xi^{a}$ in \nref{twimod}, so the $\xi^{a}$ that corresponds to an usual $SO(D)$ generator for the transformation in \nref{twigl} is
\begin{equation}
\label{rlxigen}
\xi_{a}=\frac{1}{\sqrt{2}\kappa\,I_{\text{twist}}}c_{a}\int^{\tau} \frac{d\tau' }{a(\tau')^{D+1}}~~,\text{with } I_{\text{twist}}=\int_{\text{period}}\frac{d\tau}{a(\tau)^{D+1}} 
\end{equation}
where we require $c_{a}$ to satisfy the usual $SO(D)$ commutation relations\footnote{Note that, as commented before, the mode also comes with an associated form gauge transformation.}. The factor of $\frac{1}{\sqrt{2}\kappa}$ comes because of how we defined the coordinate transformations\footnote{The reason is than an usual $SO(D)$ generator $c_{a}$ acts on the metric as $\delta g_{ab}=\mathcal{L}_{c}g_{ab}$. Therefore, we have to pick an $\xi_{a}$ defined as in \nref{diffact} that generates this transformation. For a more detailed discussion, see \cite{Law:2020cpj, Anninos:2020hfj}.} in \nref{diffact}.

The integral over these modes will then be related to the group integral over $SO(D)$, whose overall result is a group volume. However, the modes that we are discussing will contribute via their measure induced from the path integral, which is not the same as the canonical group measure. The two measures should be the same up to a normalization factor, and to find this normalization factor, we just have to compare the norm of the generators in the two norms. The $SO(D)$ generators have norm one in the canonical group norm. Therefore, let the twist mode generated from the transformation \nref{rlxigen} and the appropriate \nref{formact} transformation be $\Phi_{c}$, then the path integral and group norms are related as
\begin{equation}
\label{ntwist}
(\Phi_{c},\Phi_{c})=\mathcal{N}_{\text{twist}}^{2}(c_{a},c_{a})_{\text{canonical}}=\mathcal{N}_{\text{twist}}^{2}
\end{equation}
with $\mathcal{N}_\text{twist}$ defined as the normalization factor. The integral over these modes, $Z_{\text{twist}}$, is therefore,
\begin{equation}
Z_{\text{twist}}=\mathcal{N}_{\text{twist}}^{|SO(D)|}\text{Vol}(SO(D))_{c}
\end{equation}
where $|SO(D)|$ is the dimension of $SO(D)$, and $\text{Vol}(SO(D))_{c}$ is its canonical group volume. 

\textbf{Tangent transverse two-forms:} The eigenvalue problem for the tangent transverse two-forms $j_{ab}$ in \nref{eigeneq} does not mix with the gravitational degrees of freedom. Therefore, the spectrum of the $j_{ab}$ modes is the same as their spectrum when acted by the Hodge Laplacian for two-forms. However, the Hodge Laplacian is non-negative, and its number of zero modes is completely fixed by topology. 

To be more specific, the number of zero modes of the Hodge Laplacian for $p$-forms is the $p$-th Betti number $b_{p}$, which is a function only of the topology of the manifold. Since the manifolds we are studying have $S^{1} \times S^{D-1}$ topology, the non-zero values of $b_{p}$ are $b_{0}=b_{D}=b_{1}=b_{D-1}=1$. Therefore, the Hodge Laplacian for two-forms will have a zero if, and only if, 2 is one of these four values. Since the solutions we study have $D>2$, this is only possible if $D-1=2$, e.g, $D=3$. Therefore, the $D=3$ solution has zero modes for the $j_{ab}$ perturbations. They are of the form
\begin{equation}
j_{ab}=\frac{\bar{\epsilon}_{ab}}{a^{2}(\tau)}
\end{equation}
with $\bar{\epsilon}_{ab}$ the volume form of $S^{2}$ normalized such that $\bar{\epsilon}_{ab}\bar{\epsilon}^{ab}=2!$. The path integral over harmonic forms, however, should have a compact range because we identify them over large gauge transformations \cite{Donnelly:2016mlc}.  

To be more specific, to integrate over these modes we should go back to the $D-2=1$ form formulation, and then integrate the original fluctuation $\delta B_{a_{1}..a_{D-2}}$ with the correct identification of harmonic 1-forms under large gauge transformations. We discuss this type of zero-mode contribution in more detail in section \nref{formdet} of the Appendix. We should note, however, that the contribution from the zero mode we just discussed will always cancel an identical term in the ghost determinant \nref{ghostL}, so they are never important.

\subsection{Near sphere light modes and operator insertions}
\label{nearslight}

\begin{figure}[h!]
    \centering
    \includegraphics[width=1\linewidth]{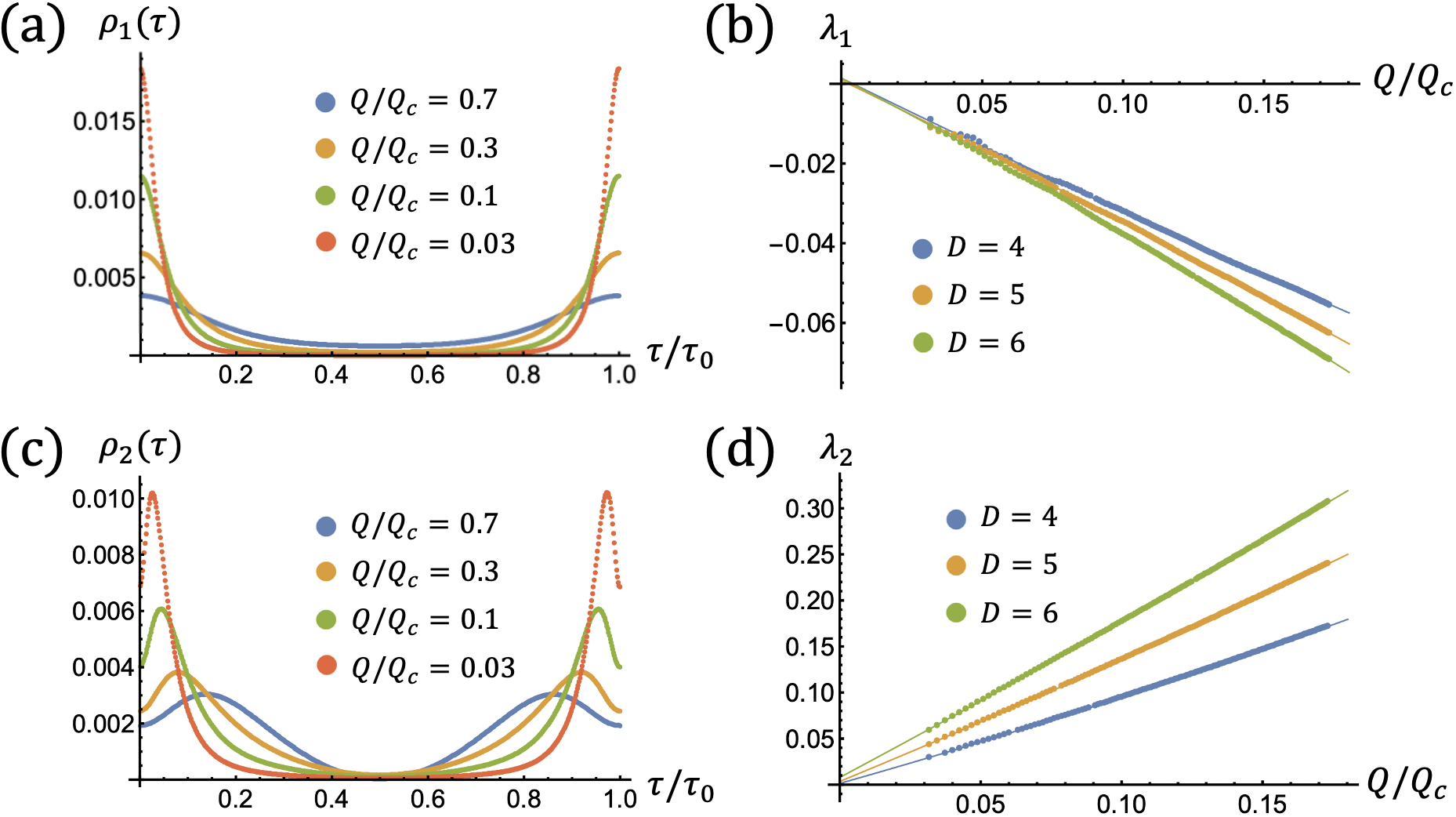}
    \caption{\textbf{(a)} The norm density $\rho_1(\tau)$ of the light negative modes at $D=5, N=1$, for various $Q$. By tuning $Q$ to the near sphere limit, the wave-function becomes localized in the wormhole mouth region. \textbf{(b)} The eigenvalue $\lambda_1$ of the light negative modes for various $Q$ and various $D$. The dots are numerical data, and the underlying lines are a linear fit of $\lambda_1$ as a function of $Q$.\textbf{(c)} The norm density $\rho_2(\tau)$ of the light positive modes at $D=5$, for various $Q$. By tuning $Q$ to the near sphere limit, the wave-function becomes localized in the wormhole mouth region. \textbf{(d)} The eigenvalue $\lambda_2$ of the light positive modes for various $Q$ and various $D$. The dots are numerical data, and the underlying lines are a linear fit of $\lambda_2$ as a function of $Q$. }
    \label{fig: light mode wavefunction and spectrum}
\end{figure}

Here we discuss the non-zero modes of the operator $M$ in \nref{eigeneq} whose eigenvalues go to zero as $Q \rightarrow 0$. More precisely, their eigenvalues behave as $|\lambda| \sim \kappa Q$ (see Figure \nref{fig: light mode wavefunction and spectrum}). These "light" modes all come from the $l=1$ scalar sector of fluctuations. For the wormhole with $N$ cycles, there are $2N$ sets of such modes, each with degeneracy $D$ due to their angular momentum number. Their eigenvalues all behave as $|\lambda|\sim \kappa Q$ as $Q\rightarrow0$. $N$ of these sets of modes have positive eigenvalues, and the other $N$ have negative eigenvalues. These modes share the feature that they localize to the wormhole throats as $Q \rightarrow 0$, as shown in Figure \nref{fig: light mode wavefunction and spectrum} for $N=1$\footnote{See Figure \nref{fig: localized wavefunc N_2_3} for examples of their wavefunctions at higher $N$.}.  For clarity, in this section, we will focus on discussing the light modes at $N=1$, and we will briefly discuss their interpretation at $N \neq 1$ later. 

We can see why the light modes are expected at $N=1$ as follows: If $Q$ is small, the solution is a sphere with a small handle. We can think of the handle as a smaller wormhole embedded into two much bigger spacetimes, like the flat space wormhole solutions. Flat space wormhole solutions have two asymptotic boundaries, which one can think of as being two infinitely distant spacetimes. These flat space solutions have two sets of exact zero modes, which correspond to moving the wormhole mouths in the ambient spacetimes by a translation. These are $D+D=2D$ zero modes, which is precisely the number of light modes we observe here.

These translations are zero modes for the flat space wormhole because the wormhole mouths are infinitely far away from each other, and thus have no interaction. However, in the sphere solution, the wormhole mouths end up in the same spacetime and can interact through the bulk, so the translations of the mouths are no longer an exact zero mode. We expect the interactions to be small if the wormholes are small compared to the ambient spacetime, so the eigenvalue for these deformations should be a light one, like we find numerically. 

Therefore, the $2D$ light modes we observe correspond to different ways of moving the two very small wormhole mouths along $S^{D}$.  Alternatively, we can also think of the wormhole mouths as local operator insertions. This picture will be useful to us later. However, before we discuss these details, we should first discuss the modes more explicitly.

Let us first discuss the light positive modes. They seem reminiscent of the light modes found for scalar instantons with small backreaction in \cite{Ivo:2025fwe}. The mechanism for the positive light modes in \cite{Ivo:2025fwe} is as follows: One considers a matter+gravity saddle where the matter backreaction is small, but that breaks some isometries of the original gravity background. At zero backreaction, these isometries either do not generate deformations, or generate zero modes of the matter fluctuations. In any case, these isometries will be lifted at non-zero backreaction, and if we add a gauge fixing term to the action, these coordinate transformations will give rise to light modes. Since the action of these modes comes mostly from the gauge fixing term \cite{Ivo:2025fwe}, they are positive and will depend strongly on the gauge fixing term we add. 

\begin{figure}[h!]
    \centering
    \includegraphics[width=1\linewidth]{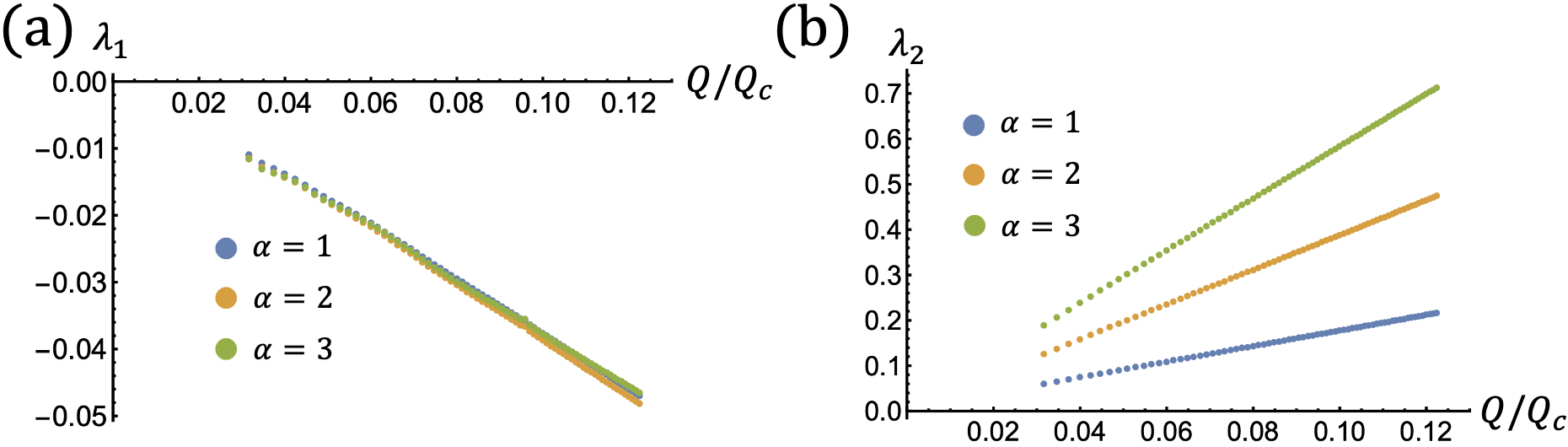}
    \caption{The eigenvalue of negative light modes $\lambda_1$ in \textbf{(a)} and the positive light modes $\lambda_2$ in \textbf{(b)}, when we change the gauge fixing parameter. To obtain this figure, we varied the gauge fixing parameters $\alpha$ and $\beta$ discussed in section \nref{altgf}, along the curve $\alpha=\frac{(D-2)}{2(\beta D-1)}$. We observe that in the small $Q$ limit, the eigenvalue of negative light modes is gauge fixing invariant, while the eigenvalue of positive light modes is not. Here we take $D=6$.}
    \label{fig: change_alpha}
\end{figure}

This is precisely the scenario we are considering here at small $Q$. That is, away from the small handle, the $Q \sim 0$ geometry is very close to that of a round sphere, but the $Q>0$ background breaks the $SO(D+1)$ symmetry of the background to $SO(D)$. There are therefore $D$ broken isometries, and each of them leads to a light positive mode of the gauge fixed action. We also checked numerically that the eigenvalue of these light positive modes depends strongly on the gauge fixing term added to the action in \nref{gfterm} (see Figure \nref{fig: change_alpha}), as expected from the proposed mechanism.

\begin{figure}[H]
    \centering
    \includegraphics[width=0.7\linewidth]{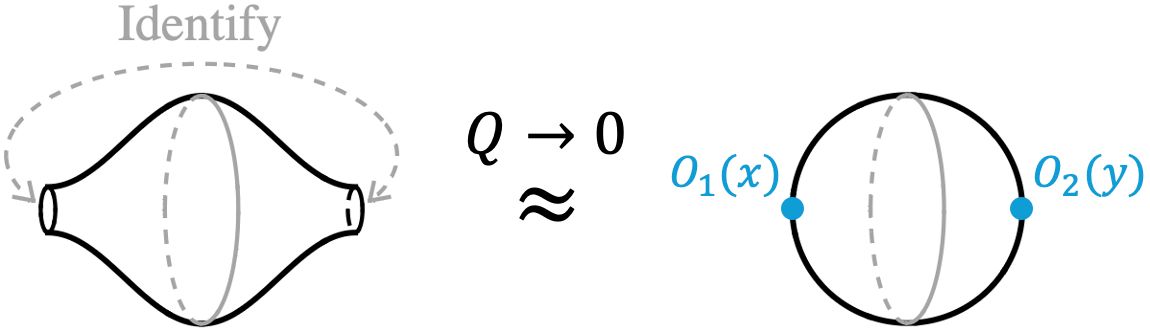}
    \caption{On the left-hand side, we have a wormhole saddle with small $\kappa Q$, which looks like a sphere with a small handle. This saddle is reproduced by the sphere with the leading effective operator insertions at antipodal positions.}
    \label{fig: opinsertionfig}
\end{figure}

Then, we are left having to explain the light negative modes. Numerically, we find that in the limit that $Q$ is small, they do not depend on the gauge fixing term we add to the action  (see Figure \nref{fig: change_alpha}). This suggests that they are physical deformations of the background. To understand these modes, it is useful to take a small tangent and review the effective description where we can think of small wormhole mouths as local operator insertions \cite{Hawking:1987mz, Coleman:1988cy, Giddings:1988cx, Klebanov:1988eh}. Using this picture, we are going to conclude that we can think of the wormhole saddle we are studying as the sphere with two operator insertions, integrated around their antipodal configuration in $S^{D}$ (see Figure \nref{fig: opinsertionfig})

More precisely, the physical picture \cite{Klebanov:1988eh, Witten:2026twr} is that at large distances one can replace the effect of a small wormhole connecting two much larger spacetimes by a sum over integrated bilocal operator insertions $O_{1}(x)O_{2}(y)$, where $x$ belongs to the first manifold, and $y$ to the second. In our context, from the perspective of the small wormhole, the two faraway asymptotes would be just two different regions of the sphere. From this picture, we would expect the full path integral over a small wormhole and its position moduli to consist of a sum of two-point functions in the sphere, integrated over all positions for the operators, as
\begin{equation}
\label{goperator0}
\sum_{ij} \mathcal{C}_{i j}\int d^{D}x\,d^{D}y\, \langle O_{i}(x)O_{j}(y)\rangle_{S^{D}}
\end{equation}
where $\langle \rangle_{S^{D}}$ denotes the expectation value of a quantity in the sphere, and the sum over $i$,$j$ is the sum over all bilocal operators consistent with the symmetries of the problem. $\mathcal{C}_{i j}$ is an undetermined EFT coefficient, which in our problem should not depend on the de Sitter scale $\lds$ since it comes from local properties of the small wormhole. Since $\mathcal{C}_{ij}$ naturally scales with the size of the wormhole, the sum in \nref{goperator0} is dominated by operators of low dimension.

However, note that while \nref{goperator0} is natural for an integral over off-shell wormholes, the classical wormhole saddle we study has the wormhole mouths in antipodal positions. This suggests that the classical wormhole reproduces the contribution of \nref{goperator0} from a saddle of the $x$,$y$ integral where the operators are antipodal to each other. Indeed, the two-point function at this configuration is an extremum, since the operators are the furthest away from each other they can be in the sphere. Therefore, to compare the operator description to the classical wormhole we are studying, we should replace the integrals in \nref{goperator0} by their contribution around the antipodal configuration of $x$,$y$. This means that we are allowed to move $x$ and $y$ together, but their distance should remain almost the antipodal distance. 

Furthermore, since in this configuration the operators are very far from each other, in units of the wormhole size, we can approximate the sum in \nref{goperator0} by its contribution from the operator of the lowest dimension.  In our context, using the $(D-2)$ form language, these operators are the magnetic operators $K_{\pm n}(x)$ that create or destroy $n$ units of flux \cite{Witten:2026twr}, which would be the flux of the wormhole solution. Therefore, we should have that
\begin{equation}
\label{goperator1}
Z_{\text{wormhole}}\approx Z_{\text{GR}}(S^{D})Z_{D-2}(S^{D})\times \mathcal{C}_{n,-n}\int d^{D}x\,d^{D}y\,\langle K_{n}(x)K_{-n}(y)\rangle_{S^{D}}\big|_{\text{near antipodal}}
\end{equation}
where we used $Z_{\text{GR}}(S^{D})$ to denote the pure gravity path integral in the sphere, and $Z_{D-2}(S^{D})$ the one-loop path integral of the $D-2$ form theory in the sphere background. Furthermore, note that we named the EFT coefficient for the $K_{\pm n}$ operators $\mathcal{C}_{n,-n}$. We should comment that since the operators $K_{\pm}(x)$ are dimensionless\footnote{The reason is that the only effect of $K_{\pm n}$ in the path integral is to enforce a different boundary condition. Therefore, they only turn an already dimensionless path integral into one with different rules (see \cite{Witten:2026twr}).}, the coefficients $\mathcal{C}_{n,-n}$ must have a length dimension of $-2D$. Since $\mathcal{C}_{n,-n}$ should not depend on $\lds$, we expect it only to depend on parameters such as $\kappa_{\text{phy}}$, $f_{\theta,\text{phy}}$, and $n$.

Before we proceed, note that even though we argued that \nref{goperator1} is correct, the antipodal configuration is not always a leading saddle. That is, for two-point functions that increase as the operators get closer, the antipodal configuration will be a local minimum instead of a local maximum. The reason is that one would be able to increase the two-point function by bringing the two operators together. Fixing one of the operators at the pode, there are $D$ independent displacements for the other operator that will decrease their distance. In our formalism for computing the one-loop determinant, these would be $D$ negative modes of the fluctuation operator. 

We can argue that, in our context, the two-point function indeed increases away from the antipodal configuration. To explain that, we make our analysis slightly simpler by using that in the scalar side of electromagnetic duality \cite{Witten:2026twr}, the operators $K_{\pm}$ are just $e^{\pm \frac{2 \pi i}{f_{\theta}}n \theta(x)}$, and that the expectation values in the two sides of the duality should match. From evaluating the two-point function explicitly, we find that
\begin{equation}
\label{2ptax}
\langle e^{\frac{2\pi i}{f_{\theta}}n \theta(x_{1})}e^{-\frac{2 \pi i}{f_{\theta}}n \theta(x_{2})}\rangle_{S^{D}}=e^{\frac{4 \pi^{2}n^{2}}{f_{\theta}^{2}}G(x_{1},x_{2})}
\end{equation}
where we dropped a divergent constant that does not depend on $x_{1}$ and $x_{2}$. The $G$ function is an $SO(D+1)$ symmetric Green's function in the sphere for the axion field $\theta$, which is the solution to
\begin{equation}
\label{gaxionmain}
-\n^{2}G(x,y)=\frac{1}{\sqrt{g}}\delta(x,y)-\frac{1}{\text{Vol}(S^{D})}
\end{equation}
and $\text{Vol}(S^{D})$ the volume of the round sphere. The right-hand side of \nref{gaxionmain} has an extra term on top of the delta function, because both sides need to be orthogonal to constant functions. Physically, this comes from the fact that the axion is a massless scalar, so its constant mode is a zero mode. We discuss the form of $G(x,y)$ further and how to derive \nref{2ptax} in Appendix \nref{axig}. The important result of that section, however, is that by expanding $x$ and $y$ around the antipodal configuration, we can see that $G(x,y)$ increases.

To be more specific, take the configuration where the points are antipodal to each other to consist of one operator insertion at $\tau_{1}=0$ and the other one at $\tau_{2}=\pi$. Then, fixing the first point and displacing the second by $\delta \tau_{2}$, we have
\begin{equation}
\label{gopgauss}
\langle e^{\frac{2 \pi i }{f_{\theta}}n \theta(x_{1})}e^{-\frac{2 \pi i }{f_{\theta}}n \theta(x_{2})}\rangle_{S^{D}}\approx  \exp\bigg(\frac{4 \pi^{2}n^{2}}{f_{\theta}^{2}}G_{\text{anti}}\bigg)\times \text{exp}\bigg(\frac{2 \pi^{2}n^{2}}{f_{\theta}^{2}D\text{Vol}(S^{D})}(\delta \tau_{2})^{2}\bigg)
\end{equation}

This implies that displacing one of the wormhole mouths with the other fixed is a wrong-sign Gaussian. More specifically, since $\delta \tau_{2}$ can be seen as a radial coordinate away from one of the podes, $(\delta \tau_{2})^{2}$ is the sum over squares of $D$ independent Euclidean displacements around the pode, such that \nref{gopgauss} is the product of $D$ wrong-sign Gaussians. Therefore, the wormhole indeed reproduces a subleading saddle in the integral over two-points function in \nref{goperator0}. They are a similar subleading saddle to, for example, the great circle geodesics for an observer in the sphere discussed in \cite{Maldacena:2024spf}. 

We also see that the Gaussian becomes shallow as $\frac{n^{2}}{f_{\theta,\text{phy}}^{2}\lds^{D-2}}\rightarrow 0$\footnote{Note, however, that the allowed values of $n$ are discrete.}. To allow for a semiclassical treatment of the antipodal saddle, we therefore need to impose $\frac{n}{f_{\theta}}\gg1$. From \eqref{eq: quantization of Q}, this means the dimensionless flux parameter $Q$ is large. This behaviour makes sense, since the integral \nref{goperator0} having a classical saddle, even if subleading, is special to de Sitter space. In other words, if $\lds$ were infinite, displacing the wormhole mouths from each other would be a zero mode. 

Another aspect of \nref{gopgauss} is that it only depends on the relative displacement of the two wormhole mouths, and it should be left unchanged if we displaced the two operators together along the $S^{D}$. Indeed, moving the operators together is a coordinate transformation, which should lead to zero modes in the gravity counterpart of the calculation. However, since we added a gauge fixing term in \nref{gfterm}, these modes obtain an action and become the $D$ light positive modes in our calculation, as we discussed earlier in the section.

Now, we discuss the contribution to \nref{goperator1} from the antipodal saddle of the two-point function in more detail. First, to make the integral in \nref{goperator1} dimensionless, we rescale $x=\lds \hat{x}$ such that the remaining position integrals in $\hat{x}$ and $\hat{y}$ are over the sphere with radius one. To perform the $x$,$y$ integral, we imagine first fixing the $x$ operator, and then integrating over $y$, which is now the displacement between the two. We should, however, integrate over the possible positions of the fixed $x$, which gives a factor of $\text{Vol}(S^{D})$. The remaining integral will then be over the Gaussians in \nref{gopgauss}.

Putting it all together, we find the contribution of the antipodal saddle to the two-point function in \nref{goperator1} to be
\begin{equation}
\label{2ptint}
\int d^{D}x\,d^{D}y\, \langle K_{n}(x)K_{-n}(y)\rangle|_{\text{antipodal saddle}} \approx \lds^{2D}\text{Vol}(S^{D})(-i)^{D}\bigg(\frac{f_{\theta,\text{phy}}^{2}D\text{Vol}(S^{D})}{2 \pi n^{2}}\bigg)^{\frac{D}{2}}\lds^{\frac{D(D-2)}{2}}
\end{equation}
where we wrote $f_{\theta}$ in terms of its dimensionful counterpart to make the overall dependence of the result in $\lds$ manifest. Note that the result \nref{2ptint} implies that the integrated two-point function has a phase of $(-i)^{D}$. We therefore have that
\begin{equation}
\begin{gathered}
Z_{\text{wormhole}}|_{\text{small }Q}\approx Z_{\text{GR}}(S^{D})\times Z_{D-2}(S^{D})\times \mathcal{C}_{n,-n}\int d^{D}x\,d^{D}y\, \langle K_{n}(x)K_{-n}(y)\rangle|_{\text{antipodal saddle}}\\\sim i^{D+2}(-i)^{D}=i^{2}
\end{gathered}
\end{equation}
where we used $\sim$ meaning "has the same phase as". Note that in the second line, we used that the phase of $Z_{\text{GR}}(S^{D})$ is famously $i^{D+2}$ (\cite{Polchinski:1988ua, Maldacena:2024spf}), and that the phase of $Z_{D-2}(S^{D})$ is one. Assuming that $\mathcal{C}_{n,-n}$ is positive, we then rederive the phase of the wormhole from the operator picture.

We should also briefly discuss the interpretation of these light modes when there are $N$ wormhole cycles. Since we can interpret this saddle as having $N$ spheres connected to their neighbouring spheres by a small handle, we can think of this solution as $N$ spheres with $2 N$ operator insertions. The light modes then follow from the $N=1$ picture we proposed. The pair of operators should be contributing by their pairwise contractions in their respective spheres, each giving a factor of \nref{2ptint}. This implies that the phase of the wormhole with $N$ cycles is the same as the phase of $N$ spheres and the appropriate operator insertions, which is $Z_\text{wormhole}\sim (i^{D+2})^{N}(-i)^{DN}=i^{2N}$, as we found in section \nref{featspec}.

\subsection{Near Einstein light modes}
\label{neareinlight}

\begin{figure}[h!]
    \centering
    \includegraphics[width=1\linewidth]{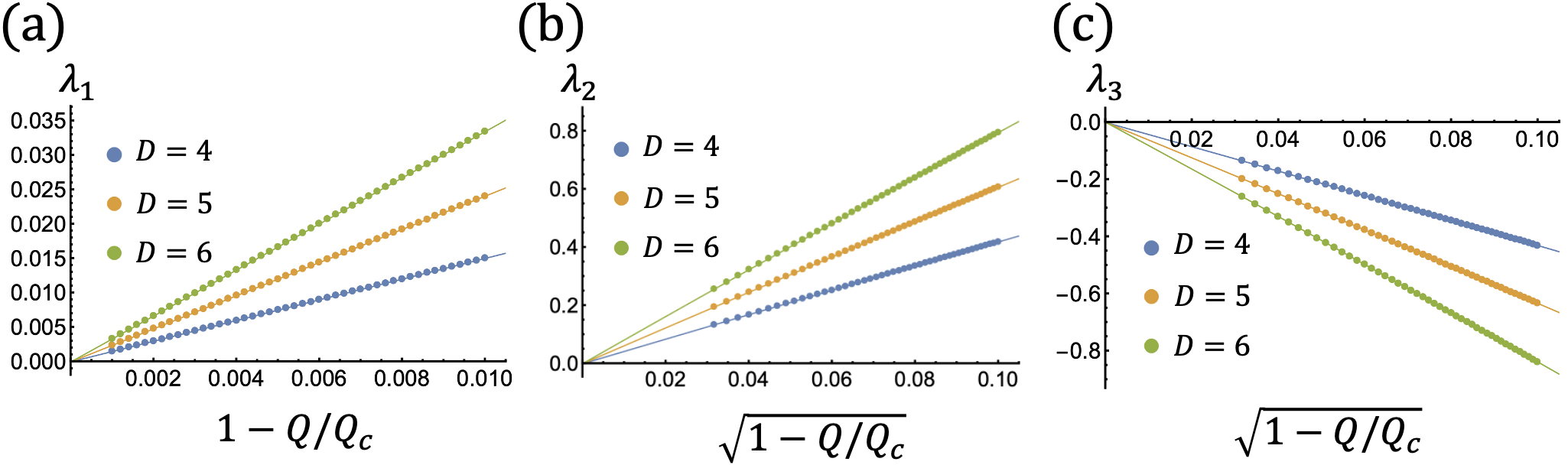}
    \caption{In \textbf{(a), (b), (c)} we plot the eigenvalue of three light modes $\lambda_1,\lambda_2,\lambda_3$ in the near Einstein limit as a function of $Q$ for various $D$. The dots are numerical data, and the underlying lines are a linear fit.}
    \label{fig: near einstein light modes}
\end{figure}

Here we discuss modes of the spectrum that go to zero as $Q \rightarrow Q_{c}$, with $Q_{c}$ the critical flux of the wormhole. We observe numerically that there are three sets of eigenvalues of this type, with different scalings as $Q \rightarrow Q_{c}$, as shown in Figure \nref{fig: near einstein light modes}. Also, these modes are all in the $l=0$ sector, so their degeneracy is one. The number of light modes and scaling persist even if the wormhole has $N$ cycles instead of one.

Parameterizing the charges as $Q^{2}=Q_{c}^{2}(1-\epsilon)$, we see one of the modes approaching zero linearly in $\epsilon$, while the other two approach as $\sqrt{\epsilon}$. One of the modes that becomes zero was negative before, which explains why the phase of the Einstein wormhole we discussed in section \nref{featspec} has an extra factor of $i$ as compared to its non-extremal counterpart.

The fact that the near-Einstein wormhole has these light modes implies that the functional determinant of the fluctuation operator, $\text{Det}' M^{-\frac{1}{2}}$, blows up as $Q \rightarrow Q_{c}$, since the other contributions are regular. In fact
\begin{equation}
\text{Det}'M^{-\frac{1}{2}}\sim (\epsilon (\sqrt{\epsilon})^{2})^{-\frac{1}{2}}\sim \epsilon^{-1}
\end{equation}
where we use $\sim$ to indicate we ignore an $O(1)$ constant, since we are just looking at the scaling with $\epsilon$.

However, we also find that the ghost determinant develops a light mode as $Q \rightarrow Q_{c}$ (see Appendix \nref{ddgh}). It comes in the scalar sector of diffeomorphisms, with $l=0$. This single light mode comes from the fact that at $Q=Q_{c}$ the solution has an extra isometry, namely the time translations along the $\tau$ direction. Therefore, there should be light ghost modes near $Q=Q_{c}$ corresponding to the fact that the near-Einstein solution weakly breaks this isometry. We furthermore find that the eigenvalue of this ghost light mode scales linearly in the near-Einstein parameter $\epsilon$. 

Therefore, since there are no other contributions that blow up as $Q \rightarrow Q_{c}$, the overall one-loop determinant of the wormhole in \nref{1lpsum} behaves as
\begin{equation}
Z_{\text{wormhole}}\sim \epsilon^{-1}\times \epsilon =1
\end{equation}
which implies that the one-loop determinant does not blow up or go to zero as $Q \rightarrow Q_{c}$. This is an interesting statement because one would naively expect the one-loop determinant to receive an enhancement near $Q=Q_{c}$ due to the existence of extra zero modes in the Einstein solution. However, this enhancement is precisely cancelled by the contribution of near isometries. 

\section{Small wormholes and operator insertions}
\label{smwhopis}

Here we discuss in more detail the comparison between the integrated two-point function in \nref{goperator1} and the gravitational path integral over the $N=1$ wormhole. To be more specific, one would like to argue that for $\kappa Q \approx 0$, the following identity should hold
\begin{equation}
\label{clid}
\mathcal{C}_{n,-n}\int d^{D}x\, d^{D}y\, \langle K_{n}(x)K_{-n}(y)\rangle|_{\text{antipodal saddle}} \approx \frac{Z_{\text{wormhole}}}{Z_{\text{GR}}(S^{D})Z_{D-2}(S^{D})}
\end{equation}
with $K_{\pm n}$ magnetic operators, dual to the scalar operators $e^{\frac{2 \pi i}{f_{\theta}}n\theta(x)}$.

That is, the path integral over the wormhole, normalized by the partition function of pure gravity and the form theory in the sphere, should be like an appropriate integral over two operator insertions that correspond to the wormhole mouths. In section \nref{nearslight} we argued that the phase of the two sides of \nref{clid} match, which is an already interesting statement. However, one could perhaps wonder if we can also match the dependence of both sides on parameters such as $\kappa$, $f_{\theta}$, $n$, and $\lds$ as well.

Of course, as we argued in section \nref{nearslight}, the left-hand side of \nref{clid} is not fixed but instead has a free constant $\mathcal{C}_{n,-n}$ that tells us how the normalization of the $x$,$y$ integration. However, $\mathcal{C}_{n,-n}$ should not depend on $\lds$ since the relation between the small wormhole and the operator insertions should not be sensitive to the ambient de Sitter. Therefore, there is a non-trivial match of powers of $\lds$ in both sides of \nref{clid} that has to happen.

To perform this matching, we need to evaluate \nref{1lpend} for when the charge $Q$ of the wormhole is very small, and divide it by the appropriate sphere answer. We will be able to do that in odd $D$ using a mix of analytical and numerical observations, while making a few reasonable assumptions about other effects for functional determinants on spacetimes with a small handle. 

In the following subsections, we will discuss how to compare the many different terms on the right-hand side of \nref{clid} between the wormhole and sphere, before we put them all together to solve for $\mathcal{C}_{n,-n}$. In doing so, we will not care about $O(1)$ factors, since we are interested only in the scaling of the answer with the parameters of the problem. 

The section is organized as follows: To compare the one-loop determinant of the wormhole with its sphere counterpart, it will be important for us to take ratios of functional determinants in both spacetimes. For some of the relevant functional determinants, there will be new eigenmodes in the wormhole spacetime that have no sphere counterpart. To compute the ratio of these determinants carefully, we will have to discuss how to consistently introduce the effect of these new modes. In section \nref{lochand} we propose a rule to do so, which we expect to hold in odd $D$ and potentially have further corrections in even $D$. However, we will see that using the rule will be important in order to find a coefficient $\mathcal{C}_{n,-n}$ that does not depend on $\lds$. In section \nref{rtwhsphere}, we compute the ratio between the one-loop determinant of the wormhole and its sphere counterpart in odd $D$, using equation \nref{1lpend} and discussing the many required steps carefully. In section \nref{matchsec} we match the ratio computed in section \nref{rtwhsphere} to the integrated two-point function in \nref{2ptint}, obtaining from it a consistent value for the EFT coefficient $\mathcal{C}_{n,-n}$ that does not depend on the de Sitter length $\lds$, as we expected from physical grounds. In section \nref{flatwh} we sketch how a similar estimate would go for a class of wormholes in flat space, and in particular what would be necessary to estimate $\mathcal{C}_{n,-n}$ in $D=4$.

\subsection{A rule for small handles}
\label{lochand}

Throughout this section, it will be important for us to compare functional determinants of differential operators $\mathcal{F}$ in a spacetime $\mathcal{M}_{0}$ versus a spacetime $\mathcal{M}$, which is almost the same as $\mathcal{M}_{0}$, but which has an small handle. In this subsection, we will propose a general principle that will act as an important computational clutch for this calculation. 

The motivation for introducing such a principle is that to compute functional determinants carefully, one would generally have to use techniques such as heat-kernel regularization, where we keep track of divergences and deal with them in a consistent way. This is an interesting future direction, but not something we studied in this paper.

However, throughout the paper, we discussed in detail how to obtain the low-lying spectrum of many relevant differential operators from numerics. We might therefore wonder if, from this data alone, we can recover any useful features of the ratio between functional determinants in the spacetimes $\mathcal{M}$ and $\mathcal{M}_{0}$. Here, we will propose a way to do that for spacetimes with an odd number of dimensions. For even $D$, we expect the rule to be a bit more complicated.

To illustrate our problem, let us define a local "path integral" associated with a self-adjoint operator $\mathcal{F}$. For us to do that, we will find it useful to keep the spacetime coordinates dimensionful, unlike in the rest of the paper. This implies that to make the path integral manifestly dimensionless, we should keep the dimensions of different terms in mind.

Assuming the operator $\mathcal{F}$ has length dimension $-2$, we will construct associated to it an dimensionless action $I_{\mathcal{F}}$
\begin{equation}
I_{\mathcal{F}}=\frac{1}{2}\int \Psi\mathcal{F}\Psi
\end{equation}
where $\Psi$ could in principle have indices that are properly contracted, but that we omit for simplicity. We will then find it convenient to define a local norm for $\Psi$ via
\begin{equation}
\label{mulocnorm}
(\Psi,\Psi)_{\mu}=\mu^{2}\int \Psi\Psi
\end{equation}
where $\mu$ has length dimension $-1$, so that the norm is dimensionless. From this norm, we can define a unique measure $D\Psi$ for $\Psi$, up to an overall normalization, which we fix by requiring the following identity to hold 
\begin{equation}
\label{psilocnorm}
\int D\Psi\,e^{-\frac{1}{2}(\Psi,\Psi)_{\mu}}=1
\end{equation}

This implies that if $\Psi_{n}$ is a orthogonal basis of $\Psi$ in the manifold such that $\int \Psi_{n}\Psi_{m}=\delta_{nm}$, then
\begin{equation}
\label{pims}
D\Psi=\prod_{n}\frac{\mu \,d\Psi_{n}}{\sqrt{2\pi}}
\end{equation}
which is a manifestly dimensionless measure.

Using all of this, we can define a path integral for $\mathcal{F}$ via
\begin{equation}
\label{pifk}
Z_{\mathcal{F}}(\mathcal{M})=\int_{\mathcal{M}} D\Psi\, e^{-\frac{1}{2}(\Psi,\tilde{\mathcal{F}}\Psi)_{\mu}}=Z_{0,\tilde{\mathcal{F}}}(\mathcal{M})\text{Det}'(\tilde{\mathcal{F}}^{-\frac{1}{2}})_{\mathcal{M}}
\end{equation}
where we defined $\tilde{\mathcal{F}}=\mu^{-2}\mathcal{F}$ to be the dimensionless version of $\mathcal{F}$, normalized by $\mu$. Note that while $\mu$ is necessary to keep things dimensionless, the calculation should be independent of $\mu$. The reason is that rescaling the measure of all modes $\Psi_{n}$ by a factor of $\mu$ will change the measure only by a $\mu$-dependent local factor, which can be absorbed into local counterterms\footnote{This is only strictly true in odd $D$ where there is no Weyl anomaly. In even $D$, there is an anomaly, and the overall factor of $\mu$ after the rescaling will be proportional to an appropriate heat kernel coefficient. This $\mu$ dependent term will combine with some higher derivative counterterms in the action.}. This fact will be very important for us later.

For now, we should go back and discuss each term of \nref{pifk}. The term $\text{Det}'(\tilde{\mathcal{F}}^{-\frac{1}{2}})$ is the product over non-zero eigenvalues of $\tilde{\mathcal{F}}$, e.g, its functional determinant with zeros removed. The factor $Z_{0,\tilde{\mathcal{F}}}$ is a path integral over zero modes of $\tilde{\mathcal{F}}$ that we assume is finite, and derived from an appropriate rule from the integration over $\Psi$\footnote{An example of such a rule is the compact range of integration for the constant mode of a period scalar.}. 

Taking the handle to be very small, we assume that all non-zero eigenvalues of the operator $\mathcal{F}$ in the spacetime $\mathcal{M}_{0}$ should carry over to $\mathcal{M}$ with minor modifications. This implies that if $\lambda_{n}^{(0)}$ is an eigenvalue of $\tilde{\mathcal{F}}$ in $\mathcal{M}_{0}$, then its eigenvalue $\lambda_{n}$ is $\mathcal{M}$ is
\begin{equation}
\label{eigshift}
\lambda_{n}=\lambda_{n}^{(0)}(1+\delta_{n})
\end{equation}
with $|\delta_{n}|\ll1$. We furthermore also assume that the path integral over the zero modes that already existed in $\mathcal{M}_{0}$ carries over with small modifications to $\mathcal{M}$. So far, these assumptions apply to all operators $\mathcal{F}$ that we discussed in this paper\footnote{In Figure \nref{fig: deviation from sphere}, we provide an explicit example of this point for scalar perturbations of the fluctuation operator $M$. To be more specific, looking at $l=0$ scalar fluctuations, we illustrate how $|\delta_n|$ approaches zero as one takes $Q\rightarrow0$. }.

However, it is also plausible that by adding the small handle, we generate new eigenmodes of $\tilde{\mathcal{F}}$ that have no $\mathcal{M}_{0}$ counterpart.  One possible type of new mode is, for example, light or zero modes of $\tilde{\mathcal{F}}$ that are exclusive to the spacetime $\mathcal{M}$ with the handle, such as the ones we discussed in section \nref{lightzero} for the fluctuation operator $M$. Here, with "light" we mean modes whose eigenvalues are very small in the scale of the handle. We call the contribution of these modes to the associated path integral \nref{pifk} to be $Z_{\text{handle},\text{low}}$. We have access to $Z_{\text{handle},\text{low}}$ from the low-lying spectrum of $\tilde{\mathcal{F}}$, so we can compute this factor explicitly. We should also mention that, in principle, the operator $\tilde{\mathcal{F}}$ could have new modes at very short wavelengths that are localized in the handle region, and therefore have no $\mathcal{M}_{0}$ counterpart. We found no such modes in the spectrum of the operators that we studied throughout the paper, but we do not rule them out. 

To compute the path integral over $\tilde{\mathcal{F}}$ in $\mathcal{M}$, we should take into account all of the factors we just discussed consistently in the right-hand side of \nref{pifk}. In principle, even the small relative deviations from the old eigenvalues, $\delta_{n}$, and the correction to the zero mode contributions, could add up to something relevant or even divergent, which we would have to regulate appropriately. However, since we know one of the novel terms to spacetime $\mathcal{M}$, namely $Z_{\text{handle},\text{low}}$, explicitly, we should include it directly in \nref{pifk}. Therefore, imagining that we can group all the other corrections to the $\mathcal{M}_{0}$ answer in an extra term, $Z_{\text{handle}}'$, it should be the case that we can write the path integral of $\tilde{\mathcal{F}}$ in $\mathcal{M}$ as
\begin{equation}
\label{zfsetup}
Z_{\mathcal{F}}(\mathcal{M})= Z_{\mathcal{F}}(\mathcal{M}_{0})\times Z_{\text{handle},\text{low}}\times Z_{\text{handle}}'
\end{equation}

The idea is that in a consistent evaluation of the functional determinant we are computing, we should be able to identify a series of terms which gives the original answer in $\mathcal{M}_{0}$, and we absorb them into the $ Z_{\mathcal{F}}(\mathcal{M}_{0})$ term. Furthermore, we include the contribution of the new low lying-handle modes explicitly in $Z_{\text{handle},\text{low}}$, and everything else in $Z_{\text{handle}}'$. That is, we think of $Z_{\text{handle}}'$ as the overall contribution from corrections to modes from $\mathcal{M}_{0}$, as well as the effect of any new handle modes at the deep UV, or any other relevant effects.

So far, \nref{zfsetup} is correct as a tautology, but we will try to use it to discuss how to consistently evaluate the ratio between $Z_{\mathcal{F}}(\mathcal{M})$ and $Z_{\mathcal{F}}(\mathcal{M}_{0})$ from the knowledge only of the low-lying handle modes we observe numerically. Our idea to argue it is as follows: Since the integral over all modes of $\Psi$ in $\mathcal{M}$ is local, it will not depend on the path integral measure we pick. We can pick a local measure for $\Psi$ in \nref{pifk}, which is rescaled over each mode of the manifold in \nref{pims} by a factor of $\tilde{\mu}_{n}=\mu_{n}\mu^{-1}$, such that we replace the old factor of $\mu$ by a new one $\mu_{n}$. As we discussed before, rescaling the contribution of each mode by $\tilde{\mu}_{n}$ should leave the path integral invariant\footnote{Again, with the caveat that the even $D$ problem has an anomaly}.

To apply this principle to \nref{zfsetup}, we just need to think about how the $\mu$ scaling will affect each term. Whatever is the contribution that produces $Z_{\mathcal{F}}(\mathcal{M}_{0})$ in \nref{zfsetup}, it must be invariant by the $\mu$ rescaling, since the path integral over $\mathcal{M}_{0}$ is already individually independent of $\mu$. Therefore, doing the $\tilde{\mu}_{n}$ rescaling to $Z_{\text{handle},\text{low}}$ and $Z_{\text{handle}}'$ should leave their product invariant.

This is a very non-trivial observation, because we derived $Z_{\text{handle},\text{low}}$ explicitly from the path integral in \nref{pifk}, whose measure depends on $\mu$. Therefore, if the spacetime with the handle has $\delta n_{\text{low}}$ new low-lying modes that have no $\mathcal{M}_{0}$ counterpart, then, $Z_{\text{handle},\text{low}}$ will behave under the $\tilde{\mu}_{n}$ scaling as $Z_{\text{handle},\text{low}}\rightarrow \tilde{\mu}_{n}^{\delta n_{\text{low}}}Z_{\text{handle},\text{low}}$. 

Therefore, if we treated $Z_{\text{handle}}'$ as a small $O(1)$ multiplicative correction, the path integral in \nref{zfsetup} would not be $\mu$ invariant and therefore not consistent. In other words, it is not consistent to think of the ratio of the path integrals in $\mathcal{M}$ and $\mathcal{M}_{0}$ as just being given by the contribution of the finitely many new handle modes, computed with a generic measure. For \nref{zfsetup} to be consistent, it must be that $Z_{\text{handle}}'$ transforms under the $\tilde{\mu}_{n}$ scaling as
\begin{equation}
\label{handrule}
Z_{\text{handle}}'\rightarrow \tilde{\mu}_{n}^{-\delta n_{\text{low}}}Z_{\text{handle}}'
\end{equation}

Equation \nref{handrule} then implies that $Z_{\text{handle}}'\sim \mu^{-\delta n_{\text{low}}}$, and since it must be dimensionless, it must come together with other dimensionful parameters of the problem. Therefore, while we cannot compute $Z_{\text{handle}}'$ exactly without performing a consistent evaluation of the functional determinant, we could argue its scaling with the parameters of the problem based on the property \nref{handrule}, and some simplicity assumptions.

Assuming the leading corrections to the path integral we are discussing to be sensitive only to the geometry at the length scale $r_{0,\text{phy}}$ of the handle, defined in our context as \eqref{rodef}, we expect the only other scale to enter $Z_{\text{handle}}'$ to be the handle scale\footnote{This seems to be a reasonable assumption for the wormhole problem we are discussing in the main text, since after rescaling the fields, the only scales of the problem are the handle size and the de Sitter length $\lds$. Other dimensionful parameters, such as $\kappa_{\text{phy}}$ and $f_{\theta,\text{phy}}$, generally only enter zero-mode contributions. Therefore, assuming that $Z_{\text{handle}}'$ is not sensitive to $\lds$, we arrive at the stated conclusion.}. Therefore, if we assume that the handle has a simple geometry, with only one dimensionful length parameter $r_{0,\text{phy}}$, then it must be that
\begin{equation}
\label{zhandpr}
Z_{\text{handle}}'\sim (\mu r_{0,\text{phy}})^{-\delta n_{\text{low}}}
\end{equation}
where $\sim $ means up to an $O(1)$ constant, which we cannot determine here without solving the problem explicitly. One might still be worried about the fact that other scales could maybe pollute the correction in \nref{zhandpr}, so we should try to get some intuition for the answer \nref{zhandpr} with an example.

The answer \nref{zhandpr} makes intuitive sense for the path integral over the wormhole degrees of freedom $\Phi$ that we discussed in \nref{z1loop}, if we assume that when $\kappa Q \sim0$ the sphere and the handle almost decouple from each other. If that is the case, the small handle degrees of freedom would contribute via the path integral of a gauge fixed flat space wormhole that ends in the pode and antipode of the sphere\footnote{As we discussed before, one should, however, treat the zero modes that move the wormhole mouths as light modes instead of zero modes. The reason is that the sphere is finite, and we are discussing the gauge fixed path integral.}. 

As we discussed, these wormholes have low-lying modes, which are the modes corresponding to moving the wormhole mouths, and the twist $SO(D)$ modes we discussed. The rest of the path integral over this small wormhole, with the boundary conditions set at their ambient spacetimes, would come from non-zero eigenvalues of order $\mu^{-2}r_{0,\text{phy}}^{-2}$, using the norm \nref{psilocnorm}. If there were no low-lying modes, the product over these non-zero modes would lead to no overall scaling with $(r_{0}\mu)$. However, due to the existence of these low-lying modes, the product over the non-zero modes will have an overall factor of $(\mu r_{0,\text{phy}})^{-\delta n_{\text{low}}}$, since the $\delta n_{\text{low}}$ low-lying modes are missing from the product.

This is an interesting point for us, because the scaling from these non-zero modes is precisely the scaling we obtain for $Z_{\text{handle}}'$ in \nref{zhandpr}, which implies that this would be the role of $Z_{\text{handle}}'$ in that context. Heuristically, one can perhaps see \nref{handrule} as telling us that $Z_{\text{handle},\text{low}}Z_{\text{handle}}'$ acts as a local path integral over the handle region. Therefore, at least pictorially, one could think of $Z_{\text{handle}}'$ as being the product over many eigenvalues natural to the handle scale, but missing a few low-lying modes that we took into account separately in $Z_{\text{handle},\text{low}}$. 

The important takeaway for us, then, is as follows: Even though by diagonalizing $\mathcal{F}$ and looking at its new low-lying eigenvalues in $\mathcal{M}$ we observe only the modes that contribute as $Z_{\text{handle},\text{low}}$, this term alone cannot give the full ratio between the path integral in $\mathcal{M}$ and the one in $\mathcal{M}_{0}$. Indeed, whatever scaling they give is not consistent unless we include a term $Z_{\text{handle}}'$ as in \nref{zfsetup}. Furthermore, assuming the only scale contributing to the effects that lead to $Z_{\text{handle}}'$ to be the handle scale $r_{0,\text{phy}}$, we can derive the scaling format of $Z_{\text{handle}}'$ in \nref{zhandpr}. 

For the purpose of our calculations in the rest of the paper where $\lds=1$, we must pick $\mu=\lds^{-1}$ for the scale in \nref{mulocnorm}, which implies that the correction in $Z_{\text{handle}}'$ is $r_{0}^{-\delta n_{\text{low}}}$ with $r_{0}=r_{0,\text{phy}}\lds^{-1}$. In the convention from the rest of the paper, we have therefore derived the rule that for handles with a single length scale $r_{0}$, in de Sitter units, it holds that
\begin{equation}
\label{rtfct}
\frac{Z_{\mathcal{F}}(\mathcal{M})}{Z_{\mathcal{F}}(\mathcal{M}_{0})}=Z_{\text{handle},\text{low}}^{(\mu)}Z_{\text{handle}}'^{(\mu)}\sim  Z_{\text{handle},\text{low}}r_{0}^{-\delta n_{\text{low}}}
\end{equation}
where $Z_{\text{handle},\text{low}}$ is computed with $\mu=\lds^{-1}$, or equivalently in units where $\lds=1$. 

That is, to compute the ratio \nref{rtfct} up to $O(1)$ contributions, we only need the path integral contribution of the low lying modes, with each of them coming with a factor of $r_{0}^{-1}$. The mnemonic for the $r_{0}^{-1}$ is that these low-lying modes are a missing would-be handle mode, with eigenvalue of order $r_{0}^{-2}$, from their path integral over non-zero modes. For more evidence for this rule, check Appendix \nref{checklochand}, where we argue that this rule is necessary for powers of $r_{0}$ to match in two quantities related by electromagnetic duality in odd $D$.

For even $D$, we expect the right-hand side of \nref{rtfct} to possibly contain further scaling corrections, involving the ratio of physical scales of the problem to the cutoff of the theory.

\subsection{Ratio between wormhole and sphere}
\label{rtwhsphere}

In this section, we discuss the ratio between the one-loop path integral of the gravity+forms modes around the wormhole, in the limit that $\kappa Q$ is very small, and their analogue path integral in the sphere. To be more specific, we will not worry about $O(1)$ numerical contributions to the ratio, but instead only its dependence on parameters of the problem, such as $n$, $\kappa$, and $f_{\theta}$. Also, we will need to use the rule we developed in section \nref{lochand}, so we restrict ourselves to computing the ratio for odd $D$.

To understand the ratio, we can use the simplified formula \nref{1lpend} for the path integral, which applies for both manifolds\footnote{Note, however, that $Z_{\text{iso}}$ for the sphere will be related to a division by $SO(D+1)$ isometries instead of $SO(D)$.}. For the reader's convenience, we rewrite the formula below
\begin{equation}
Z_{1-\text{loop}}=Z_{\text{twist}}Z_{\text{iso}}|\text{Det}'(\delta_{\xi}P)|\frac{\text{Det}'(M^{-\frac{1}{2}})\text{Det}'(\Delta_{0}^{-\frac{1}{2}})}{\text{Det}'(\Delta_{D-2}^{-\frac{1}{2}})}\text{det}^{\frac{1}{2}}\bigg(\frac{f_{\theta}^{2}}{2 \pi}\Gamma_{0}\bigg){}\text{det}^{-\frac{1}{2}}\bigg(\frac{f_{\theta}^{2}}{2 \pi}\Gamma_{1}\bigg)
\end{equation}
where we already used that the anomaly term $S_{\text{ano,em}}$ in \nref{1lpend} is zero for odd $D$.

We briefly discussed many of those terms in section \nref{prelim}, where we introduced the formula, but we will briefly review their meaning when we discuss them here. First, we discuss the contribution from the ratio between
\begin{equation}
\label{z1term}
\frac{Z_{\text{twist}}\text{Det}'(M^{-\frac{1}{2}})}{\text{Det}'(\Delta_{D-2}^{-\frac{1}{2}})}=\frac{\text{det}^{\frac{1}{2}}\bigg(\frac{2\pi}{f_{\theta}^{2}}\Gamma_{D-2}\bigg)Z_{\text{twist}}\text{Det}'(M^{-\frac{1}{2}})}{\text{det}^{\frac{1}{2}}\bigg(\frac{2\pi}{f_{\theta}^{2}}\Gamma_{D-2}\bigg)\text{Det}'(\Delta_{D-2}^{-\frac{1}{2}})}
\end{equation}
where in the second line we included the $\Gamma$ matrix contribution in the numerator and denominator. $\Gamma_{p}$ is the inner product matrix for a basis of harmonic $p$-forms of the manifold\footnote{More specifically, a basis for the set of harmonic $p$-forms that have integer flux around non-contractible $p$-dimensional surfaces. See \cite{Donnelly:2016mlc} or Appendix \nref{formdet} for more details.}. The idea is that these harmonic forms are zero modes of a path integral over $p$-forms weighted by the $p$-form Hodge Laplacian $\Delta_{p}$. The determinant of the $\Gamma_{p}$'s with these given prefactors is the contribution from these zero modes for a theory with coupling constant $f_{\theta}$.

Therefore, we can interpret the denominator of \nref{z1term} as a gauge fixed local path integral over $D-2$-forms, weighted with $\Delta_{D-2}$ as fluctuation operator. We can also interpret the numerator of \nref{z1term} as a local path integral, since this is precisely the gauge fixed path integral over $\Phi$ with fluctuation operator $M$ that we discussed in section \nref{genset}. In particular, the two zero-mode contributions $Z_{\text{twist}}$ and $\text{det}^{\frac{1}{2}}\big(\frac{2\pi}{f_{\theta}^{2}}\Gamma_{D-2}\big)$ are the two we discussed exist for this path integral in section \nref{zeromod}.

Since these are both local path integrals, and we want to compare their answers to their counterpart in $S^{D}$, we can use the rule we developed in section \nref{lochand}, that is, equation \nref{rtfct}. To apply the rule, we just need to know the new low-lying modes in the spectrum of the fluctuation operators $M$ and $\Delta_{D-2}$, e.g, the modes with no $S^{D}$ counterpart. For the operator $M$, these are the modes we reviewed in section \nref{lightzero}, e.g, they are: $2D$ light modes associated with moving the wormhole mouths, and $|SO(D)|$ twist zero modes. Furthermore, if $D=3$, there is one extra harmonic 1-form zero mode, which will enter as the non-zero contribution $\text{det}^{\frac{1}{2}}\big(\frac{2\pi}{f_{\theta}^{2}}\Gamma_{D-2}\big)$. Note that this term is equal to one if $D \neq 3$, since then $\Gamma_{D-2}$ would be trivial.

For the path integral in the denominator of \nref{z1term}, there are only new zero modes in $D=3$, which are precisely the same extra modes that appear in the numerator at $D=3$. We can therefore conclude that
\begin{equation}
\begin{gathered}
\label{z1ratio}
\bigg(\frac{Z_{\text{twist}}\text{Det}'(M^{-\frac{1}{2}})}{\text{Det}'(\Delta_{D-2}^{-\frac{1}{2}})}\bigg)_{\text{wormhole}}\bigg(\frac{\text{Det}'(M^{-\frac{1}{2}})}{\text{Det}'(\Delta_{D-2}^{-\frac{1}{2}})}\bigg)_{S^{D}}^{-1}\\\sim \frac{\text{det}^{\frac{1}{2}}\bigg(\frac{2\pi}{f_{\theta}^{2}}\Gamma_{D-2}\bigg)Z_{\text{twist}}Z_{\text{light}}r_{0}^{-(2D+|SO(D)|+x)}}{\text{det}^{\frac{1}{2}}\bigg(\frac{2\pi}{f_{\theta}^{2}}\Gamma_{D-2}\bigg)r_{0}^{-x}}=Z_{\text{twist}}Z_{\text{light}}r_{0}^{-(2D+|SO(D)|)}
\end{gathered}
\end{equation}
where $x=1$ if $D=3$ and is zero otherwise. $Z_{\text{light}}$ is the contribution from the $2D$ light modes associated with moving the wormhole mouths. 

Now, we need to discuss the contribution from the other terms to the ratio. At least scaling-wise, we point out that both $\text{Det}'(\Delta_{0}^{-\frac{1}{2}})$ and $\text{det}^{\frac{1}{2}}\big(\frac{f_{\theta}^{2}}{2 \pi}\Gamma_{0}\big)$ are small deformations of their sphere counterparts. The reason is that we observe numerically that the spectrum of $\Delta_{0}$ is a small deformation of the sphere one, with no new low-lying modes. Also, the $\Gamma_{0}$ contribution was explained in the Appendix \nref{formdet}, where we showed that it is just a simple function of the volume of the manifold.

Therefore, we just need to focus on the other terms of \nref{1lpend}, namely the $\Gamma_{1}$ term, $Z_{\text{iso}}$, and the functional determinant of the de Donder gauge condition \nref{dedodgh}. The $\Gamma_{1}$ term is only non-trivial for the wormhole, since the sphere does not have harmonic 1-forms. $\Gamma_{1}$ is non-trivial for the wormhole due to the $S^{1}$ factor. The contribution from the division over isometries, $Z_{\text{iso}}$, is different between the sphere and the wormhole because their isometry groups are different. Namely, the sphere isometry group is $SO(D+1)$ while the wormhole's is $SO(D)$. This means that the division by isometries in the sphere has $D$ more generators. The contribution from the $SO(D)$ isometries, however, is about the same in both the wormhole and the sphere, and they cancel out with each other in the ratio. We call the contribution of these residual isometries in the sphere $Z_{\text{iso,transl}}$, since we can think of the $D$ broken isometries as translations along the sphere.

We then have to discuss the contribution from the functional determinant of the de Donder condition \nref{dedodgh}. The only difference between the functional determinant in the wormhole is that it has extra light modes as compared to the sphere answer. These are not new modes; they are just previous zero modes of the functional that are lifted by the wormhole geometry. The reason is that the zero modes of $\delta_{\xi}P$ are the isometries of the background, and since the wormhole has fewer isometries, it has fewer zero modes. Since the wormhole only weakly breaks the isometries, these previous zero modes become light modes. We call the contribution of these ghost light modes $Z_{\text{light,gh}}$. Note that none of these three new contributions that we discussed involves new handle modes, so they do not come with extra factors of $r_{0}^{-1}$.

Putting all these points together, we conclude that the ratio between the one-loop path integral over the wormhole and its sphere counterpart is
\begin{equation}
\label{z1step}
\bigg(\frac{Z_{\text{wormhole}}}{Z_{\text{GR}}(S^{D})Z_{\text{D-2}}(S^{D})}\bigg)_{\text{1-loop}}\sim Z_{\text{twist}}Z_{\text{light}}r_{0}^{-(2D+|SO(D)|)}\text{det}^{-\frac{1}{2}}\bigg(\frac{f_{\theta}^{2}}{2 \pi}\Gamma_{1}\bigg)\frac{Z_{\text{light,gh}}}{Z_{\text{iso,transl}}}
\end{equation}
where, again, we are doing equalities up to $O(1)$ constants. 

Now, we discuss how to directly evaluate all these terms. The simpler terms to describe are the ones that come from a couple of light eigenvalues, which are $Z_{\text{light}}$ and $Z_{\text{light},\text{gh}}$. The $Z_{\text{light}}$ term is the path integral contribution of $2D$ light eigenvalues that go as $\kappa Q$, and it therefore goes as $(\kappa Q)^{-\frac{2D}{2}}=(\kappa Q)^{-D}$. Similarly, the light ghost eigenvalues also scale as $\kappa Q$, and since there are $D$ of them, they contribute as $Z_{\text{light,gh}}\sim (\kappa Q)^{D}$. Therefore, these two contributions cancel each other up to $O(1)$ factors. 

To evaluate the isometry contribution $Z_{\text{iso},\text{transl}}$, we can use that they are given by a normalization factor $\mathcal{N}_{\text{iso}}^{-1}$, discussed in \nref{niso}, per isometry generator, and a division by an appropriate group volume. Since we only care about the dependence on the parameters of the solution, the only thing that matters to us is that $\mathcal{N}_\text{iso}^{-1}\sim \kappa$, such that the $D$ isometries will contribute as $Z_{\text{iso,transl}}\sim\kappa^{D}$. Note that $\mathcal{N}_{\text{iso}}$ is also a function of the norm of the relevant isometries, but these will be $O(1)$. 

Since the $Z_{\text{light}}$, $Z_{\text{light},\text{gh}}$ and $Z_{\text{iso},\text{transl}}$ terms are all related to the motions of the wormhole mouths and translation isometries, we find convenient to group them together with the $r_{0}^{-2D}$ factor to obtain
\begin{equation}
\frac{Z_{\text{light}}Z_{\text{light,gh}}}{Z_{\text{iso,transl}}}r_{0}^{-2D}\sim \kappa^{-D}r_{0}^{-2D}
\end{equation}

We are then left to explain the twist contribution and the harmonic 1-form contribution in the $\Gamma_{1}$ term. To discuss $Z_{\text{twist}}$, we can use the result for it introduced in section \nref{zeromod}, where we noted that this contribution splits into a group volume factor and a normalization factor. The group volume factor is $O(1)$, so we only care about the normalization factor. To obtain this normalization factor, we need the norm for the twist modes $\Phi_{c}$ generated locally by the coordinate transformation \nref{rlxigen} and an appropriate form gauge transformation.  

These modes are of the form
\begin{equation}
\label{rlxih}
h_{ab}=\frac{2n_{(a}c_{b)}}{\sqrt{2}\kappa \,I_{\text{twist}}}\frac{1}{a^{D+1}}
\end{equation}
with $I_{\text{twist}}$ as in \nref{rlxigen} and $c_{a}$ an isometry of $SO(D)$ which respects the usual Lie algebra. The $\Phi_{c}$ mode also has an associated $b_{ab}$. We do not know the analytic form of $b_{ab}$ explicitly, but numerically we observe that its contribution to the norm is of the same order as the contribution of the metric deformation $h_{ab}$, which is all the information we need. Therefore, using \nref{ntwist}, up to an $O(1)$ factor the normalization factor $\mathcal{N}_{\text{twist}}$ is
\begin{equation}
\mathcal{N}_{\text{twist}}^{2}\sim \frac{1}{\kappa^{2}I_{\text{twist}}^{2}}\int \frac{d\tau}{a^{D+1}}=\frac{1}{\kappa^{2}I_{\text{twist}}}\sim \frac{r_{0}^{D}}{\kappa^{2}}
\end{equation}
where we used that for small $\kappa Q$ the integral $I_{\text{twist}}$ is dominated by the region \nref{smwhreg} of the integral where $a$ is small, and both $a$ and $d\tau$ are of order $r_{0}$. We have one such factor for each isometry generator, and there are $|SO(D)|=\frac{D(D-1)}{2}$ of them. Also, this contribution we just discussed comes together with a factor of $r_{0}^{-|SO(D)|}$ since each of these modes is a new handle mode of the fluctuation operator $M$ in \nref{actm}. Putting all these points together, we therefore have that
\begin{equation}
Z_{\text{twist}}r_{0}^{-|SO(D)|}\sim \bigg(\frac{r_{0}^{D}}{\kappa^{2}}\bigg)^{\frac{|SO(D)|}{2}}r_{0}^{-|SO(D)|}=\bigg(\frac{r_{0}^{D-2}}{\kappa^{2}}\bigg)^{\frac{|SO(D)|}{2}}
\end{equation}

The last factor we need to discuss is the contribution from harmonic 1-forms, e.g, the determinant of the $\Gamma_{1}$ matrix. We discussed this in more detail in section \nref{formdet}, where we derived equation \nref{gamma1cont} for this contribution. More specifically, we can evaluate it in the regime we are interested in as
\begin{equation}
\text{det}^{-\frac{1}{2}}\bigg(\frac{f_{\theta}^{2}}{2 \pi}\Gamma_{1}\bigg)=\bigg(\frac{f_{\theta}^{2}}{2\pi}\text{Vol}(S^{D-1})\bigg(\int_{\text{loop}}\frac{d\tau'}{a^{D-1}(\tau')}\bigg)^{-1}\bigg)^{-\frac{1}{2}}\sim (f_{\theta}^{2}r_{0}^{D-2})^{-\frac{1}{2}}
\end{equation}
since for small $\kappa Q$ the relevant integration along $\tau$ will be dominated by the handle region \nref{smwhreg}. Putting all of these results together, we derive that
\begin{equation}
\label{zratiowh}
\bigg(\frac{Z_{\text{wormhole}}}{Z_{\text{GR}}(S^{D})Z_{{D-2}}(S^{D})}\bigg)_{\text{1-loop}}\sim r_{0}^{-2D}\kappa^{-D}\bigg(\frac{r_{0}^{D-2}}{\kappa^{2}}\bigg)^{\frac{|SO(D)|}{2}}(f_{\theta}^{2}r_{0}^{D-2})^{-\frac{1}{2}}
\end{equation}
as we wanted to find.

\subsection{Operator matching}
\label{matchsec}

In the previous section, we obtained the ratio \nref{zratiowh} between the wormhole one-loop path integral and its sphere counterpart at odd $D$. We would now like to match this normalized wormhole answer with the result for the integrated two-point function that we discussed in \nref{2ptint}. For this, we also need to consider the contribution from the action in the ratio between the wormhole and the sphere. 

In fact, the action contribution is the most relevant one to this ratio in the semiclassical limit. The change in the Einstein-Hilbert+matter action was discussed in section \nref{onactsec} where we derived equation \nref{onactflat}. While this term is the most relevant change in the action, taking GR as an EFT, there are, in principle, a series of higher derivative terms in the action, suppressed by higher powers of $\kappa$. In fact, these terms are necessary for our one-loop determinant calculation to have a well-defined finite answer in the end, since the one-loop divergences are absorbed into local counterterms in the action.

The structure of the counterterms, however, is strongly dimension-dependent, and we cannot write a general formula for their variation that is simple for any $D$. We therefore call the contribution from the variation of these extra terms $\Delta S_{\text{ct}}$, which is a term that needs to be explored for the specific $D$ one is interested in.

Putting all of these results together, we obtain
\begin{equation}
\begin{gathered}
\label{match}
\mathcal{C}_{n,-n}\sim \kappa^{-D}\bigg(\frac{r_{0}^{D-2}}{\kappa^{2}}\bigg)^{\frac{|SO(D)|}{2}}(f_{\theta}^{2}r_{0}^{D-2})^{-\frac{1}{2}}r_{0}^{-2D}\lds^{-2D}\bigg(\frac{n}{f_{\theta}}\bigg)^{D}\exp\bigg(-\frac{2 \pi^{2} n}{\kappa _{\text{phy}}f_{\theta,\text{phy}}}\sqrt{\frac{D-1}{D-2}}-\Delta S_{\text{ct}}\bigg)\\
=r_{0,\text{phy}}^{-2D}\bigg(\frac{n}{\kappa_{\text{phy}}f_{\theta,\text{phy}}}\bigg)^{\frac{(|SO(D)|+2D)}{2}}\frac{1}{(n \kappa_{\text{phy}}f_{\theta,\text{phy}})^{\frac{1}{2}}}\exp\bigg(-\frac{2 \pi^{2} n}{\kappa _{\text{phy}}f_{\theta,\text{phy}}}\sqrt{\frac{D-1}{D-2}}-\Delta S_{\text{ct}}\bigg)\\
=\bigg(\frac{n \kappa_{\text{phy}}}{f_{\theta,\text{phy}}}\bigg)^{-\frac{2D}{(D-2)}}\bigg(\frac{n}{\kappa_{\text{phy}}f_{\theta,\text{phy}}}\bigg)^{\frac{(|SO(D)|+2D)}{2}}\frac{1}{(n \kappa_{\text{phy}}f_{\theta,\text{phy}})^{\frac{1}{2}}}\exp\bigg(-\frac{2 \pi^{2} n}{\kappa _{\text{phy}}f_{\theta,\text{phy}}}\sqrt{\frac{D-1}{D-2}}-\Delta S_{\text{ct}}\bigg)
\end{gathered}
\end{equation}
where we made the $\lds$ dependence explicit by rescaling the couplings in the second line, and used equation \nref{rodef}. From this, we see that the constant $\mathcal{C}_{n,-n}$ is indeed independent of $\lds$, as expected physically! We see this, in particular, as evidence for the rule \nref{rtfct} that we proposed in section \nref{lochand}. Indeed, if that rule were slightly different, we would have found overall factors of $\lds$ in the right-hand side of \nref{match}\footnote{To be more specific, our main assumption to derive \nref{rtfct} was that the only scale entering \nref{zhandpr} was $r_{0,\text{phy}}$. For the problem we are discussing, the only other scale that could have entered \nref{zhandpr} was $\lds$. If any power of $\lds$ also appeared in the right-hand side of \nref{zhandpr}, there would be extra factors of $\lds$ in the right side of \nref{match}.}. 

Now, we should discuss the one-loop contribution in \nref{match}. Because $\mathcal{C}_{n,-n}$ has a length dimension of $-2D$, it makes sense that it contains a term $r_{0,\text{phy}}^{-2D}$, since this coefficient naturally inherits the scale of the wormhole. Moreover, $\mathcal{C}_{n,-n}$ has an enhancement term given by the ratio of $n$ and the dimensionless combination $\kappa f_{\theta}$. We have one such term for each zero mode of the handle, if we view it as a flat space wormhole with $2D+|SO(D)|$ zero modes. In a way, this zero-mode contribution is the one we would expect for an instanton with action $\frac{n}{\kappa_{\text{phy}} f_{\theta,\text{phy}}}$, which is the action of the wormhole, and this amount of zero modes. Other than that, $\mathcal{C}_{n,-n}$ also received another non-zero contribution $(n \kappa f_{\theta})^{-\frac{1}{2}}$ from the integral over harmonic $1$-forms in $\Gamma_{1}$. This term is a bit less intuitive physically, but was forced into the calculation by the ghost determinant result in \nref{ghostL}.

\subsection{Comments on flat space wormholes and $D=4$}
\label{flatwh}

One would like to compute the coefficients $\mathcal{C}_{n,-n}$ for wormholes in our universe. To do so, we would have to extend our calculation from the previous section to $D=4$, where many anomalies that we ignored would be relevant.

We expect this would make a rigorous calculation of the ratio of the wormhole path integral to its pure sphere counterpart quite complicated. Therefore, at least to understand the scaling of $\mathcal{C}_{n,-n}$ with parameters of the wormhole, it would be more natural to consider a simpler problem. The simpler problem would be the path integral directly of the "handle" part of the geometry, where we treat it as a flat space wormhole solution with appropriate boundary conditions at its asymptotic regions. 

More specifically, we consider the wormhole as a contribution to the path integral with boundary conditions at large ambient spacetimes, and we compute its ratio to that of a contribution with no wormhole. We furthermore assume that the scales of the ambient spacetimes do not change the finite part of the ratio. That is, the ratio to the solution with no wormhole is there just to regulate the contributions from the asymptotic far-away regions of the geometry. This is a limit of the de Sitter problem we discussed, where we take $\lds\rightarrow \infty$.

What makes this problem simpler than the previous one we studied is that it has only one scale, namely the physical scale $r_{0,\text{phy}}$ of the handle region \nref{smwhreg}. Also, all the wormhole moduli that we discussed in the last section would be zero modes of this solution, since the position moduli are no longer lifted. This is related to the fact that, from the effective operator side, this wormhole and its moduli reproduce the integral of the effective operators over the entire ambient manifolds, since nothing makes them localize to specific positions.

Let us discuss first how the path integral over a generic $SO(D)$ symmetric wormhole of this type would go, for general $D$. For this calculation, we could imagine rescaling all coordinates by appropriate factors of $r_{0,\text{phy}}$, and absorbing this scale into the couplings of the theory as in equation \nref{coupscale}. Because of the Weyl anomaly, we would then have to restore the scale of the coordinates appropriately later, which will introduce extra contributions that are a power law on the ratio of $r_{0,\text{phy}}$ to a distance cutoff of the theory. With this normalization, the eigenvalues of fluctuations around the wormhole should be dimensionless numbers of order $1$. Therefore, the power law scaling of the path integral with the parameters of the wormhole will come solely from the integral over moduli and other zero modes, as well as the Weyl anomaly that will be sensitive to the coordinate rescaling. 

We should first discuss the twist and position moduli, because they should be present for any $SO(D)$ symmetric wormhole solution\footnote{A physical interpretation of the twist moduli, for example, is that it imposes angular momentum conservation between the wormhole mouths.}. Each of these moduli will contribute with a factor of $\kappa_{\text{phy}}^{-1}$. To see that, note that the twist moduli do so because a similar argument to that of section \nref{zeromod} should apply here\footnote{The difference is that these modes now parameterize a relative twist between the two asymptotic wormhole mouths.}. The $\kappa_{\text{phy}}^{-1}$ is present for the position moduli because their contribution is similar to that of a path integral over coordinate transformations, which also comes with a factor of $\kappa_{\text{phy}}^{-1}$ for each mode. 

Since the only other scale of the problem is $r_{0,\text{phy}}$, these contributions must come\footnote{We thank Juan Maldacena for discussions about these contributions.} from the dimensionless combination of $r_{0,\text{phy}}$ and $\kappa_{\text{phy}}^{-1}$. Therefore, these modes contribute as
\begin{equation}
Z_{\text{moduli}}\sim \bigg(\frac{r_{0,\text{phy}}^{D-2}}{\kappa_{\text{phy}}^{2}}\bigg)^{\frac{(2D+|SO(D)|)}{2}}\times (\text{moduli volume})
\end{equation}
with $(\text{moduli volume})$ an appropriate integral over the moduli. This integral will give the canonical group volume of $SO(D)$, from the twist modes, times a contribution from the position moduli. The position moduli volume will be an integral over positions in units of the wormhole scale, because of how we rescaled the coordinates. Therefore, relating these to physical displacements will give a factor of $r_{0,\text{phy}}^{-1}$ for each integral, so we obtain
\begin{equation}
\label{zmoduli}
Z_{\text{moduli}}\sim \bigg(\frac{r_{0,\text{phy}}^{D-2}}{\kappa_{\text{phy}}^{2}}\bigg)^{\frac{2D+|SO(D)|}{2}}r_{0,\text{phy}}^{-2D}\int d^{D}x\,d^{D}y
\end{equation}
with the integral being over displacements of the two wormhole mouths in the ambient spacetimes. Note that this reproduces the first two factors of our equation \nref{match}. Now, depending on the wormhole solution we study, there can be further zero-mode and moduli contributions, which will depend on the ratio of some coupling of the theory to the scale of the wormhole. An example was the last power law factor for the odd $D$ wormhole that we obtained in equation \nref{match}. In general, we call the contribution from extra zero modes $Z_{\text{zero}}$.

Putting these points together, let us define $Z_{\text{wormhole}}$ to be the path integral over the wormhole in the large spacetime and $Z_{\text{flat}}$ to be a similar contribution with no wormhole. Then, the ratio of these path integrals for some generic matter coupling should go as
\begin{equation}
\label{zflat0}
\frac{Z_{\text{wormhole}}}{Z_{\text{flat}}}\sim r_{0,\text{phy}}^{-2D}\bigg(\frac{r_{0}^{D-2}}{\kappa^{2}}\bigg)^{\frac{(2D+|SO(D)|)}{2}}\times Z_{\text{zero}}\times Z_{\text{anomaly}}\times e^{-S_{\text{wh}}-\Delta S_{\text{ct}}}\int d^{D}x\, d^{D}y
\end{equation}
with $S_{\text{wh}}$ the action of the wormhole from the Einstein Hilbert term and the matter theory, and $\Delta S_{\text{ct}}$ the contribution from higher derivative corrections. $Z_{\text{anomaly}}$ will be a contribution from the Weyl anomaly involving a power of the ratio of $r_{0,\text{phy}}$ to a physical cutoff scale. The log of this term is proportional to a heat kernel coefficient, which can be written as a local integral over background fields. Therefore, redundancies in $\log Z_{\text{anomaly}}$ having to do with changing the cutoff can be absorbed into local counterterms. However, the term involving $\log r_{0,\text{phy}}$ is a dependence on the physical scale of the wormhole, which cannot be absorbed by counterterms, being therefore physical. 

Picking the cutoff of the theory to be about the Planck scale,  $M_{\text{pl}}=\kappa_{\text{phy}}^{-\frac{2}{(D-2)}}$, we then obtain that the scaling of equation \nref{zflat0} with $r_{0,\text{phy}}$ goes as as
\begin{equation}
\label{zflat1}
\frac{Z_{\text{wormhole}}}{Z_{\text{flat}}}\sim r_{0,\text{phy}}^{-2D}\bigg(\frac{r_{0,\text{phy}}^{D-2}}{\kappa_{\text{phy}}^{2}}\bigg)^{\frac{(2D+|SO(D)|)}{2}}\times Z_{\text{zero}}\times \bigg(\frac{r_{0,\text{phy}}}{\kappa^{\frac{2}{(D-2)}}_{\text{phy}}}\bigg)^{a_{1}}\times e^{-S_{\text{wh}}-\Delta S_{\text{ct}}}\int d^{D}x\, d^{D}y
\end{equation}
where $a_{1}$ is related to appropriate heat kernel coefficients. More specifically, we expect the coefficient $a_{1}$ to be an overall contribution coming from the fluctuation operators of all fields and ghosts around the wormhole. In particular, $a_{1}$ should be sensitive to the matter background that we used to stabilize the wormhole. 

Note also that our argument to compute \nref{zflat1} was very general, so it is plausible that we can also treat it as a rule to integrate over the $x$ and $y$ positions of the wormhole off-shell, even if these positions modes obtain an action. That is, in \nref{zflat1} we wrote the effect of the path integral over the wormhole region, leaving an $x$ and $y$ integral to be done last. If we insert external operator insertions in the two ambient spacetimes, we still should receive this contribution, assuming that the insertions do not affect the path integral over the wormhole region at leading order. However, the right-hand side of \nref{zflat1} would have an extra term, which is the expectation value of the operator insertions in the connected spacetime. This expectation value will generally depend on $x$ and $y$. Let us call the ambient spacetimes $\mathcal{X}$ and $\mathcal{Y}$ respectively. Then, let $\mathcal{A}(w)$ be an operator insertion with $w \in \mathcal{X}$, and $\mathcal{B}(z)$ be an insertion with $z \in \mathcal{Y}$. Moreover, let us call the connected spacetime with the wormhole $\mathcal{M}_{x,y}$, where the subscript indicates its position moduli. Then, for the analogous path integral with operator insertions, there should be an extra term\footnote{For simplicity, we assume that the $SO(D)$ twist acts trivially on these operators. If they are affected by the twist, what enters the path integral is an appropriate average over $SO(D)$.} $\langle \mathcal{A}(w)\mathcal{B}(z)\rangle_{\mathcal{M}_{x,y}}$ in the right-hand side of \nref{zflat1}. Note that we again use $\langle \rangle$ to mean expectation value in a given spacetime.

Since the effect of the wormhole can be replaced by a sum over effective local operator insertions, at least for physics far enough away from the mouths, we should have that
\begin{equation}
\begin{gathered}
\label{zflat2}
r_{0,\text{phy}}^{-2D}\bigg(\frac{r_{0,\text{phy}}^{D-2}}{\kappa_{\text{phy}}^{2}}\bigg)^{\frac{(2D+|SO(D)|)}{2}}\times Z_{\text{zero}}\times\bigg(\frac{r_{0,\text{phy}}}{\kappa^{\frac{2}{(D-2)}}_{\text{phy}}}\bigg)^{a_{1}}\times e^{-S_{\text{wh}}-\Delta S_{\text{ct}}}\int d^{D}x\, d^{D}y \,\langle \mathcal{A}(w)\mathcal{B}(z)\rangle_{\mathcal{M}_{x,y}} \\
\sim \sum_{i j}\mathcal{C}_{ij} \int d^{D}x\,d^{D}y\, \langle \mathcal{A}(w)\mathcal{O}_{i}(x)\rangle_{\mathcal{X}} \langle\mathcal{B}(z)\mathcal{O}_{j}(y)\rangle_{\mathcal{Y}}
\end{gathered}
\end{equation}
with $\mathcal{C}_{ij}$ the EFT coefficients that we discussed before. We could imagine applying this setup to compute the $\mathcal{C}_{n,-n}$ term from the last section, for the $(D-2)$ form wormholes analyzed in this paper, but in general $D$. To do so, we need to consider a generic configuration where the leading effective operators $K_{\pm n}$, that we discussed before, dominate the right-hand side of \nref{zflat2}. An issue is that we need the magnetic flux to flow appropriately away from the wormhole. An option is to have no operators $\mathcal{A}$,$\mathcal{B}$, but to have boundary conditions for the path integrals over $\mathcal{X}$ and $\mathcal{Y}$ requiring a certain amount of flux flowing to infinity. If $\mathcal{X}$ and $\mathcal{Y}$ are big but finite, we can instead just imagine that we pick the operators $\mathcal{A}$ and $\mathcal{B}$ to be similar source operators $K_{\pm n}$ of opposite sign, such that there is no overall magnetic flux in $\mathcal{X}$ or $\mathcal{Y}$. Furthermore, we can pick $w$ and $z$ to be very far from $x$ and $y$. Since we expect $\langle K_{-n}(x)K_{n}(w)\rangle_{\mathcal{X}}$, and its counterpart for $\mathcal{Y}$, to approach a constant\footnote{This is the case because this is what we expect for the axion two point function $\langle e^{\frac{2 \pi i}{f_{\theta}}n \theta(x)}e^{-\frac{2 \pi i}{f_{\theta}}n \theta(y)}\rangle$ dual to it. We expect that because the Green's function of $\theta$ should go to zero as we take the separation of $x$ and $y$ to infinity.} as the points become very far from each other, we can extract $\mathcal{C}_{n,-n}$ from the earlier expression in \nref{zflat1}. More specifically, we should obtain $\mathcal{C}_{n,-n}$ from equation \nref{zflat1} without its position integrals, and applied to the $(D-2)$ form wormhole in flat space.

We now should discuss how we would expect the calculation of \nref{zflat1} to go for the $D=4$ wormholes sourced by either $D-2=2$ forms or scalar fields. First, we should discuss if they have a contribution from zero modes, e.g, in $Z_{\text{zero}}$. We would expect such a factor for the $2$-form theory because their gauge redundancies consist of $1$-forms and $0$-forms, and the wormhole has both harmonic zero and one forms. For the scalar theory, there should be just zero modes associated with harmonic $0$-forms. It is unclear if the harmonic $0$-forms are relevant in either case, since they were not relevant for our de Sitter computation in section \nref{rtwhsphere}.

Then, the last step to obtain the power law corrections in $r_{0,\text{phy}}$ would be to compute the anomaly coefficient $a_{1}$ in \nref{zflat1}. In the form theory case, this would involve computing the anomaly coefficient of the fluctuation operator involving gravity+forms, as well as from the chain of functional determinants coming from the ghosts. Some of these ghost determinants can be simplified using electromagnetic duality as we discussed in Appendix \nref{ghostdet}, but it would be nice to compute these coefficients directly without any trick. For the scalar case, we would only need to discuss the anomaly from the fluctuation operator of the gravity+scalar theory and the diffeomorphism ghosts. This could be a simpler case to study, provided one can construct a well-defined saddle from the scalar description.

As a last comment, in $D=4$ the effect of higher derivative corrections $\Delta S_{\text{ct}}$ will come at $O(1)$ in the $G_{N}$ expansion. We expect they would just affect the $O(1)$ scaling of \nref{zflat1}, and perhaps provide subleading corrections in $r_{0,\text{phy}}$. For example, for the $2$-form wormhole, two such counterterms are $R^{2}$ and $R_{ab}R^{ab}$, which we can evaluate in the flat space wormhole geometry, \nref{smwhreg}, as
\begin{equation}
\int R_{ab}R^{ab}\sim \int R^{2}\sim r_{0,\text{phy}}^{4}\int dx \varrho^{3}\bigg(\frac{n^{4}\kappa_{\text{phy}}^{4}}{f_{\theta,\text{phy}}^{4}r_{0,\text{phy}}^{12}\varrho^{12}}\bigg)\sim r_{0,\text{phy}}^{-8}\bigg(\frac{n \kappa_{\text{phy}}}{f_{\theta,\text{phy}}}\bigg)^{4}\sim 1
\end{equation}
where we wrote everything in terms of physical coordinates, and $\sim $ meant we only kept scalings with parameters of the wormhole, but not $O(1)$ constants. We also used that $\varrho$, defined as in \nref{smwhreg}, goes from $1$ to $\infty$, so the integral is convergent. As a last step, we used the expression for $r_{0,\text{phy}}$ in \nref{rodef} to conclude that this counterterm does not depend on $n$.

In any case, understanding the power law corrections in \nref{zflat1} for $D=4$ wormholes in either form or scalar descriptions would be very interesting. The main motivation is perhaps because they would be one-loop corrections to global symmetry violation effects in gravity \cite{Kallosh:1995hi, Hsin:2020mfa, Bah:2022uyz} that, as far as we are aware, were not computed before\footnote{We thank Yiming Chen for discussion on this aspect.}. Whether or not these corrections might be phenomenologically relevant will depend on how $f_{\theta,\text{phy}}$ compares to $\kappa_{\text{phy}}^{-1}=M_{\text{pl}}$. This ratio, however, also controls the size of the action from the 2-form side (see \nref{onactflat}), so we would need $n\times M_{\text{pl}}f_{\theta,\text{phy}}^{-1}$ to be somewhat big for the semiclassical description to be justified. In this regime, the one-loop corrections we discussed would be subleading. However, perhaps there are interesting models where $f_{\theta,\text{phy}}^{-1}M_{\text{pl}}$ is big enough that we have a semiclassical description, but the scales are still comparable in a way that we would obtain interesting corrections. To estimate the smallest values of $f_{\theta,\text{phy}}^{-1}M_{\text{pl}}$ for which the expression \nref{zflat1} would be a good approximation, we would have to estimate the size of two-loop effects.

\section{Einstein wormhole}
\label{einwh}

In this section, we discuss the one-loop spectrum of the wormhole with maximum charge $Q=Q_{c}$. The background we will study is therefore that of a cylinder with a line element
\begin{equation}
ds^{2}=d\tau^{2}+\frac{(D-2)}{(D-1)}d\Omega_{D-1}^{2 }~~,~~\tau \sim \tau+L
\end{equation}
where $L$ is the length of the wormhole, which we fix by hand. The length $L$ of the Einstein wormhole is arbitrary, unlike its non-extremal counterpart. However, it is useful to note that by continuing the wormhole solution with $Q<Q_{c}$ and $N$ cycles to $Q \rightarrow Q_{c}$, the limiting length of the resulting Einstein wormhole would be $L_{N}=\frac{2\pi}{\sqrt{2(D-1)}}N$ (see Appendix \nref{eincl}). Therefore, it is also convenient to parameterize the length of the generic Einstein wormhole as
\begin{equation}
\label{lN}
L=\frac{2\pi}{\sqrt{2(D-1)}}N
\end{equation}
with $N$ an arbitrary number, which is an integer only if the wormhole is a limit of a non-extremal solution. Moreover, we will see that if $N$ is chosen to be an integer, the one-loop spectrum will have special features. 

We organize this section as follows: In subsection \nref{phyinst} we will review the spectrum of physical linearized fluctuations of the wormhole, using the KK analysis of \cite{Hinterbichler:2013kwa}, and discuss its physical instabilities. The conclusion will be that there is a single physically unstable mode of the wormhole, which corresponds to perturbing the scale factor $a(\tau)$ slightly. Using this, we will be able to predict the phase for the path integral around this wormhole in \nref{phein}. 

In subsection \nref{anaspecein}, we discuss the Euclidean path integral around the wormhole directly, using the usual formalism discussed in section \nref{fluc}, but with a slight modification. We will see that the phase computed from the Euclidean path integral precisely matches the one from the physical instabilities discussed in subsection \nref{phyinst}. This fact is reminiscent of the relation between physical instabilities and negative modes discussed in \cite{Ivo:2025yek}.

\subsection{Physical instabilities}
\label{phyinst}

In this section, we review the physical instabilities of the Einstein wormhole in Lorentzian signature, using the KK reduction discussion of \cite{Hinterbichler:2013kwa}. For that, we will think of the Einstein wormhole solution as a product manifold $I_{\tau} \times S^{D-1}$, with $I_{\tau}$ a one-dimensional manifold that stands for the $\tau$ direction of the wormhole. We do not assume anything about the structure of $I_{\tau}$ yet. 

Taking $I_{\tau}$ as a base manifold, we can KK reduce the linearized fluctuations by viewing $S^{D-1}$ as the extra dimension. Since $I_{\tau}$ is a one-dimensional manifold, the resulting KK spectrum will consist solely of harmonic oscillators with various frequencies. As discussed at the end of section 6.6 of \cite{Hinterbichler:2013kwa}, the spectrum of physical fluctuations of the gravity+form theory will consist of three sets of modes\footnote{There are also some remaining form fluctuations. However, they are non-negative, so they will not be important for us.}. In almost their notation\footnote{What we call $F_{n}$ is their $F_{a}$, and what we call $\hat{\phi}_{0}$ and $\hat{\phi}_{I}$ are their $\phi_{I}$ and $\phi_{0}$, respectively.}, the modes are as follows: A volume modulus, $\hat{\phi}_{0}$, higher modes of the $S^{D-1}$ Laplacian, $F_{n}$, and eigenmodes of the tensor Laplacian in $S^{D-1}$, $\hat{\phi}_{I}$. 

The last two scalars consist only of stable modes. The reason is that the $\hat{\phi}_{I}$ modes, which correspond to tensor deformations of the internal manifold, are never unstable for sphere compactifications. The $F_{n}$ modes correspond to perturbations which are scalar harmonics of $S^{D-1}$, with eigenvalue above the conformal Killing value \cite{Hinterbichler:2013kwa}. These fluctuations will consist of a complicated combination of gravity and form fluctuations, whose format is not very illuminating. While these modes are not immediately stable, by inspecting their action in \cite{Hinterbichler:2013kwa} we can see that they are stable.

Therefore, the only unstable perturbations of the Einstein wormhole will have to come from the scalar $\hat{\phi}_{0}$. The $\hat{\phi}_{0}$ modes are fluctuations in the size of the $S^{D-1}$, which do not depend on the $S^{D-1}$ coordinates. To be more concrete, they are modes of the form
\begin{equation}
h_{ab}=\frac{1}{(D-1
)\sqrt{a_{c}^{D-1}\text{Vol}(S^{D-1})}}\gamma_{ab}\hat{\phi}_{0}(\tau)
\end{equation}
and their Euclidean action is
\begin{equation}
\label{phi0act}
I_{E}=-\frac{(D-2)}{2(D-1)}\int d\tau\, \hat{\phi}_{0}(\tau)(-\partial_{\tau}^{2}-2(D-1))\hat{\phi}_{0}(\tau)
\end{equation}

Equation \nref{phi0act} makes clear that, in Euclidean signature, $\hat{\phi}_{0}$ is a field with a negative sign kinetic term. More specifically, we can think of the action of $\hat{\phi}_{0}$ as being minus the action of an inverted harmonic oscillator with negative squared frequency $\omega^{2}=-2(D-1)$. 

The fact that $\phi_{0}$ has the same equation of motion as an inverted harmonic oscillator implies that the Einstein wormhole saddle has a linearized instability in Lorentzian signature. This is no surprise; the Einstein static universe is famously unstable \cite{Eddington:1930zz} against this type of deformation. 

Following \cite{Ivo:2025yek}, we expect these instabilities to induce a phase for the Euclidean path integral of the theory if $\tau$ is a circle. However, the $\hat{\phi}_{0}$ mode has the opposite action of an inverted harmonic oscillator, and the analysis is slightly different. Taking the background to be the Einstein wormhole of length $L$, and Fourier transforming $\hat{\phi}_{0}$ as $\hat{\phi}_{0}=\sum_{n}\hat{\phi}_{0,n}e^{\frac{2 \pi i n}{L}\tau}$, we see that the $\phi_{n,0}$ modes are non-negative for every $n$ such that
\begin{equation}
\lfloor N\rfloor \geq n \geq -\lfloor N \rfloor 
\end{equation}
where we used $N$ defined from the length of the wormhole via \nref{lN}.

Therefore, for all other values of $n$, the spectrum of $\hat{\phi}_{0}$ is negative, such that $\hat{\phi}_{0}$ is a mostly negative field with $1+2 \lfloor N \rfloor$ non-negative modes. If all modes of $\hat{\phi}_{0}$ were negative, $\hat{\phi}_{0}$ would induce no overall phase, but since we have $1+2 \lfloor N\rfloor$ missing negative modes the phase of the one-loop determinant is
\begin{equation}
\label{phein}
Z_{\text{wh}}\sim (-i)^{-(1+2\lfloor N\rfloor)}=i^{1+2 \lfloor N\rfloor}
\end{equation}

Note that the number of negative modes jumps whenever $N$ crosses an integer. This is an interesting point because if $N$ is an integer, we can think of the Einstein solution as the $Q \rightarrow Q_{c}$ limit of a non-extremal solution with $N$ cycles. Therefore, $n_{-}$ just changes when we cross such solutions. 

Moreover, the zero modes at the crossing will be modes with $n=N$. These modes are of the form 
\begin{equation}
\hat{\phi}_{0}\sim \cos\bigg(\frac{\sqrt{2(D-1)}}{2\pi}\tau\bigg)
\end{equation}
which is interestingly enough the same as the correction in \nref{neareina} to the scale factor $a(\tau)$ near extremality. Therefore, we have a linearized zero mode at these special values of the length, which corresponds to changing the background metric to that of the non-extremal wormhole. Of course, the flux of the solution will still be maximal, so the solution is not deformed into a non-extremal one.

In the following section, we will double-check the result \nref{phein} for the phase by computing the phase explicitly from the Euclidean path integral. 

\subsection{Analytic features of the spectrum*}
\label{anaspecein}

\subsubsection{Preliminary discussion}
\label{prelimein}

Here, we discuss a strategy to compute the number of negative modes, which is slightly different from the eigenvalue method discussed in section \nref{genset}\footnote{Thus the asterisk in the name of this subsection}. There, we defined a local norm and carefully discussed how to solve for the eigenvalues of the fluctuation operator defined in that norm. 

Instead of doing that, here we will write the quadratic action directly in terms of the field decomposition in \nref{repsod}. Then, we use the fact that the measure of the path integral is, up to overall positive constants, a product measure of all these fields. 

For example, for the scalars, the measure would roughly be of the form
\begin{equation}
\label{simpmeas}
\int D\psi\, DA\,D'f\,D'\eta\,D'\chi
\end{equation}
where we used the $D'$ notation again, to mean that we only integrate over the fields that have non-trivial fluctuations. For example, we exclude the $l=0$ modes of $f$, $\eta$ and $\chi$, and the $l=1$ modes of $\chi$.

Then, we will start doing field redefinitions that leave a trivial Jacobian for the measure following from \nref{locnorm}, until our quadratic action is diagonal. With these steps complete, we will then read off the spectrum of the diagonalized action. The eigenvalues obtained this way will generally differ from the ones obtained from the method in \nref{genset}. However, since the field redefinitions we will perform have a trivial Jacobian, the two methods should give the same functional determinant in the end. In particular, conclusions obtained from both methods, such as the phase of the partition function, should match\footnote{This is a reasonable expectation if all fields in the path integral are originally integrated along the real axis. This expectation, however, might fail if the original contours of integration for different fields do not match. In that case, field redefinitions will generally change the contours of integration non-trivially.}.

Since the background geometry is a pure cylinder, the symmetry group is enhanced from $SO(D)$ to $U(1) \times SO(D)$, and we can expand all the scalars into fields with a definite $SO(D)$ eigenmode and a given momentum $k$ along the $\tau$ direction. Because the $\tau$ direction is a circle, we need to impose the quantization condition
\begin{equation}
k_{n}=\frac{2n\pi}{L}
\end{equation}
with $n$ an integer.

Using this fact, we can write the fields $\Phi_{l}(\tau)$ for a given spherical harmonic in section \nref{repsod} as a sum of exponentials $e^{\pm i k_{n}\tau}$. To avoid introducing complex coefficients for these exponentials, we will instead expand the fields in terms of sines and cosines. For convenience, we will refer to the $SO(D)$ representation of a field, its momentum $k$, and its sine/cosine sector via a collective index $I$. 

With these points in mind, we will expand the scalar fields as
\begin{equation}
\begin{gathered}
\label{scexpein}
\psi=\sum_{n,l}(\psi_{n,l,+}\cos (k_{n}\tau)+\psi_{n,l,-}\sin (k_{n}\tau))Y_{l} ~,~ A=\sum_{n,l}(A_{n,l,+}\cos (k_{n}\tau)+A_{n,l,-}\sin (k_{n}\tau))Y_{l}\\
f=\sum_{n,l}(f_{n,l,+}\sin (k_{n}\tau)-f_{n,l,-}\cos (k_{n}\tau))Y_{l}~,~\eta=\sum_{n,l}(\eta_{n,l,+}\cos (k_{n}\tau)+\eta_{n,l,-}\sin(k_{n}\tau))Y_{l}\\
\chi=\frac{1}{\sqrt{2(D-1)}}\sum_{n,l}(\chi_{n,l,+}\cos (k_{n}\tau)+\chi_{n,l,+}\sin(k_{n}\tau))Y_{l}
\end{gathered}
\end{equation}
with $Y_{l}$ the scalar spherical harmonic discussed in section \nref{repsod}. For convenience, we pick the normalization of $Y_{l}$ such that, for $n \neq 0$
\begin{equation}
\int d^{D}x\sqrt{g}\, Y_{l}^{2}\cos^{2}(k_{n}\tau)=1
\end{equation}

Note also that in \nref{scexpein} we used an index $\pm$ to indicate whether the mode consists of mostly cosines or mostly sines, with the motivation that the $\pm$ modes only couple to other $\pm$ modes. The $+$ and $-$ modes have the same action, however.

For transverse tangent vectors $u,v,x$ and $\omega$ we have a similar decomposition, but instead
\begin{equation}
\begin{gathered}
\label{tvexpein}
v_{a}=\sum_{n,l}(v_{n,l,+}\cos(k_{n}\tau)+v_{n,l,-}\sin(k_{n}\tau))Y_{l,a} ~,~ x_{a}=\sum_{n,l}(x_{n,l,+}\cos(k_{n}\tau)+x_{n,l,-}\sin(k_{n}\tau))Y_{l,a}\\
u_{a}=\frac{1}{\sqrt{2(D-1)}}\sum_{n,l}(u_{n,l,+}\sin(k_{n}\tau)-u_{n,l,-}\cos(k_{n}\tau))Y_{l,a}\\\omega_{a}=\frac{1}{\sqrt{2(D-1)}}\sum_{n,l}(\omega_{n,l,+}\cos(k_{n}\tau)+\omega_{n,l,-}\sin(k_{n}\tau))Y_{l,a}
\end{gathered}
\end{equation}
with $Y_{l,a}$ as in section \nref{repsod}. For convenience, we also fix the normalization of $Y_{l,b}$ by requiring that, for $n \neq 0$
\begin{equation}
\int d^{D}x\sqrt{g}\, Y_{l,a}Y^{l,a}\cos^{2}(k\tau)=1
\end{equation}

We can also do a similar expansion for the tangent transverse two-forms $j_{ab}$ and the tangent transverse traceless modes $\phi_{ab}$. Namely, for the harmonics $Y_{l,[ab]}$ and $Y_{l,(ab)}$ introduced in section \nref{repsod} we write
\begin{equation}
\begin{gathered}
\label{tsexpein}
j_{ab}=\sum_{n,l}(j_{n,l,+}\cos(k_{n}\tau)+j_{n,l,-}\sin(k_{n}\tau)Y_{l,[ab]}\\
\phi_{ab}=\sum_{n,l}(\phi_{n,l,+}\cos(k_{n}\tau)+\phi_{n,l,-}\sin(k_{n}\tau))Y_{l,(ab)}
\end{gathered}
\end{equation}
where, for convenience, we also impose normalization conditions. Namely, for $n \geq 0$ we fix the normalization
\begin{equation}
\int d^{D}x\sqrt{g}\,Y_{l,[ab]}Y^{l,[ab]}\cos^{2}(k_{n}\tau)=\int d^{D}x\sqrt{g}\,Y_{l,(ab)}Y^{l,(ab)}\cos^{2}(k_{n}\tau)=1
\end{equation}

A small detail relevant to the next section is that, because of our normalization for spherical harmonics, the fields with $n=0$ in all sectors should have an overall factor of $2$ in their action. To obtain this fact from the equations for the action that we write in the next section, one should treat the $\pm$ variables for the $n=0$ fields as independent equal variables, even though only one of the $\pm$ modes exists at $n=0$.

\subsubsection{Spectrum}
\label{einspec}

In this section, we use the field decomposition and overall strategy described in section \nref{prelimein} to compute the quadratic action of the wormhole. From this, we will discuss the number of zero and negative modes of the fluctuation spectrum.

\textbf{Scalars:} The quadratic action for the scalars $\{\psi, A, f, \eta, \chi\}$ is, using the decomposition \nref{scexpein},
\begin{equation}
\begin{gathered}
I_{E}^{(2)}=\frac{1}{2}\sum_{I}\bigg[\frac{D(D-2)}{2}\bigg(-\lambda_{I}+\frac{4(D-1)^{2}}{D}\bigg)\psi_{I}^{2}\\+\frac{D}{(D-1)}\bigg(\lambda_{I}+\frac{2(D-1)(D-2)}{D}\bigg)A_{I}^{2}-4(D-1)(D-2)\psi_{I}A_{I}\\+\frac{\lambda_{I}\lambda_{l}}{(D-1)}\chi_{I}^{2}+2(D-2)\lambda_{l}\psi_{I}\chi_{I}-4\lambda_{l}A_{I}\chi_{I}+2\lambda_{l}\lambda_{I}f_{I}^{2}+4k \lambda_{l}\chi_{I}f_{I}\\+\frac{(D-2)}{(D-1)}\lambda_{l}\bigg(\lambda_{l}-\frac{(D-1)^{2}}{(D-2)}\bigg)(\lambda_{I}-2(D-1))\eta_{I}^{2}\bigg]
\end{gathered}
\end{equation}
where we defined
\begin{equation}
\label{lmbdlI}
\lambda_{l}=\frac{(D-1)l(l+D-1)}{(D-2)}~~,~~ \lambda_{I}=\lambda_{l}+k^{2}
\end{equation}

Again, the collective index refers to $k$, the angular momentum $l$, and the $\pm$ expansions of section \nref{prelimein}. One should also remember that some fields do not exist for some $I$, in which case they should be set to zero by hand. To obtain a less cumbersome version of the action, it is convenient to define the fields $\tilde{A}$ and $\tilde{\psi}$ via
\begin{equation}
\tilde{A}=\frac{(D-2)}{2}\psi-A ~~~~,~~~~ \tilde{\psi}=(D-1)\psi-A
\end{equation}

After that, we complete squares until we obtain an action that is a sum of squares, as
\begin{equation}
\begin{gathered}
\label{diagactsc}
I_{E}^{(2)}=
\frac{1}{2}\sum_{I}\bigg[\frac{(D-2)}{(D-1)}(2(D-1)-\lambda_{I})\tilde{\psi}_{I}^{2}+\frac{\lambda_{l}}{(D-1)}(\lambda_{I}-2(D-1))\chi_{I}^{2}\\+2 \lambda_{I}\bigg(\tilde{A}+\frac{\lambda_{l}}{\lambda_{I}}\chi_{I}\bigg)^{2}+2\lambda_{l}\lambda_{I}\bigg(f_{I}+\frac{k}{\lambda_{I}}\chi_{I}\bigg)^{2}+\frac{(D-2)}{(D-1)}\lambda_{l}\bigg(\lambda_{l}-\frac{(D-1)^{2}}{(D-2)}\bigg)(\lambda_{I}-2(D-1))\eta_{I}^{2}\bigg]
\end{gathered}
\end{equation}

We see that the action \nref{diagactsc} becomes diagonal in the resulting fluctuations as long as we redefine $\tilde{A}_{I} \rightarrow \tilde{A}_{I}+\frac{\lambda_{l}}{\lambda_{I}}\chi_{I}$ and $f \rightarrow f+\frac{k}{\lambda_{I}}\chi_{I}$. Note that these redefinitions have a trivial Jacobian. 

The redefined $\tilde{A}_{I}$ and $f_{I}$ fields are manifestly non-negative. Their only zero mode is the $\tilde{A}$ mode at $k=l=0$. This is the mode corresponding to length fluctuations of the wormhole, which we do not integrate over since we study the wormhole at fixed length. 

To understand the number of negative modes, we can therefore restrict ourselves to studying $\tilde{\psi}_{I}$, $\chi_{I}$, and $\eta_{I}$. The actions of these fields are all proportional to the same differential operator
\begin{equation}
\label{gdiff}
G=-\n^{2}-2(D-1)=\frac{(D-1)}{(D-2)}l(l+D-2)+\frac{4n^{2}\pi^{2}}{L^{2}}-2(D-1)
\end{equation}
with the proportionality constant non-negative for $\chi$ and $\eta$, and negative for $\tilde{\psi}$. A relevant feature of \nref{gdiff} is that this operator is positive for $l \geq 2$. Therefore, since $\eta_{I}$ is only defined for $l \geq 2$, the action of $\eta_{I}$ is positive definite. 

Therefore, to study the number of negative modes, we can restrict ourselves to $\tilde{\psi}_{I}$ and $\tilde{\chi}_{I}$. Since the $\tilde{\psi}_{I}$ action has a minus sign relative to the $\tilde{\chi}_{I}$ one, a positive mode of $\tilde{\psi}$ will map to a negative of $\tilde{\chi}$ and vice versa. Thus, if there are no zero modes in their spectrum, these modes always contribute with one overall negative mode per label $I$, if they both exist for the given $I$.

Since the operator $G$ has no zero modes for $l \geq 2$, any overall phase from these modes will come from the $l=0$ or $l=1$ sectors. At $l=1$, we have that
\begin{equation}
G|_{l=1}=\frac{4n^{2}\pi^{2}}{L^{2}}-\frac{(D-1)(D-3)}{(D-2)}
\end{equation}

Therefore, at $l=1$ there are zero modes of $G$ only if the value of $L$ is fine-tuned, or if $D=3$. For $D=3$, we can understand the zero mode at $n=0$ as the fact that the de Donder gauge condition is inadequate in the presence of $S^{2}$ factors, since it leaves some residual gauge transformations \cite{Ivo:2025yek, Ivo:2025xek}. This zero mode, therefore, is lifted for different gauge fixing conditions. We can also wonder if the zero modes at specific values of $L$ are gauge artifacts. In section \nref{einghost} we argue this is the case, by explicitly showing that there are ghost zero modes if $D=3$ or if $L$ is equal to these special discrete lengths. This, therefore, implies that the gauge fixing condition is not appropriate in these scenarios.

Having established this, we can conclude that the $l=1$ sector has no zero modes as long as one is careful about picking an adequate gauge fixing condition for the given value of $L$. Therefore, the $l=1$ modes do not contribute with an overall phase. 

The phase contribution of the scalar sector must therefore come only from the $l=0$ modes. In this sector, the field $\chi_{I}$ does not exist, and therefore $\psi_{I}$ is the only field with negative modes. To determine the phase contribution of this field, we therefore only need to count the number of negative modes from the action of $\psi_{I}$ at $l=0$. Up to overall positive constants, the action of $\psi_{I}$ is $-G$, therefore we have to analyze the negative modes of
\begin{equation}
-G|_{l=0}=-\bigg(\frac{4n^{2}\pi^{2}}{L^{2}}-2(D-1)\bigg)=-\frac{4\pi^{2}}{L^{2}}(n^{2}-N^{2})
\end{equation}

We have already done this in section \nref{phyinst}. As a reminder, all the modes of $G$ with $|n|>N$ are negative, with the other modes being zero or positive. The $\psi_{I}$ mode is therefore mostly negative, with some number of missing negative modes, corresponding to the modes with $|n| \leq N$. There are $1+2\lfloor N\rfloor$ such modes, so the contribution of the scalars to $n_{-}$, defined as in \nref{nminus}, is
\begin{equation}
n_{-} \supset -(1+2\lfloor N\rfloor)
\end{equation}

\textbf{Transverse vectors:} We now discuss the action for the transverse vectors $\{v,u,x,\omega\}$ and diagonalize it appropriately, using the decomposition in \nref{tvexpein}. The action for these modes is
\begin{equation}
\begin{gathered}
I_{E}^{(2)}=\sum_{I}\bigg[\bigg(\lambda_{I}+\frac{(D-1)(D-3)}{(D-2)}\bigg)\bigg(v_{I}^{2}+\frac{u_{I}^{2}}{2(D-1)}+\frac{\omega_{I}^{2}}{2(D-1)}\bigg(\lambda_{l}+\frac{(D-1)(D-3)}{(D-2)}\bigg)\bigg)\\-2k u_{I}v_{I}
+2\bigg(\lambda_{l}+\frac{(D-1)(D-3)}{(D-2)}\bigg)\omega_{I} v_{I}+\bigg(\lambda_{I}-\frac{(D-1)^{2}}{(D-2)}\bigg)\bigg(\lambda_{l}-\frac{(D-1)^{2}}{(D-2)}\bigg)x_{I}^{2}\bigg]
\end{gathered}
\end{equation}
where we have to remember that the $x_{I}$ field corresponding to $l=1$ transverse vector harmonics does not exist. 

We see from this action that we can decouple the fields by performing a $v_{I}$ dependent field redefinition in both $u_{I}$ and $\omega_{I}$. In other words, we can rewrite the action as
\begin{equation}
\begin{gathered}
I_{E}^{(2)}=\sum_{I}\bigg[\bigg(\lambda_{I}+\frac{(D-1)(D-3)}{(D-2)}\bigg)\bigg(\frac{1}{2(D-1)}\bigg(u_{I}-\frac{2(D-1)k}{\lambda_{I}+\frac{(D-1)(D-3)}{(D-2)}}v_{I}\bigg)^{2}\\+\frac{1}{2(D-1)}\bigg(\lambda_{l}+\frac{(D-1)(D-3)}{(D-2)}\bigg)\bigg(\omega_{I}+\frac{2(D-1)}{\lambda_{I}+\frac{(D-1)(D-3)}{(D-2)}}v_{I}\bigg)^{2}\bigg)\\
+\bigg(\lambda_{I}-\frac{(D-1)^{2}}{(D-2)}\bigg)v_{I}^{2}+\bigg(\lambda_{I}-\frac{(D-1)^{2}}{(D-2)}\bigg)\bigg(\lambda_{l}-\frac{(D-1)^{2}}{(D-2)}\bigg)x_{I}^{2}\bigg]
\end{gathered}
\end{equation}

Therefore, we see that the redefined $u_{I}$ and $v_{I}$ fields are manifestly positive. The possibly zero and negative fields are only $v_{I}$ and $x_{I}$. However, $x_{I}$ only exists for angular momentum $l>1$, in which case its action is also manifestly positive. 

The $v_{I}$ field has a non-negative action as well; however, it has zero modes at $l=1$ and $k=0$. These are the transverse vector zero modes we discussed in section \nref{zeromod}, but they have an analytic form in the Einstein solution, given by
\begin{equation}
v_{I}=c~~,~~ \omega_{I}=-c~~,~~ u_{I}=0
\end{equation}
with $c$ a constant and $I$ corresponding to $l=1$ and $k=0$. 

Also, note that, since the tangent transverse vectors have no negative modes, they do not contribute to $n_{-}$.

\textbf{Tangent transverse two-forms:} The action for the $j_{ab}$ fields, expanded as \nref{tsexpein}, is
\begin{equation}
I_{E}^{(2)}=\frac{1}{2}\sum_{I}\bigg(\lambda_{I}+2\frac{(D-1)(D-4)}{(D-2)}\bigg)j_{I}^{2}
\end{equation}
which is non-negative after we use that $l \geq 1$ for these modes. It can have zero modes only if $D=3$, in which case the $k=0$, $l=1$ mode is a zero mode. This is the transverse two-form zero mode described in section \nref{zeromod}, corresponding to harmonic two-forms in the 3d solution. In particular, this sector of modes does not contribute to $n_{-}$.

\textbf{Tangent transverse traceless symmetric tensor:} The action for the $\phi_{ab}$ fields, expanded as \nref{tsexpein}, is
\begin{equation}
I_{E}^{(2)}=\frac{1}{2}\sum_{I}\lambda_{I}\phi_{I}^{2}
\end{equation}
which is manifestly positive since the $\phi_{I}$ fields are only defined for $l \geq 2$. This implies, in particular, that $\lambda_{I}\geq\lambda_{l}>0$ for all modes. Note also that these modes do not contribute to $n_{-}$.

\textbf{Summary:} In summary, the negative modes of the wormhole come solely from the scalar sector, which are the fields $\{\psi, A, f, \eta, \chi\}$ in the decomposition \nref{decompf}. There are infinitely many such negative modes. More precisely, there is one negative mode per scalar basis mode of the manifold, plus a finite number $n_{-}$ given by
\begin{equation}
n_{-}=-1-2\big\lfloor N\big\rfloor~~\text{, with}~~N=\frac{\sqrt{2(D-1)}L}{2\pi} 
\end{equation}
and $L$ is the length of the $\tau$ direction. The phase of the manifold is therefore given by $(-i)^{n_{-}}=i^{1+2\lfloor N \rfloor}$. The phase is that of a $Q<Q_{c}$ wormhole with $\lfloor N \rfloor$ cycles, with one extra factor of $(-i)^{-1}=i$. The result for the phase is quite interesting, because it exactly matches the phase one expects from the physical instabilities of the Einstein wormhole in \nref{phein}. This is another example of the relation between physical instabilities and the phase of the path integral, discussed in \cite{Ivo:2025yek}. However, it comes with the caveat that the action from the physical instabilities of the Einstein wormhole has an overall minus sign in Euclidean signature. Therefore, it contributes to the phase in the opposite way that an unstable harmonic oscillator of the same frequency would. We should also comment that the relation between the phase of the path integral and the phase from physical instabilities that we found is an even more extreme realization of the idea in \cite{Ivo:2025yek}. The reason is that in \cite{Ivo:2025yek}, the phase from the physical instabilities and the phase of the path integral were related up to a non-zero offset. In our context, there is no such offset.  

Now we discuss the zero modes of the Einstein wormhole background, starting with scalar zero modes. One of the scalar zero modes is the constant $\tilde{A}$ mode of the wormhole, which corresponds to changing its length. Since we fix the length of the wormhole, we do not include this mode, keeping in mind that we integrate over it later. There are also scalar zero modes when $N$ is an integer, which are related to deforming the wormhole metric to its non-extremal counterpart. These modes are zero at the linearized level, but might not be at higher orders in the fluctuation size. 

The other zero modes are at the tangent transverse vector sector and the tangent transverse two-form sectors. They are the same ones described in section \nref{zeromod}, so they should be exact zero modes with compact range, which therefore have a finite contribution. 

\section{Discussion}
\label{disc}

In this paper, we discussed the path integral around de Sitter axion wormholes at one-loop, emphasizing different interesting points about it. Here, we discuss interesting future directions and raise a few questions about some of them.

For example, one of the most interesting things that we discussed was the wormhole solutions that look like a sphere with a small handle, and their effective description as a sphere with two operator insertions. In particular, we found that the phase of this solution was the phase of the sphere times a phase that came from integrating over the positions of the operator insertions.

The mechanism for this extra phase we discussed was as follows: The classical wormhole that we discussed corresponds to the sphere with operator insertions in antipodal points. Moreover, integrating over fluctuations of the saddle consists of integrating over the usual sphere degrees of freedom and the small fluctuations of the operators' positions around the antipodal configuration. If the two-point function of these operators decreases with distance, these fluctuations will increase the two-point function. Therefore, by integrating over position fluctuations, we will be effectively integrating over $D$ wrong-sign Gaussians. From this point, the phase of this solution should be the phase of the sphere times the phase of the integral over $D$ wrong-sign Gaussians, which is $i^{D+2}(-i)^{D}=i^{2}$. 

Since this is a very general mechanism, one might wonder if this is a general feature of solutions of this form. That is, one could wonder the following: Take a Euclidean wormhole solution in flat space, sourced by some general matter background. We should be able to, from this solution, find a wormhole solution for the theory with a small enough positive cosmological constant, by gluing the asymptotic region of these wormholes to a sphere of appropriate radius. 

Once we require these wormholes to end up in a finite sphere, their position will be required to extremize the action, if we want them to be a classical solution. It is plausible that this implies that the wormhole mouths must end up in antipodal points in the sphere. If that is the case, we expect the analysis from before to apply. That is, if the action decreases as we bring the wormhole mouths closer, we expect that there should be $D$ extra negative modes in the path integral. Therefore, it seems that in the absence of any extra phase, the phase of this "on-shell wormhole with sphere asymptote" should be $i^{D+2}(-i)^{D}=i^{2}$. The $(-i)^{D}$ is there if, and only if, the leading bilocal operator insertion that describes the wormhole has a two-point function that increases as we displace the operators away from the antipodal configuration.

\begin{figure}
    \centering
    \includegraphics[width=0.35\linewidth]{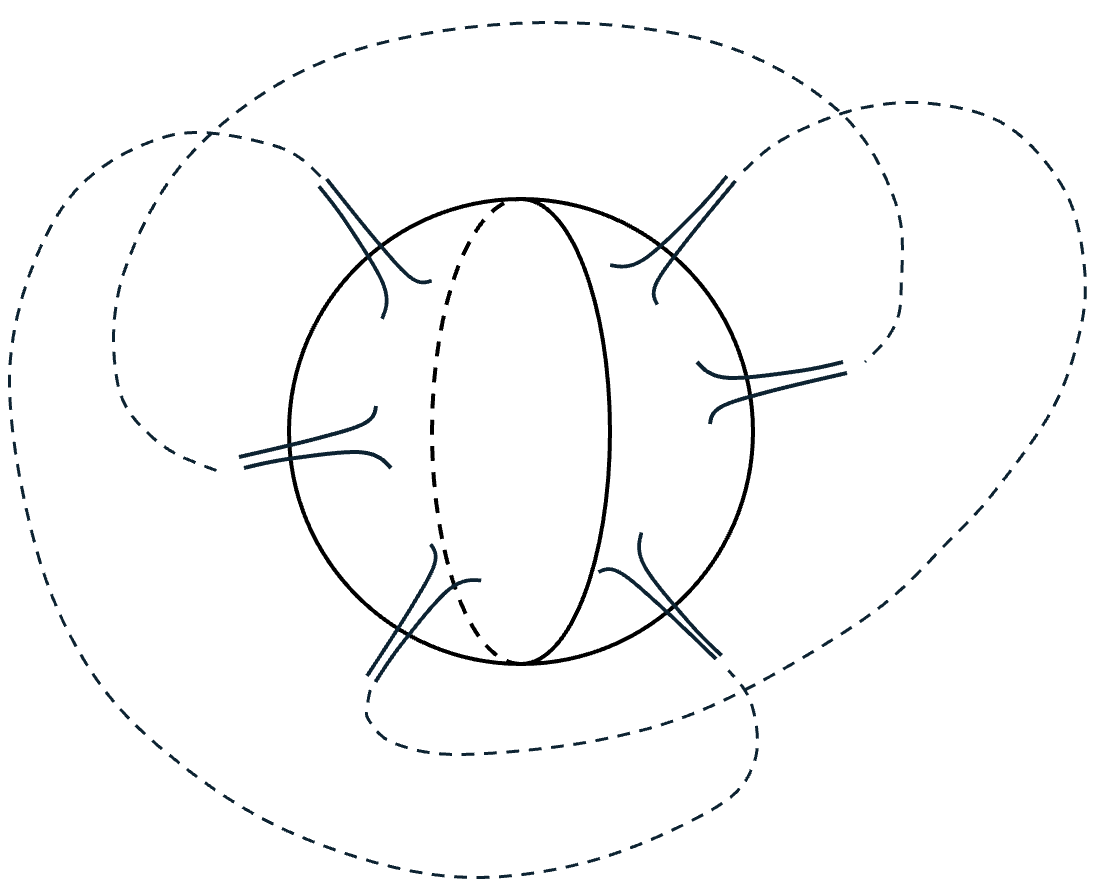}
    \caption{An illustration of a wormhole solution that has multiple pairs of handles on a sphere. The dashed lines denote identification.}
    \label{fig: multi handle}
\end{figure}

It would be perhaps interesting to explore this idea further by studying different Euclidean wormhole solutions in spacetimes with a positive cosmological constant, with different matter content. In particular, it would be interesting to study solutions consisting of spheres with multiple handles (see Figure \nref{fig: multi handle}) and whether their phase can be understood from the effective operator picture. 

Also, note that while in this paper we focused on studying axion wormholes in de Sitter space, our formalism likely extends in a straightforward way to $SO(D)$ symmetric wormholes in different spacetimes. An obvious future direction would be to, for example, study similar wormholes in AdS using a similar setup to ours. 

In particular, the stability of flat space or AdS wormholes, defined (or inspired) from canonical gravity, was studied in detail in the literature \cite{Marolf:2021kjc, Loges:2022nuw, Liu:2023jvm, Marolf:2025evo}. An interesting question would be if, in AdS, the phase of the path integral as defined in \cite{Polchinski:1988ua, Maldacena:2024spf, Ivo:2025yek} is connected to the stability of these saddles defined from the usual criteria. It would be interesting to understand, for example, if wormholes in AdS, that are stable according to usual criteria, also have a phase equal to one. This does not seem to be the case for saddles in de Sitter, but perhaps AdS is different.  

In fact, it is reasonable to stipulate from our discussion in the main text that the counterparts of the wormholes we discussed in flat space have a phase equal to one. The reasoning is that the only extra phase we obtained from the wormholes in de Sitter came from $D$ negative modes associated with the position moduli of the wormhole mouths. However, in flat space, these would be zero modes. Therefore, in the absence of other non-trivial effects, we would expect no overall phase\footnote{We thank Juan Maldacena for discussions about this point}. For the same reason, one would expect no phase for these wormholes in AdS.



Another comment is that, throughout the paper, we discussed not only the de Sitter wormhole solution with one cycle, but its $ N$-cycle generalizations at the same time. However, including all the $N$-cycle wormholes in the gravitational path integral leads to apparent paradoxes, since by increasing $N$ we decrease the action \nref{onact}, and therefore increase $e^{-I_{E}}$. Therefore, the sum over all these wormholes seems to be a divergent sum \cite{Blommaert:2025bgd}. See also \cite{Klebanov:1988eh, Horowitz:2025zpx} for similar points.

However, by studying their one-loop determinant, we learned that the wormholes with $N$ cycles have a phase of $(-1)^{N}$. This implies that, if we were supposed to sum over all these wormhole solutions, the sum would be an oscillating sum, with slightly better convergent properties. Furthermore, as we discussed in the main body of the paper, these wormholes might have some discrete symmetries that need to be taken into account later. This is relevant here because the wormhole with $N$ cycles has a cyclic permutation symmetry which appears in the division by diffeomorphisms, so we should divide their final contribution by $N$. 

Therefore, the sum over wormholes with $N$ cycles would be as
\begin{equation}
\label{rel-sum}
Z_{Q}=\sum_{N=1}^{\infty}\frac{(-1)^{N}}{N}|Z_{N,\text{wormhole}}|
\end{equation}
with $Z_{N,\text{wormhole}}$ the path integral around the wormhole with $N$ cycles, without the discrete symmetry factors as we discussed in the main body of the paper. From \nref{rel-sum}, it seems that depending on the behaviour of $Z_{N,\text{wormhole}}$ with $N$, we might be able to regularize the sum. 

To illustrate this, let us consider a hypothetical scenario where the path integral over wormholes with $N$ cycles is just $(Z_{1})^{N}$, with $Z_{1}$ the path integral of the $N=1$ wormhole. This is an inspired guess for the overall contribution of $Z_{Q}$, at least for small $\kappa Q$. The reason is that the spectrum that we observed for the wormholes with $N$ cycles was $N$ copies of the sphere one (see Appendix \nref{supdata}), with a couple of extra light and zero modes. Therefore, up to how zero and light modes contributions might change with $N$, the naive contribution from the wormholes with $N$ cycles at small $\kappa Q$ would be $(Z_{1})^{N}$. If this were the case, we could rewrite \nref{rel-sum} and regularize it as
\begin{equation}
\label{rel-sum2}
Z_{Q}\overset{?}=\sum_{N=1}^{\infty}\frac{(-1)^{N}}{N}|Z_{\text{1}}|^{N}=-\log(1+|Z_{\text{1}}|)\approx -\log(|Z_{\text{1}}|)
\end{equation}
where in the last line we regularized this sum by noticing it is the expansion of $-\log(1+x)$. Therefore, the sum with the relative sign in \nref{rel-sum2}, and regularized as we described, would be much smaller than the contribution of a single wormhole. Of course, our hypothetical scenario used an idealization of the path integral around the wormhole with $N$ cycles. Further corrections in $N$ should, and very likely will, change the structure of the regularized sum we obtain in the end.

However, it would be interesting if an argument along similar lines solved the paradox associated with summing over the $N$-cycle wormholes. Another resolution for the paradox would be to include only the $N=1$ wormhole in the sum, and not its higher oscillator counterparts (which will be a point argued for more carefully in \cite{Blommaert:2026}). This would be possible if the steppest descent contour that defines the path integral does not cross these saddles. For a recent discussion of this point for wormholes in AdS, see \cite{Held:2026huj, Held:2026bbo}. See also \cite{Betzios:2026rbv, Lavrelashvili:2026zsw} for other recent papers about oscillatory wormholes.

As a final point, we discuss the estimate of the EFT coefficient $\mathcal{C}_{n,-n}$ of the leading effective operator that reproduces the effect of a small wormhole with flux $n$, at odd $D$, that we did in the main body of the paper. To perform this estimate, one has to take an appropriate ratio between the path integral over the sphere with a small wormhole and its counterpart with no wormhole. To perform this ratio, we did not do a proper heat kernel analysis of the one-loop determinant. Instead, we did two things: First, we evaluated the contribution from some new zero and light modes to the path integral in the spacetime with the wormhole. Then, we argued how the rest of the ratio of functional determinants had to scale with the parameters of the problem, according to a proposed rule \nref{rtfct}. 

Since the problem for the functional determinant only had two parameters, the scale of the handle $r_{0,\text{phy}}$ and the de Sitter scale $\lds$, the relevant expression for the ratio of functional determinants, \nref{rtfct}, could not have been much different in odd $D$. The reason is that, in odd $D$, factors involving an extra regularization scale $\Lambda_{\text{cutoff}}$ would not enter power law corrections. Furthermore, we learned that if the rule we proposed for computing functional determinants had been any different, the EFT coefficients $\mathcal{C}_{n,-n}$ would have depended on $\lds$, which is not reasonable physically. 

Therefore, in a way, we were able to bypass a more first-principle calculation of the functional determinants at odd $D$. However, the situation at even $D$ is different. The reason is that at even $D$, because of the Weyl anomaly, one expects power law corrections to the one-loop determinant involving the ratio of physical scales and the cutoff scale. Taking the cutoff to be, for example, the Plank scale, it is plausible that the final expression for $\mathcal{C}_{n,-n}$ will end up being more complicated than its odd $D$ counterpart.

It seems then that it would be interesting to study the one-loop corrections to $\mathcal{C}_{n,-n}$ using more conventional heat-kernel techniques. The first reason for doing so is that we would like to obtain the same answer for $\mathcal{C}_{n,-n}$ from a more honest calculation, and in a way that gives us control over $O(1)$ contributions. And last, but not least, we would like to estimate these coefficients for wormholes in our universe, which has $D=4$ dimensions!

\subsection*{Acknowledgments}

We would like to specially thank Juan Maldacena for many useful insights and discussions, in particular for pushing us to include section \nref{smwhopis}. We would also like to thank Shoaib Akhtar, Beatrix Muehlmann, Yiming Chen, David Kolchmeyer, Jonah Kudler-Flam, Xiaoyi Liu, Xiao-Liang Qi, Douglas Stanford, Zimo Sun, Zhencheng Wang, and Edward Witten for discussions. We also thank Jonah Kudler-Flam, Juan Maldacena and Zimo Sun for comments on an earlier version of this draft. HT is supported by Shoucheng Zhang Graduate Fellowship Program. 

\appendix

\section{Ghost determinant}
\label{ghostdet}

Here, we discuss the details of the necessary ghost determinant for the gauge fixing procedure discussed in section \nref{genset}. The gauge fixing conditions we want to discuss are the following
\begin{equation}
P_{a}=-\sqrt{2}\n^{b}\tilde{h}_{ab}~~,~~ L_{a_{1}a_{2}..a_{D-3}}=-\n^{c}\delta B_{ca_{1}...a_{D-3}}
\end{equation}

To discuss the ghost determinant associated with these gauge fixing conditions, we need to write the variation of $P_{a}$ and $L_{a_{1}...a_{D-3}}$ with respect to the gauge transformations \nref{diffact} and \nref{formact}. However, the coordinate transformation \nref{diffact} will affect both the $P_{a}$ and $L_{a_{1}...a_{D-3}}$ conditions, while the form one \nref{formact} affects only $L_{a_{1}...a_{D-3}}$. This means that the map that takes one from gauge transformations to the gauge conditions is lower triangular. 

This seems to suggest that perhaps the ghost determinant should not depend on the variation of the $L_{a_{1}...a_{d-3}}$ condition with $\xi^{a}$, since a lower triangular matrix could be made diagonal by appropriate redefinitions. In section \nref{decgag} we argue that such a diagonalization is possible, in such a way that after that, the redefined gauge transformations act on the gauge conditions as
\begin{equation}
\label{ghostfunce}
P_{a}=-\n^{2}\xi_{a}-R_{ab}\xi^{b}~~,~~ L_{a_{1}a_{2}..a_{D-3}}=-(D-2)\n^{c}\n_{[c}C_{a_{1}...a_{D-3}]}
\end{equation}

After doing so, the ghost determinant problem factorizes into one for $P_{a}$ and one for $L_{a_{1}...a_{D-3}}$, which we will analyze in the following subsections. In subsection \nref{ddgh}, we discuss the ghost determinant for the de Donder condition, and in subsection \nref{formdet}, the equivalent problem for the form condition.

We should also mention that the ghost determinant for the $L_{a_{1}...a_{D-3}}$ gauge condition is not well-defined by itself, because it has residual zeros corresponding to exact $D-3$ forms. This implies that the gauge transformations $C_{a_{1}...a_{D-3}}$ have gauge redundancies, which themselves might have further gauge redundancies. Therefore, to compute the form ghost determinant, one has to introduce "ghosts-for-ghosts" repeatedly until there is no gauge redundancy left. With "form ghost determinant", we will therefore generally mean the ghost determinant obtained from this full procedure. However, we are going to be able to bypass doing this full calculation by using electromagnetic duality, as discussed in section \nref{formdet}.

\subsection{Decoupling the ghost determinants}
\label{decgag}

In this section we argue that we can redefine the gauge transformations \nref{diffact} and \nref{formact} in such a way that only $\xi^{a}$ affects the $P^{a}$ gauge condition and only the form gauge transformation $C_{a_{1}...a_{D-3}}$ affects $L_{a_{1}...a_{D-3}}$. To do so, we first use a different but equivalent formulation of the gauge transformation and gauge condition for the forms. 

The $L_{a_{1}..a_{D-3}}$ gauge condition is a $D-3$ form, and the gauge transformation of $\delta B$ is the $d$ of a $(D-3)$ form. Alternatively, using the Hodge star and some redefinitions, we can think of the gauge condition $L$ as a $3$-form $\tilde{L}_{a_{1}a_{2}a_{3}}$, which is related to $b^{ab}$ from the main text via
\begin{equation}
\tilde{L}_{abc}=3\n_{[a}b_{bc]}
\end{equation}

Also, since the gauge variation $\delta B_{a_{1}..a_{D-2}}$ is the $d$ of a $D-3$ form, the gauge variation of $b_{a_{1}a_{2}}$ is the co-differential, $\deltaco$, of a $3$ form as
\begin{equation}
\label{formact2}
b_{ab}=(\deltaco C)_{ab}=-\n^{c}\tilde{C}_{cab}
\end{equation}
with $\tilde{C}_{cab}$ an appropriate $3$ form gauge parameter. From here, we will also find it more convenient to define the action of the coordinate transformations as a combination of \nref{diffact} and \nref{formact2} with $\tilde{C}_{abc}=3\xi_{[a}\tilde{B}_{bc]}$ such that the fields change under it as
\begin{equation}
\label{diffact2}
h_{ab}=\frac{1}{\sqrt{2}}(\n_{a}\xi_{b}+\n_{b}\xi_{a})~~,~~  b^{ab}=\kappa(\tilde{F}^{a}\xi^{b}-\tilde{F}^{b}\xi^{a})
\end{equation}

Therefore, the gauge conditions transform with the gauge parameters as
\begin{equation}
\label{ghostfunc}
P_{a}=-\n^{2}\xi_{a}-R_{ab}\xi^{b}~~,~~ \tilde{L}_{abc}=-3\n_{[a}\n^{d}\tilde{C}_{bc]d}-6\tilde{F}_{[a}\n_{b}\xi_{c]}
\end{equation}

We can further simplify \nref{ghostfunc} by noting we can expand $\xi_{a}$ as
\begin{equation}
\xi_{a}=n_{a}m+\mcD_{a}s+V_{a}
\end{equation}
with $V_{a}$ a tangent transverse vector. The part of the transformation generated by the $m$ and $s$ scalars is annihilated in the $L_{a_{1}..a_{d-3}}$ gauge condition in \nref{ghostfunc}, so only $V_{a}$ mixes with the $\tilde{C}$ gauge transformation in \nref{ghostfunc}. 

A trick we can do is the following: We can further redefine the gauge transformation parameter $\tilde{C}_{abc}$ by an $V_{a}$ dependent 3-form as
\begin{equation}
\label{credef}
\tilde{C}_{abc}\rightarrow \tilde{C}_{abc}+6n_{[a}\mcD_{b}U_{c]}
\end{equation}
in such a way to cancel the contribution of $V_{a}$ in \nref{ghostfunc}. This is well defined as long as we can always find the necessary $U_{a}$ for a given $V_{a}$. This is possible here since the gauge condition $\n_{[a}\n^{d}C_{bc]d}$ acting on the modes $6 n_{[a}\mcD_{b}U_{c]}$ has a trivial kernel. More explicitly, let $U_{c}=U Y_{l,c}$, with $Y_{l,c}$ a transverse vector spherical harmonic defined as in \nref{tvharm}, then we have that
\begin{equation}
-3\n_{[a}\n^{d}\tilde{C}_{bc]d}=-6n_{[a}\mcD_{b}Y_{l,c]}(\mathcal{O}U)
\end{equation}
with
\begin{equation}
\label{opv}
(\mathcal{O}U)=-U''-(D-3)HU'+\frac{(l(l+D-2)+(D-3))}{a^{2}}U-(D-4)(H^{2}+H')U
\end{equation}

Then, if $\mathcal{O}$ has no zero modes, one can always solve for a $U$ to cancel the $V$ contribution in \nref{ghostfunc}. We check numerically that $\mathcal{O}$ has no zero modes, but we can also argue that analytically. The argument goes as follows: The operator that maps $\tilde{C}_{abc}$ to the gauge condition $\tilde{L}_{abc}$ in \nref{ghostfunc} is explicitly $d\deltaco$. Therefore, the operator $\mathcal{O}$ only has zeros if $d \deltaco$ restricted to 3-forms of the type $6 n_{[a}\mcD_{b}U_{c]}$ has zeros. Since $d$ and $\deltaco$ are adjoints of each other, this is only possible if these 3-forms are annihilated by $\deltaco$. However, one can check that these three forms are already annihilated by $d$, which implies that $d\deltaco$ acting on these forms only has zero modes if there are three-forms of this type annihilated by both $d$ and $\deltaco$. 

Therefore, $6 n_{[a}\mcD_{b}U_{c]}$ will be annihilated by the gauge condition if and only if it is a harmonic 3-form. For the wormholes we are studying, there are only harmonic 3-forms in $D=3$ or $D=4$. If $D=3$, the harmonic 3-form is $\epsilon_{abc}$ times a constant, which cannot be written as $6 n_{[a}\mcD_{b}U_{c]}$ if we require $U$ to be globally well defined. In $D=4$, the harmonic 3-form is completely tangent, so it cannot be of the form $6 n_{[a}\mcD_{b}U_{c]}$ since it has no normal components. Therefore, $\mathcal{O}$, as an operator acting on transverse vectors, has no zero modes, and one can always invert the relation to find $U$ for a given $V$. 

This implies that to compute the ghost determinant, we can work instead with the functional determinant of the map \nref{ghostfunce}, which is what we focus on doing in the rest of this section. 

\subsection{de Donder ghosts}
\label{ddgh}

\begin{table}[t]
\centering
\renewcommand{\arraystretch}{1.3} 
\setlength{\tabcolsep}{5pt}  

\begin{tabular}{|c|c|c|c|c|c|}
\hline
Spectrum & Sector & $l$ & $n_\text{zero}$ &
\makecell[c]{$n_\text{light},$\\[-2pt] $Q\to 0$} &
\makecell[c]{$n_\text{light},$\\[-2pt] $Q\to Q_c$}
\\ \hline

\multirow{5}{*}{\makecell[c]{de Donder\\ghost}}
& \multirow{3}{*}{\makecell[c]{Scalars}}
  & $l=0$   &  &  & $1$ \\ \cline{3-6}
& & $l=1$   &  & $N\times D$ &  \\ \cline{3-6}
& & $l\ge 2$&  &  &  \\ \cline{2-6}

& \multirow{2}{*}{\makecell[c]{Tangent transverse\\vectors}}
  & $l=1$   & $1\times \frac{D(D-1)}{2}$ &  &  \\ \cline{3-6}
& & $l\ge 2$&  &  &  \\
\hline

\multirow{2}{*}{$\Delta_{0}$}
& \multirow{2}{*}{Scalars}
  & $l=0$   & $1$ &  &  \\ \cline{3-6}
& & $l\ge 1$&  &  &  \\
\hline

\multirow{3}{*}{\makecell[c]{$\Delta_2$}}
& \makecell[c]{Tangent transverse\\vectors}
  & $l\geq1$   &  &  &  \\ 
\cline{2-6}

& \multirow{2}{*}{\makecell[c]{Tangent transverse\\two-forms}}
  & $l=1$   & $1$, iff $D=3$ &  &  \\ \cline{3-6}
& & $l\ge 2$&  &  &  \\
\hline
\end{tabular}

\caption{A summary of several important features of ghost spectrum for general charge $0<Q<Q
_c$ and general dimension $D\geq3$, highlighting on the number of zero modes $n_\text{zero}$ and light modes $n_\text{light}$ in two limits. For the form ghost part, we show the relevant part of the Hodge Laplacian of 0-forms and $(D-2)$-forms. Notice that $\Delta_{D-2}$ has the same spectrum as $\Delta_2$, so numerically we solve $\Delta_2$ instead, which is consistent with our general strategy of using a two-form to parametrize a $(D-2)$-form. To make the visualization clearer, we leave the cell blank if there are no such modes.}
\label{tab: ghost}

\end{table}

In this section, we discuss the ghost determinant associated with inserting the gauge condition $P_{a}=-\sqrt{2}\n^{b}\tilde{h}_{ab}$ in the path integral with an appropriate gauge fixing term. First, we note that the division by differmorphism in the path integral \nref{z1loop} is the division by the local path integral
\begin{equation}
Z_{\xi}=\frac{1}{\int D\xi} 
\end{equation}
of diffeomorphisms $\xi^{a}$ around the background, where we fix the measure of $\xi^{a}$ by defining the local norm
\begin{equation}
\label{xinorm}
(\xi,\xi)=\int \xi_{a}\xi^{a}
\end{equation}
and imposing that
\begin{equation}
\int D\xi\, e^{-\frac{1}{2}(\xi,\xi)}=1
\end{equation}

With this setup in mind, we can remove the integral over $\xi$ by inserting into the path integral an identity of the form
\begin{equation}
\label{dDid}
\int D'\omega\,e^{-\frac{1}{2}(\omega,\omega)}\int D'\xi\,\delta'(P^{a}(\xi)-\omega^{a})\, \text{Det}'(\Delta_{gh,dD})=1
\end{equation}
where the norm of $\omega$ is the same as \nref{xinorm}, and the measure $D\omega$ is normalized the same way as the $\xi$ measure. The $'$ in many expressions stands for the fact that, because the background is $SO(D)$ invariant, the local gauge condition will vanish if $\xi$ is an $SO(D)$ generator. Furthermore, the gauge conditions we are interested in will be orthogonal to these $SO(D)$ generators, since they are orthogonal to isometries. Therefore, one should only integrate over $\omega$ in the image of $P$, such that the delta function condition has solutions, and only integrate over $\xi$ that change $P$ non-trivially. The factor of $\text{Det}'(\Delta_{gh, dD})$ is functional determinant of $P^{a}$ with respect to $\xi^{a}$, with zero modes removed.

Inserting the identity into the path integral cancels most of the division by the gauge group in $Z_{\xi}$, except for the $\xi$ that are missing from \nref{dDid}. These missing $\xi$ are precisely the isometries of the background, whose group action we should still divide over. The strategy to compute the path integral over these modes is similar to the one we discussed for the twist modes in \nref{zeromod}. 

The strategy goes as follows: Assuming the $\xi^{a}$ isometries to close into an appropriate $SO(D)$ algebra, we rescale the $\xi$ such that they respect the canonical Lie algebra\footnote{One might perhaps expect the bracket to become complicated because of the way we deformed the gauge transformations in \nref{diffact2} and \nref{credef}. However, we do not expect this point to affect the one-loop answer.} $[\xi_{a},\xi_{b}]=f_{abc}\xi^{c}$ of $SO(D)$. Because of how we defined coordinate transformations in \nref{diffact}, we have to rescale the coordinate transformations as $\xi^{a}\rightarrow\frac{\xi^{a}}{\sqrt{2}\kappa}$, with $\xi^{a}$ an isometry generator with the usual $SO(D)$ action. Integrating over such $\xi$ will then be related to the group integral over $SO(D)$. The difference is that in the path integral, we integrate over $\xi$ with the path integral measure, which is not the group measure. Restricted to the isometries, these measures should be the same up to a normalization factor, which we can fix by computing the norm of an isometry $\xi$ using the canonical group norm and the path integral one. Since an $SO(D)$ generator is required to have norm one in the canonical group norm, the normalization is
\begin{equation}
\label{niso}
\frac{1}{2\kappa^{2}}(\xi^{a},\xi^{a})=\mathcal{N}_{\text{iso}}^{2}(\xi^{a},\xi^{a})_{\text{can}}=\mathcal{N}_{\text{iso}}^{2}
\end{equation}
where $\mathcal{N}_{\text{iso}}$ is the normalization factor for these isometry modes, $(\cdot ,\cdot)$ is the usual path integral norm for $\xi$ and $(\cdot,\cdot)_{\text{can}}$ is the canonical group norm. For the wormhole, the contribution from the division by these isometries is therefore
\begin{equation}
Z_{\text{iso}}=\frac{1}{(\int D\xi)_{\text{iso}}}=\frac{\mathcal{N}_{\text{iso}}^{-|SO(D)|}}{\text{Vol}(SO(D))_{c}}
\end{equation}
with $|SO(D)|$ the dimension of $SO(D)$ and $\text{Vol}(SO(D))_{c}$ its canonical group volume. There should also be some residual discrete gauge symmetries that we have to divide by, but we imagine these contributions are included later in the calculation.

Other than that, inserting the identity \nref{dDid} into the path integral adds the gauge fixing term for $P^{a}$ in \nref{gfterm} into the action, and it also inserts the functional determinant for $P$. For the gauge condition and gauge transformations we defined to arrive at \nref{ghostfunce}, the relevant functional determinant will be that of
\begin{equation}
\label{dedodgh}
P_{a}=-\n^{2}\xi^{a}-R_{ab}\xi^{b}
\end{equation}
which maps the appropriately defined coordinate transformations to the gauge condition $P^{a}=-\sqrt{2}\n^{b}\tilde{h}_{ab}$. The ghost determinant for the de Donder gauge condition will therefore be
\begin{equation}
\label{zdD}
(Z_{\text{ghost}})_{\text{de Donder}}=Z_{\text{iso}}\times \text{Det}'(-\delta_{a}^{b}\n^{2}-\tensor{R}{^b_a})
\end{equation}

To solve for the eigenvalue spectrum of this operator, we will use the $SO(D)$ invariance of the background to simplify the problem, by expanding the coordinate transformation $\xi^{a}$ as
\begin{equation}
\xi_{a}=n_{a}m+\mcD_{a}s+V^{a}
\end{equation}
with $V^{a}$ a tangent transverse vector. $SO(D)$ invariance implies that the eigenvalue equation for the scalars will decouple from the one for the transverse vectors. One can then proceed by expanding all these fields into appropriate spherical harmonics as
\begin{equation}
m=\sum_{l=0}^{\infty}m_{l}(\tau)Y_{l}~~,~~ s=\sum_{l=1}^{\infty}s_{l}(\tau)Y_{l}~~,~~ V_{a}=\sum_{l=1}^{\infty}V_{l}(\tau)Y_{l,a}
\end{equation}
with the spherical harmonics defined as in \nref{scharm} and \nref{tvharm}. Using this decomposition, we can solve for the spectrum of \nref{dedodgh} (see Table \nref{tab: ghost}). We solve for it by writing the eigenvalue equation for the functional determinant of $P^{a}$ sector by sector. With the following notation in mind
\begin{equation}
H=\frac{a'}{a}~~,~~\n_{\tau}^{2}=\partial_{\tau}^{2}+(D-1)H\partial_{\tau}~~,~~\tll=l(l+D-2)
\end{equation}
the eigenvalue problems are as discussed below.

\textbf{Scalars:} The eigenvalue equation for the scalar sector of \nref{dedodgh} is, using the notation \nref{noteig}
\begin{equation}
\begin{gathered}
\label{ghostsceq}
\lambda m_{l}=\bigg[-\n_{\tau}^{2}m_{l}+\frac{\tilde{\lambda}_{l}}{a^{2}}m_{l}+(D-1)(2H^{2}+H')m_{l}\bigg]-\frac{2H \tll}{a^{2}}s_{l}~~,\\
\lambda s_{l}=\bigg[-\n_{\tau}^{2}s_{l}+2H s_{l}'+\frac{\tll}{a^{2}}s_{l}+2\bigg((D-1)H^{2}+H'-\frac{(D-2)}{a^{2}}\bigg)s_{l}\bigg]-2H m_{l}
\end{gathered}
\end{equation}
where we should remember that the field $s_{l}$ only exists for $l \geq 1$, so it should be excluded from the eigenvalue equations for $l=0$. The norm for the fields $m$ and $s$, inherited from the local norm \nref{xinorm}, is
\begin{equation}
(\xi,\xi)=\int (m^{2}+s(-\mcD^{2})s)
\end{equation}

\begin{figure}[t]
    \centering
    \includegraphics[width=1\linewidth]{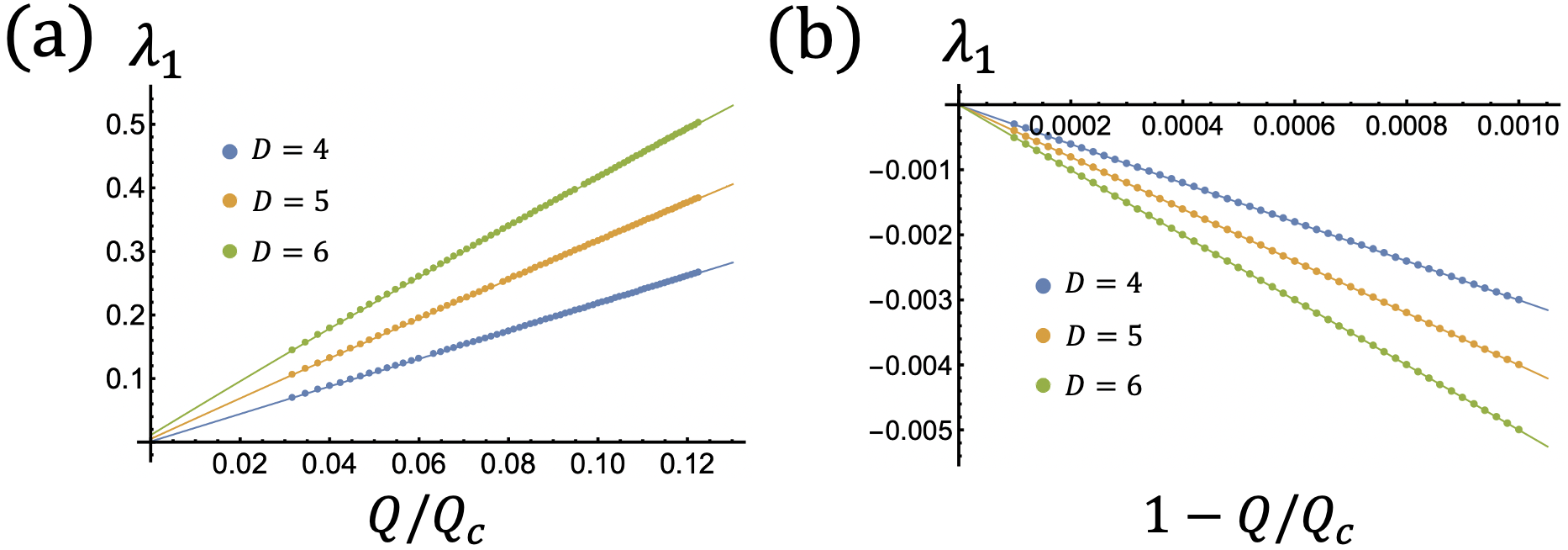}
    \caption{Illustration of the scaling of eigenvalues of light mode in de Donder ghosts. We choose cycle number $N=1$. \textbf{(a)} The eigenvalue $\lambda_1$ of the light mode in the near sphere limit in $l=1$ sector, for various $D$. \textbf{(b)} The eigenvalue $\lambda_1$ of the light mode in the near Einstein in $l=0$ sector, for various $D$. Notice that although this mode is negative,  we only care about its absolute value, since eventually we only take the absolute value of ghost determinant.}
    \label{fig: ghost light modes}
\end{figure}

We should briefly comment on the spectrum we observe in solving \nref{ghostsceq}. For simplicity, let us discuss $N=1$ solutions first. Near $\kappa Q \approx 0$, the spectrum of this equation is very similar to the sphere one, but some zero modes of the $S^{D}$ answer are lifted to the light modes of the wormhole. These are the $SO(D+1)$ isometries of $S^{D}$ that are not in $SO(D)$. There are $D$ of these modes, and they are all in the $l=1$ sector. We find numerically that their eigenvalue scales as $\lambda \sim \kappa Q$, see figure \nref{fig: ghost light modes}(a) for an illustration. For the solution with $N$ cycles, there are $N$ sets of these light modes, which also have the $\lambda \sim \kappa Q$ scaling. 

Another relevant aspect of the spectrum is that near $Q \approx Q_{c}$ it contains a light mode at the $l=0$ sector. This mode corresponds to the time translation isometry of the $Q=Q_{c}$ solution. For $Q^{2}=Q_{c}^{2}(1-\epsilon)$, this mode is a light mode that scales linearly in $\epsilon$(see figure \nref{fig: ghost light modes}(b))

\textbf{Transverse vectors:} The eigenvalue equation for the transverse vector sector of \nref{dedodgh} is
\begin{equation}
\label{ghosttveq}
\lambda V_{l}=\bigg[-\n_{\tau}^{2}+\frac{\tll}{a^{2}}+\bigg(DH^{2}+H'-\frac{(D-1)}{a^{2}}\bigg)\bigg]V_{l}
\end{equation}
where we have to remember that, since $V_{l}$ is defined from transverse vector harmonics, it only exists for $l \geq 1$. 

We should briefly discuss the spectrum we obtain in solving \nref{ghosttveq}. Near $\kappa Q \approx 0$, the spectrum is very close to that of the sphere, with no new modes. Like the sphere, it contains zero modes in the $l=1$ transverse vector harmonics, corresponding to the $SO(D)$ isometries. We find no new zero or light modes as we increase $Q$.

\subsection{Form ghost determinant}
\label{formdet}

Here we discuss the ghost determinant related to the $L_{a_{1}...a_{D-3}}$ gauge condition in \nref{ghostfunce}. Fortunately, to deal with this calculation, we can leverage the fact that the path integral over $p$-forms was studied in detail in the literature \cite{Donnelly:2016mlc}. To do so, we proceed as follows: We imagine we fix the wormhole background geometry, which we will refer to as a manifold $\mathcal{M}$, and we study the path integral over the $D-2$ form theory in this manifold. The theory is described by the matter action in \nref{action}, and we can gauge fix it by adding the gauge fixing term for $L_{a_{1}..a_{D-3}}$ in \nref{gfterm} and inserting the appropriate ghost determinant. This ghost determinant will be the same one we are interested in evaluating in this subsection, since the background geometry is the same. 

However, we know that the full path integral for the $D-2$ form theory in this manifold, including the sum over $D-1$ form instantons, is equal to\cite{Donnelly:2016mlc}
\begin{equation}
\label{dminus2pf}
Z_{D-2,f_{\theta}}(\mathcal{M})=(Z_{\text{inst}})_{D-2,f_{\theta}}\times(Z_{\text{ghost}})_{D-2,f_{\theta}}\times \text{Det}'(\Delta_{D-2}^{-\frac{1}{2}})\times (Z_{\text{zero}})_{D-2,f_{\theta}} 
\end{equation}
where the subscript $(D-2,f_{\theta})$ stands for the fact that these quantities are computed in the $D-2$ form theory with coupling constant $f_{\theta}$. Let us explain where every term comes from in this formula: $Z_{\text{inst}}$ is the sum over all the instantons of the theory as
\begin{equation}
\label{zinst}
Z_{\text{inst}}=\sum_{\text{instanton}}e^{-I_{\text{inst}}} 
\end{equation}

Furthermore, for every such term in the sum over instantons, there will be an associated one-loop path integral, which will not depend on the instanton solution. In fact, the path integral around them will be one-loop exact since the $D-2$ form theory is quadratic. The one-loop determinant will be computed with the same gauge fixing strategy we discussed in the main text \nref{genset}. Therefore, they will consist of a ghost determinant and the path integral over the gauge-fixed action around the instanton. 

The non-zero modes of the gauge fixed action are the contribution $\text{Det}'(\Delta_{D-2}^{-\frac{1}{2}})$, but there will be associated zero modes. The zero modes will be harmonic $D-2$ forms, that might or might not exist for the solution. Assuming they exist, it is convenient to expand them in an appropriate basis of harmonic $D-2$ forms $w_{I}$ of the background. Let $w_{I}$ be a basis for the set of harmonic $D-2$ forms with integer flux around the non-contractible $D-2$ dimensional surfaces of $\mathcal{M}$. Then, we can write the most general $D-2$ form zero mode as
\begin{equation}
\delta B_{a_{1}...a_{D-2}}|_{\text{zero mode}}=\sum_{I} \beta_{I}w_{I} 
\end{equation}

However, we should identify these $D-2$ flat connections under large gauge transformations, which implies that we should restrict $\beta_{I}$ to the interval $\big[0,\frac{2\pi}{f_{\theta}}\big)$. For more details, see \cite{Donnelly:2016mlc}\footnote{The $\beta_{I}$ interval in their notation in $2 \pi$, but ours differs by a factor of $f_{\theta}$ since we rescaled the $B$ fields to be canonically normalized in \nref{action}.}. The path integral over these zero modes is then, using the norm we fixed in \nref{locnorm} and \nref{normnor}
\begin{equation}
\label{zd2zero}
(Z_{\text{zero}})_{D-2,f_{\theta}}=\text{det}^{\frac{1}{2}}\bigg(\frac{2 \pi}{f_{\theta}^{2}}\Gamma_{D-2}\bigg)
\end{equation}
where we defined $\Gamma_{D-2}$ to be the inner product matrix of the basis $w_{I}$. For later convenience, we generally define the matrix $\Gamma_{p}$ for an associated basis of harmonic $p$-forms with integer flux to be
\begin{equation}
(\Gamma_{p})_{IJ}=\frac{1}{p!}\int (\omega_{I})_{a_{1}..a_{p}}(\omega_{J})^{a_{1}..a_{p}} 
\end{equation}

Now that we explained what every term stands for in \nref{dminus2pf}, we are going to use a trick to compute $(Z_{\text{ghost}})_{D-2}$, which is the ghost determinant we are interested in. The idea is that electromagnetic duality implies that \nref{dminus2pf} is equal to the path integral over $0$ forms in this manifold, with coupling constant $f_{\theta} \rightarrow \frac{2 \pi}{f_{\theta}}$. These are the scalars with period $f_{\theta}$ that we reviewed in section \nref{prelim}. Therefore, the following identity is true
\begin{equation}
\begin{gathered}
(Z_{\text{inst}})_{D-2,f_{\theta}}\times(Z_{\text{ghost}})_{D-2,f_{\theta}}\times \text{Det}'(\Delta_{D-2}^{-\frac{1}{2}})\times \text{det}^{\frac{1}{2}}\bigg(\frac{2 \pi}{f_{\theta}^{2}}\Gamma_{D-2}\bigg)\\
=(Z_{\text{inst}})_{0,\frac{2\pi}{f_{\theta}}}\times \text{Det}'(\Delta_{0}^{-\frac{1}{2}})\times \text{det}^{\frac{1}{2}}\bigg(\frac{f_{\theta}^{2}}{2 \pi}\Gamma_{0}\bigg)e^{S_{\text{ano,em}}}
\end{gathered}
\end{equation}
where the $0$-form theory does not have a ghost determinant, and $S_{\text{ano}}$ is an anomaly in the duality that exists if $D$ is even. It is given in our context by
\begin{equation}
\label{sano}
S_{\text{ano,em}}=\begin{cases*}
    0~~,~~~~~~~~~~~~~~~~~~~~~\text{for odd $D$}\\
    \frac{1}{2}\chi(M)\log\big(\frac{2\pi}{f_{\theta}^{2}}\big)~~,~~\text{for even $D$}
    \end{cases*}
\end{equation}
with $\chi(M)=\sum_{k}(-1)^{k}b_{k}$, and $b_{k}$ the $k$-th betti numbers of the manifold. Note that we wrote \nref{sano} adapted to our choice of measure for the path integral in the main text, but it will generally depend on the measure of the path integral non-trivially (see \cite{Donnelly:2016mlc}).

We then use the result in \cite{Donnelly:2016mlc} that
\begin{equation}
\label{zinstdual}
(Z_{\text{inst}})_{0,\frac{2\pi}{f_{\theta}}}=\frac{1}{\text{det}^{\frac{1}{2}}\bigg(\frac{f_{\theta}^{2}}{2 \pi}\Gamma_{1}\bigg)}(Z_{\text{inst}})_{D-2,f_{\theta}}
\end{equation}
to conclude that
\begin{equation}
\label{ghostL}
(Z_{\text{ghost}})_{D-2,f_{\theta}}=\frac{1}{\text{det}^{\frac{1}{2}}\bigg(\frac{2 \pi}{f_{\theta}^{2}}\Gamma_{D-2}\bigg)\text{Det}'(\Delta_{D-2}^{-\frac{1}{2}})}{}\text{Det}'(\Delta_{0}^{-\frac{1}{2}})\text{det}^{\frac{1}{2}}\bigg(\frac{f_{\theta}^{2}}{2 \pi}\Gamma_{0}\bigg)\text{det}^{-\frac{1}{2}}\bigg(\frac{f_{\theta}^{2}}{2 \pi}\Gamma_{1}\bigg)e^{S_{\text{ano,em}}}
\end{equation}
which is the ghost determinant for the form gauge condition $L_{a_{1}..a_{D-3}}$ in \nref{ghostfunce}, that we wanted to find. A convenient way to read \nref{ghostL} is as follows: The first term cancels the gauge fixed one-loop path integral over $D-2$ forms, and the next two come from the one-loop path integral over the dual massless scalar with coupling $\frac{2 \pi}{f_{\theta}}$. The last term is a bit less intuitive, but it is necessary to relate the sum over instantons on the two sides of electromagnetic duality. The reason is that relating the sum over instantons in \nref{zinstdual} involves a Poisson resummation which introduces this type of factor, so this term is needed.

We should also mention that the last two terms of \nref{ghostL} are simple to compute for the wormhole we have. Namely, we know that any manifold has a single harmonic zero form. The basis zero form $w_{0}$ needs to have unit flux, which implies that this zero form is the constant function equal to one, and
\begin{equation}
\label{harm0form}
\text{det}^{\frac{1}{2}}\bigg(\frac{f_{\theta}^{2}}{2 \pi}\Gamma_{0}\bigg)=\bigg(\frac{f_{\theta}^{2}}{2\pi}\int 1\bigg)^{\frac{1}{2}}=\bigg(\frac{f_{\theta}^{2}\text{Vol}(\mathcal{M})}{2\pi}\bigg)^{\frac{1}{2}}
\end{equation}
where $\text{Vol}(\mathcal{M})$ is the volume of the manifold in question. 

For the $\Gamma_{1}$ factor, note that not every manifold has a harmonic 1-form. For example, the sphere $S^{D}$ does not. However, the wormhole we are studying does have one, and only one, harmonic 1-form. It corresponds to the following form
\begin{equation}
\label{harm1form}
(w_{1})_{a}=\bigg(\int_{\text{loop}}\frac{d\tau'}{a^{D-1}(\tau')}\bigg)^{-1}\frac{(d\tau)_{a}}{a^{D-1}(\tau)}
\end{equation}
where the prefactor in \nref{harm1form} is so that $w_{1}$ has an unit loop integral around the $S^{1}$ direction, which is its associated flux. Its contribution to the path integral is therefore 
\begin{equation}
\label{gamma1cont}
\text{det}^{-\frac{1}{2}}\bigg(\frac{f_{\theta}^{2}}{2 \pi}\Gamma_{1}\bigg)=\bigg(\frac{f_{\theta}^{2}}{2\pi}\text{Vol}(S^{D-1})\bigg(\int_{\text{loop}}\frac{d\tau'}{a^{D-1}(\tau')}\bigg)^{-1}\bigg)^{-\frac{1}{2}}
\end{equation}

Another relevant aspect of \nref{ghostL} is that the factor containing the determinant of $\Gamma_{D-2}$ always cancels an identical term in the path integral over fluctuations of the wormhole. The reason is that the modes that contribute as in \nref{zd2zero} persist in the gravity+form computation, as we discussed in section \nref{zeromod}.

We should also briefly comment on the spectrum of the Hodge Laplacians $\Delta_{0}$ and $\Delta_{D-2}$ that appear in \nref{ghostL} (see Table \nref{tab: ghost}). Numerically\footnote{To study the spectrum of $\Delta_{D-2}$, or equivalently $\Delta_{2}$, numerically, we have to solve \nref{eigeneq} with the metric degrees of freedom removed. $\Delta_{0}$ is just the usual scalar Laplacian.}, we found that for $\kappa Q \approx 0$ the spectrum of these operators is a small deformation of their sphere spectrum, unless $D=3$, in which case $\Delta_{D-2}$ has a new zero mode due to the harmonic 1-forms. Increasing the charge from $Q \approx 0$ to $Q=Q_{c}$, we see no qualitatively new behaviour. 

\section{Eigenvalue equations}
\label{eigensec}

In this section, we explicitly write the eigenvalue equation \nref{eigeneq} for the fluctuations in terms of the decomposition \nref{decompf}, and the further decomposition of its fields into harmonics. These are the equations that we use solve the eigenvalue problem numerically, as discussed in more detail in Appendix \nref{numdeta}. For convenience, we define the following notation
\begin{equation}
\label{noteig}
H=\frac{a'}{a}~~,~~\n_{\tau}^{2}=\partial_{\tau}^{2}+(D-1)H\partial_{\tau}~~,~~\tll=l(l+D-2)
\end{equation}

Since the eigenvalue problem for different $SO(D)$ representations decouples, we will write them separately, starting with the scalars. Also, since considering the eigenvalue equation in different norms is easy, we write the eigenvalue equations for the following family of norms
\begin{equation}
(\Phi,\Phi)=\int h_{ab}h^{ab}+\frac{(\textcolor{red}{p_{1}}-1)}{D}\int h^{2}+\textcolor{blue}{p_{2}}\int b_{ab}b^{ab}
\label{eq: De Witt parameter}
\end{equation}
which are a two-parameter family generalization of \nref{locnorm}.

\textbf{Scalars:} The eigenvalue equation for the scalars of \nref{decompf}, using decomposition \nref{scexp}, is 
\begin{equation}
\begin{gathered}
\label{eigensc}
\textcolor{red}{p_{1}}\lambda \psi_{l}=\frac{(D-2)}{2}\bigg[\n_{\tau}^{2}\psi_{l}-\frac{\tilde{\lambda}_{l}}{a^{2}}\psi_{l}+\bigg(2(D-1)+\frac{2\kappa^{2}q(D-2)}{D}\bigg)\psi_{l}\bigg]\\-\frac{2\kappa^{2}q(D-2)}{D}A_{l}+\frac{(D-2)\sqrt{2q}\kappa}{D}\frac{\tilde{\lambda}_{l}}{a^{2}}\chi_{l}~~,\\
\lambda A_{l}=\bigg[-\n_{\tau}^{2}A_{l}+\frac{\tll}{a^{2}}A_{l}+\bigg(\frac{2D}{a^{2}}-2(D-1)+\frac{2\kappa^{2}q(D^{2}-6D+4)}{D(D-2)}\bigg)A_{l}\bigg]\\-4 H \frac{\tll f_{l}}{a^{2}}-\frac{2\kappa^{2}q(D-1)(D-2)}{D}\psi_{l}-\frac{2\sqrt{2q}\kappa(D-1)}{D}\frac{\tll\chi_{l}}{a^{2}}~~,\\
\lambda f_{l}=\bigg[-\n_{\tau}^{2}f_{l}+\frac{\tll}{a^{2}}f_{l}+2Hf_{l}'+\bigg(\frac{(D+1)}{a^{2}}-2(D-1)+\frac{\kappa^{2}q(D-5)}{(D-2)}\bigg)f_{l}\bigg]\\
-\frac{2HD}{(D-1)}A_{l}-\frac{2H(D-2)}{(D-1)}\bigg(\frac{\tll-(D-1)}{a^{2}}\bigg)\eta_{l}-\sqrt{2 q}\kappa(\chi_{l}'+(D-3)H\chi_{l})~~,\\
\lambda \eta_{l}=\bigg[-\n_{\tau}^{2}\eta_{l}+\frac{\tll}{a^{2}}\eta_{l}+4H\eta_{l}'+\bigg(-\frac{4}{a^{2}}-2(D-3)+\frac{4\kappa^{2}q}{(D-1)(D-2)}\bigg)\eta_{l}\bigg]
-4Hf_{l}~~,\\
\textcolor{blue}{p_{2}}\lambda \chi_{l}=(-\n_{\tau}^{2}\chi+\frac{\tll}{a^{2}}\chi_{l}+2H\chi_{l}'-(D-3)H'\chi_{l})\\+\sqrt{2q}\kappa\bigg[\frac{(D-2)}{2}\psi_{l}-A_{l}+f_{l}'-(D-1)H f_{l}\bigg]
\end{gathered}
\end{equation}
where we should remember that for some values of $l$ the fields are not defined. That is, one should delete the eigenvalue equations for $f_{l}$ and $\chi_{l}$ if $l=0$, and set them to zero in the other equations. Analogously, one should delete the eigenvalue equation for $\eta_{l}$ and set them to zero in the other equations if $l=0$ or $l=1$.

Also, note that the norm for the fields $\{\psi, A, f, \eta, \chi\}$ is not a flat norm because they are inherited from the local norm for the fluctuation fields. Their norm is as follows
\begin{equation}
\begin{gathered}
\label{normsc}
(\Phi,\Phi)=\int \bigg[\textcolor{red}{p_{1}}D\psi^{2}+\frac{D}{(D-1)}A^{2}+2f(-\mcD^{2})f\\+\frac{(D-2)}{(D-1)}\eta(-\mcD^{2})\bigg(-\mcD^{2}-\frac{(D-1)}{a^{2}}\bigg)\eta+2\textcolor{blue}{p_{2}}\chi(-\mcD^{2})\chi\bigg]
\end{gathered}
\end{equation}

\textbf{Tangent transverse vectors:} Analogously, the eigenvalue equation for the transverse vectors using the decompositions \nref{decompf} and \nref{tvexp} is
\begin{equation}
\begin{gathered}
\label{eigentv}
\lambda v_{l}=\bigg[-\n_{\tau}^{2}+\frac{\tll}{a^{2}}+\bigg(\frac{(D+1)}{a^{2}}-D+\frac{\kappa^{2}q(D^{2}-6D+4)}{(D-1)(D-2)}\bigg)\bigg]v_{l}\\
-2H\frac{(\tll-(D-1))}{a^{2}}x_{l}+\sqrt{2q}\kappa\frac{(\tll+(D-3))}{a^{2}}\omega_{l}-\sqrt{2q}\kappa(u_{l}'+(D-2)Hu_{l})~~,\\
\lambda x_{l}=\bigg[-\n_{\tau}^{2}x_{l}+\frac{\tll}{a^{2}}x_{l}+2 H x_{l}'+\bigg(-\frac{2}{a^{2}}-(D-2)+\frac{\kappa^{2}q}{(D-1)(D-2)}\bigg)x_{l}\bigg]-2Hv_{l}~~,\\
\lambda \textcolor{blue}{p_{2}}u_{l}=\bigg[-\n_{\tau}^{2}+\frac{\tll}{a^{2}}+\bigg(\frac{(D-3)}{a^{2}}+(D-2)-\frac{(D-2)\kappa^{2}q}{(D-1)}\bigg)\bigg]u_{l}\\-2H\frac{(\tll+(D-3))}{a^{2}}\omega_{l}
+\sqrt{2q}\kappa(v_{l}'-(D-2)Hv_{l})\\
\lambda \textcolor{blue}{p_{2}}\omega_{l}=\bigg[-\n_{\tau}^{2} \omega_{l}+\frac{\tll}{a^{2}}\omega_{l}+2H\omega_{l}'+\bigg(\frac{2}{a^{2}}+(D-4)-\frac{3\kappa^{2}q}{(D-1)(D-2)}\bigg)\omega_{l}\bigg]\\-2Hu_{l}
+\sqrt{2q}\kappa v_{l}~~,
\end{gathered}
\end{equation}
where again we have to remember that all these fields are only defined for $l$ at least 1. In particular, the $x_{l}$ mode only exists for $l \geq 2$, and needs to be excluded from the eigenvalue equation when $l=1$. Also, note that the norm for the fields $\{v_{a}, x_{a},u_{a}, \omega_{a}\}$ follow from \nref{locnorm} and is equal to
\begin{equation}
\begin{gathered}
(\Phi,\Phi)=\int \bigg[2v_{a} v^{a}+2x_{a}\bigg(-\mcD^{2}-\frac{(D-2)}{a^{2}}\bigg)x^{a}+2\textcolor{blue}{p_{2}}u_{a} u^{a}+2\textcolor{blue}{p_{2}} \omega_{a}\bigg(-\mcD^{2}+\frac{(D-2)}{a^{2}}\bigg)\omega^{a}\bigg]
\end{gathered}
\end{equation}

\textbf{Tangent transverse two-forms:} The eigenvalue equation for the transverse two-forms, using the decompositions \nref{decompf} and \nref{t2fexp}, is
\begin{equation}
\label{eigent2f}
\textcolor{blue}{p_{2}}\lambda j_{l}=\bigg(-\n_{\tau}^{2}+\frac{\tll}{a^{2}}+2(D-3)-\frac{4\kappa^{2}q}{(D-1)(D-2)}\bigg)j_{l}
\end{equation}
where $j_{l}$ is only defined for $l \geq 1$.

\textbf{Tangent transverse traceless symmetric tensors:} The eigenvalue equation for the tangent transverse symmetric tensors is, using the decompositions \nref{decompf} and \nref{t2texp}
\begin{equation}
\label{eigent2t}
\lambda \phi_{l}=\bigg(-\n_{\tau}^{2}+\frac{\tll}{a^{2}}\bigg)\phi_{l}
\end{equation}
where $\phi_{l}$ is only defined for $l \geq 2$. An interesting point is that equation \nref{eigent2t} is the eigenvalue equation for the scalar Laplacian, so the eigenmodes of $\phi_{l}$ are those of the scalar Laplacian with $l \geq 2$. However, the degeneracies are different, since the degeneracy of the eigenvalues of $\phi_{l}$ for a given $l$ is that of the spherical harmonics $Y_{l,(ab)}$, instead of the scalar spherical harmonic $Y_{l}$.

\subsection{Alternative gauge fixing terms}
\label{altgf}

Here, we discuss how the eigenvalue equations would change if we added different gauge fixing terms to the action instead of the one in \nref{gfterm}. What we investigate, in particular, is what would have happened if we had added
\begin{equation}
I_{gf}(\alpha,\beta)=\alpha \int (\n^{b}h_{ab}-\beta \n_{b}h)(\n^{c}h_{ab}-\beta\n^{c}h)+\frac{1}{6}\int(db)_{abc}(db)^{abc}
\label{eq: general gauge fixing}
\end{equation}
with $\alpha \neq 1$ and $\beta \neq \frac{1}{2}$ to the action instead of \nref{gfterm}. The motivation is that we would like to make sure the prescription we used to compute the phase in \nref{featspec} is not gauge fixing dependent. Since the phase we obtained came only from the scalar sector, here we restrict ourselves to studying what would have changed in the scalar sector, for simplicity. 

For alternative gauge fixing terms, to reproduce the correct eigenvalue equation, we need to add the following term to the right-hand side of \nref{eigensec}
\begin{equation}
\begin{gathered}
\label{eigeneqgfch}
\lambda \psi_{l} \supset -\frac{(4\alpha(\beta D-1)^{2}-(D-2)^{2})}{2 D}\bigg[\psi_{l}''+(D-1)H\psi_{l}'-\frac{\tll}{a^{2}}\psi_{l}\bigg]\\
+\frac{(2\alpha(\beta D-1)-(D-2))}{D}\bigg[A_{l}''+(2D-1)HA_{l}'+D(H'+(D-1)H^{2})A_{l}\\+\frac{\tll A_{l}}{a^{2}(D-1)}
-\frac{2\tll}{a^{2}}(f_{l}'+(D-2)Hf_{l})+\frac{(D-2)}{a^{4}(D-1)}\tll\big(\tll-(D-1)\big)\eta_{l}\bigg],\\
\lambda A_{l} \supset \frac{2(1-\alpha)(D-1)}{D}\bigg[A_{l}''+(D-1)HA_{l}'+D(H'-H^{2})A_{l}-\frac{\tll A_{l}}{a^{2}(D-1)^{2}}\\
-\frac{\tll}{a^{2}}\bigg(\frac{(D-2)}{(D-1)}f_{l}'-4Hf_{l}\bigg)-\frac{(D-2)}{a^{4}(D-1)^{2}}\tll\big(\tll-(D-1)\big)\eta_{l}\bigg]\\
+\frac{(D-1)(2\alpha(\beta D-1)-(D-2))}{D}\bigg[\psi_{l}''-H\psi_{l}'+\frac{\tll\psi}{a^{2}(D-1)}\bigg]~~,~~\\
\lambda f_{l} \supset (1-\alpha)\bigg[f_{l}''+(D-3)Hf_{l}'-\frac{\tll}{a^{2}}f_{l}+(D-1)H'f_{l}-2(D-1)H^{2}f_{l}\\
+\frac{(D-2)}{(D-1)}A_{l}'+\frac{(D^{2}-D+2)}{(D-1)}HA_{l}-\frac{(D-2)}{a^{2}(D-1)}\big(\tll-(D-1)\big)(\eta_{l}'-4H\eta_{l})\bigg]\\
+(2\alpha(\beta D-1)-(D-2))(\psi_{l}'-H\psi_{l})~~,~~\\
\lambda \eta_{l} \supset 2(1-\alpha)\bigg[f_{l}'+(D-1)Hf_{l}-\frac{A_{l}}{(D-1)}-\frac{(D-2)}{a^{2}(D-1)}\big(\tll-(D-1)\big)\eta_{l}\bigg]\\
+(2\alpha(\beta D-1)-(D-2))\psi_{l}
\end{gathered}
\end{equation}
where we, again, have to remember that some fields are only defined if $l$ is large enough.

Note that in the correction \nref{eigeneqgfch}, some terms vanish if we impose
\begin{equation}
\label{specialgf}
\beta D-1=\frac{(D-2)}{2\alpha}
\end{equation}

Therefore, sometimes when we discuss changing gauge fixing parameters, we will do so along the curve \nref{specialgf}. This is, of course, just one of the many families of gauge fixing parameters we could study, but it can be a convenient one to use.

\section{Details on the numerics}
\label{numdeta}
In this section we describe the spectral collocation method used to solve the eigenvalue problems in the various $SO(D)$ sectors derived in Appendix~\nref{eigensec}, and hence to obtain the one-loop spectrum discussed in the main text. We also provide additional numerical data for completeness.

\subsection{Spectral collocation method}
In Appendix~\nref{eigensec}, the full eigenvalue problem \eqref{actm}, \eqref{eigeneq} in the main text is decomposed into sectors labelled by $SO(D)$ representations and, within each sector, by the angular momentum quantum number $l$. The remaining equations are one-dimensional, multi-component, second-order ODEs in the $\tau$ direction. Without loss of generality, we write them in the unified form as follows. Let $\Phi_i(\tau)$ depend only on $\tau$, where $i=1,\ldots,N_\text{f}$ labels the components. The eigenvalue equations then take the general form
\begin{equation}
M\Phi=\lambda\Phi,\ M:=M^{(2)}(\tau)\partial_\tau^2+M^{(1)}(\tau)\partial_\tau+M^{(0)}(\tau)
\end{equation}
For each fixed $\tau$, the matrices $M^{(2)},M^{(1)},M^{(0)}$ are $N_\text f\times N_\text f$. Moreover, $\Phi(\tau)$ (and hence the coefficients in $M$) is periodic in $\tau\in[0,T]$, where $T$ denotes the period.

The spectral collocation method is motivated by discretizing $\tau\in[0,T]$ into $n_\text{grid}$ points $\{\tau_{k},\,k=1,\ldots,n_\text{grid}\}$ and replacing derivatives by finite-difference approximations, e.g.
$\partial_\tau\Phi(\tau_k)\approx(2\Delta t)^{-1}[\Phi(\tau_{k+1})-\Phi(\tau_{k-1})]$ and
$\partial^2_\tau\Phi(\tau_k)\approx(\Delta t)^{-2}[\Phi(\tau_{k+1})+\Phi(\tau_{k-1})-2\Phi(\tau_{k})]$.
Here $\Delta t:=\frac{T}{n_\text{grid}}$ is the lattice spacing, and periodicity implies $\Phi(\tau_{k})\equiv\Phi(\tau_{k+n_\text{grid}})$. With these replacements, the differential operators are represented by $n_\text{grid}\times n_\text{grid}$ differentiation matrices,
\begin{equation}
\partial_\tau\rightarrow(2\Delta t)^{-1}\begin{bmatrix}
0 & -1 & & &  1\\
1 & 0 & -1 & &  \\
 & 1 & \ddots & \ddots &  \\
& &  \ddots &0 & -1  \\
 -1& & &   1 & 0
\end{bmatrix},\ \partial^2_\tau\rightarrow(\Delta t)^{-2}\begin{bmatrix}
-2 & 1 & & &  1\\
1 & -2 & 1 & &  \\
 & 1 & \ddots & \ddots &  \\
& &  \ddots &-2 & 1  \\
 1& & &   1 & -2
\end{bmatrix}
\end{equation}
Such naive finite-difference choice above generally has relatively large discretization errors. In spectral collocation, one instead uses the optimal differentiation matrices constructed in \cite{Trefethen2000Spectral}, which minimize the discretization error for fixed $n_\text{grid}$,
\begin{equation}
\partial_\tau\rightarrow\mathcal{D}_1,\ \partial_\tau^2\rightarrow\mathcal{D}_2 
\end{equation}
where the explicit forms of $\mathcal D_{1,2}$ (first and second derivatives) are given in equations~(3.10) and (3.12) of \cite{Trefethen2000Spectral}. After this replacement, the operator $M$ is represented by an $N_\text fn_\text{grid}\times N_\text fn_\text{grid}$ matrix, which can be diagonalized to obtain the eigenvalues and eigenvectors.

\subsection{Supplementary numerical data}
\label{supdata}
The main text presents several general statements about the one-loop spectrum as functions of parameters such as $N$, $D$, and $Q$. Since an exhaustive scan over the full parameter space is infeasible, we instead study representative regimes. For completeness, we record here the parameter choices used in the numerical analysis.
\begin{itemize}
\item For the (physical and ghost) spectra and for the phase counting in \eqref{whphase}, we considered $D=3,4,5,6$ and cycle number $N=1,2,3,4$, for various charges in the range $0<Q<Q_c$. In each case we checked angular momenta $l=0,1,2,3$. (The main qualitative features occur in the $l=0,1$ sectors; correspondingly, we expect no special phenomena at higher $l$ and found $l\leq 3$ sufficient.) In these regimes we also verified that, in the small-$Q$ limit, the spectrum reduces to $N$ copies of the sphere spectrum plus additional light modes.
\item We tested gauge invariance of several quantities, including the phase counting in \eqref{whphase} and the existence of the light modes and their charge scaling, as discussed in Section~\nref{lightzero}. The general gauge fixing term depends on two parameters $\alpha,\beta$ defined in \eqref{eq: general gauge fixing}. We first fixed $\beta=\frac{1}{D}+\frac{D-2}{2\alpha D}$ as in \eqref{specialgf}, and varied $\alpha\in[0.1,3]$. We then fixed $\alpha=1$ and varied $\beta\in[0,0.9]$. For all these choices, the phase counting as well as the existence and scaling of the light modes were unchanged.
\item In \eqref{eq: De Witt parameter} we introduced a two-parameter family of ultralocal norms, parametrized by $p_1$ and $p_2$. We explored $p_1\in[1.05,2]$ and $p_2\in[0.5,2]$ (with the constraint $p_2>0$ required for positive definiteness of norm). Again, we found that the phase counting and the existence and scaling of the light modes do not depend on $p_1$ and $p_2$.
\end{itemize}

We now present additional numerical results illustrating the discussion in the main text.

\begin{figure}[H]
    \centering
    \includegraphics[width=1\linewidth]{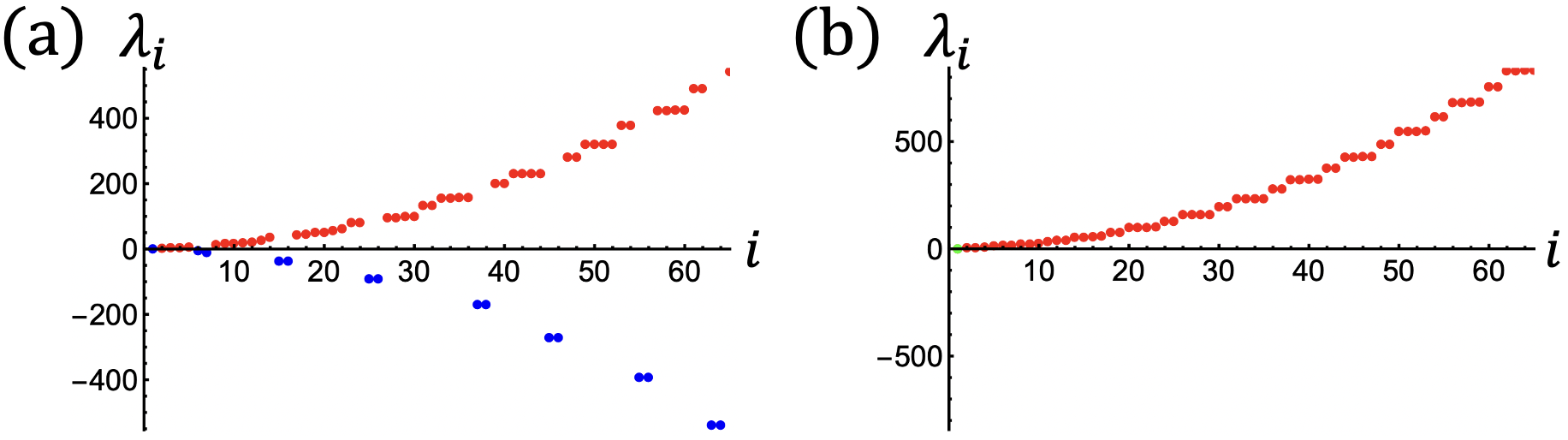}
    \caption{An example of the low-lying spectrum, with parameters $D=5$, $\ Q/Q_c=0.7$, and $\ N=1$. We use $n_\text{grid}=600$ grid points, which is sufficient for convergence of the low-lying eigenvalues. Positive/zero/negative modes are colored red/green/blue. Panel \textbf{(a)} shows the scalar sector with angular momentum $l=0$ and panel \textbf{(b)} the tangent transverse vector sector with angular momentum $l=1$. A zero mode is visible in this particular vector sector, as explained in section~\nref{zeromod}; numerically it has eigenvalue $\sim 10^{-7}$.}
    \label{fig: example of spectrum}
\end{figure}

First, As a concrete example of the numerical output, figure \nref{fig: example of spectrum} shows the physical spectrum in the scalar and vector sectors. We use different colors to denote positive, zero, and negative modes. The horizontal axis labels the eigenvalue index $i$ for $\lambda_i$, with eigenvalues ordered by increasing absolute value. The index $i$ can be interpreted as a ``radial'' quantum number, i.e.
roughly the momentum along the $\tau$ direction, which explains why at large $i$ the eigenvalues grow quadratically with $i$. One also observes that the infinite tower of negative modes responsible for the ``conformal-mode problem'' appears only in the scalar sector, as expected.

\begin{figure}[H]
    \centering
    \includegraphics[width=1\linewidth]{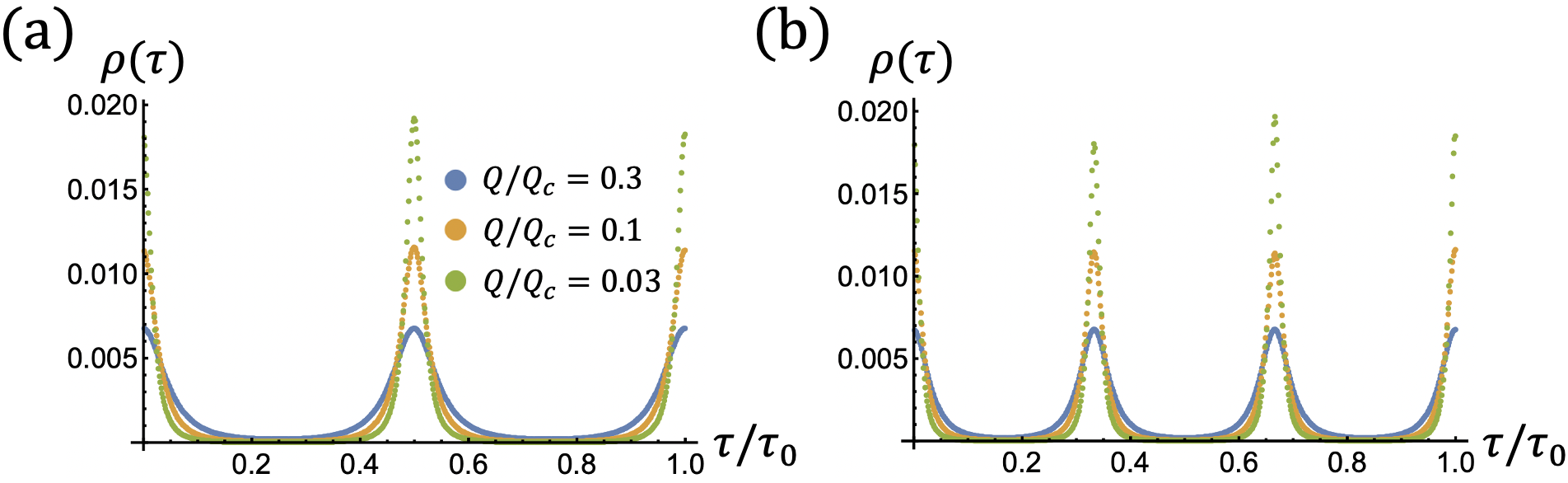}
    \caption{The norm density $\rho(\tau)$ of the light negative modes at $D=5$, for various $Q$ and $N$. As $Q$ is tuned toward the sphere limit, the wavefunction becomes localized near the wormhole mouths. For general $N$ there are $N$ sets of negative light modes; here we plot only the one with the smallest absolute eigenvalue. The two panels correspond to different cycle numbers: \textbf{(a)} $N=2$ and \textbf{(b)} $N=3$.}
    \label{fig: localized wavefunc N_2_3}
\end{figure}

Second, we show examples of localized modes for general $N$-cycles. For general $N$, the $l=1$ scalar sector contains $2N$ sets of light modes, each with multiplicity $D$ due to angular momentum. As in the $N=1$ case, their eigenvalues scale as $|\lambda|\sim\kappa Q$ as $Q\rightarrow0$, and their wavefunctions are localized; see Fig.~\nref{fig: localized wavefunc N_2_3}. Among these $2N$ eigenvalues, $N$ are positive and $N$ are negative.

\begin{figure}[H]
    \centering
    \includegraphics[width=1\linewidth]{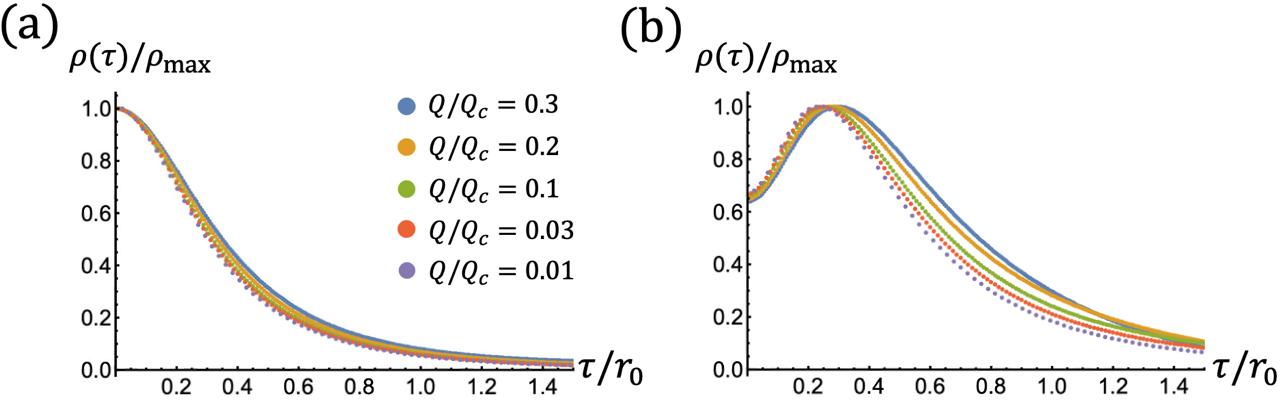}
    \caption{The norm density $\rho(\tau)$ of near-sphere light modes at $D=5, \ N=1$, for various $Q$. The horizontal axis is $\tau/r_0$, where $r_0$ is the wormhole mouth size defined in \eqref{rodef}. We observe that the curves exhibit data collapse. }
    \label{fig: data collapse}
\end{figure}

Next, we demonstrated in the main text that the two light modes in the near-sphere limit are localized near the wormhole mouth; see \nref{fig: light mode wavefunction and spectrum}, where we plot their norm density $\rho(\tau)$ as a function of $\tau$. Here we further show that the corresponding localization length is set by the wormhole mouth size $r_0$ defined in \eqref{rodef}. As shown in \nref{fig: data collapse}, plotting $\rho(\tau)$ against $\tau/r_0$ for several small values of $Q$ yields a collapse of the curves.

\begin{figure}[H]
    \centering
    \includegraphics[width=0.8\linewidth]{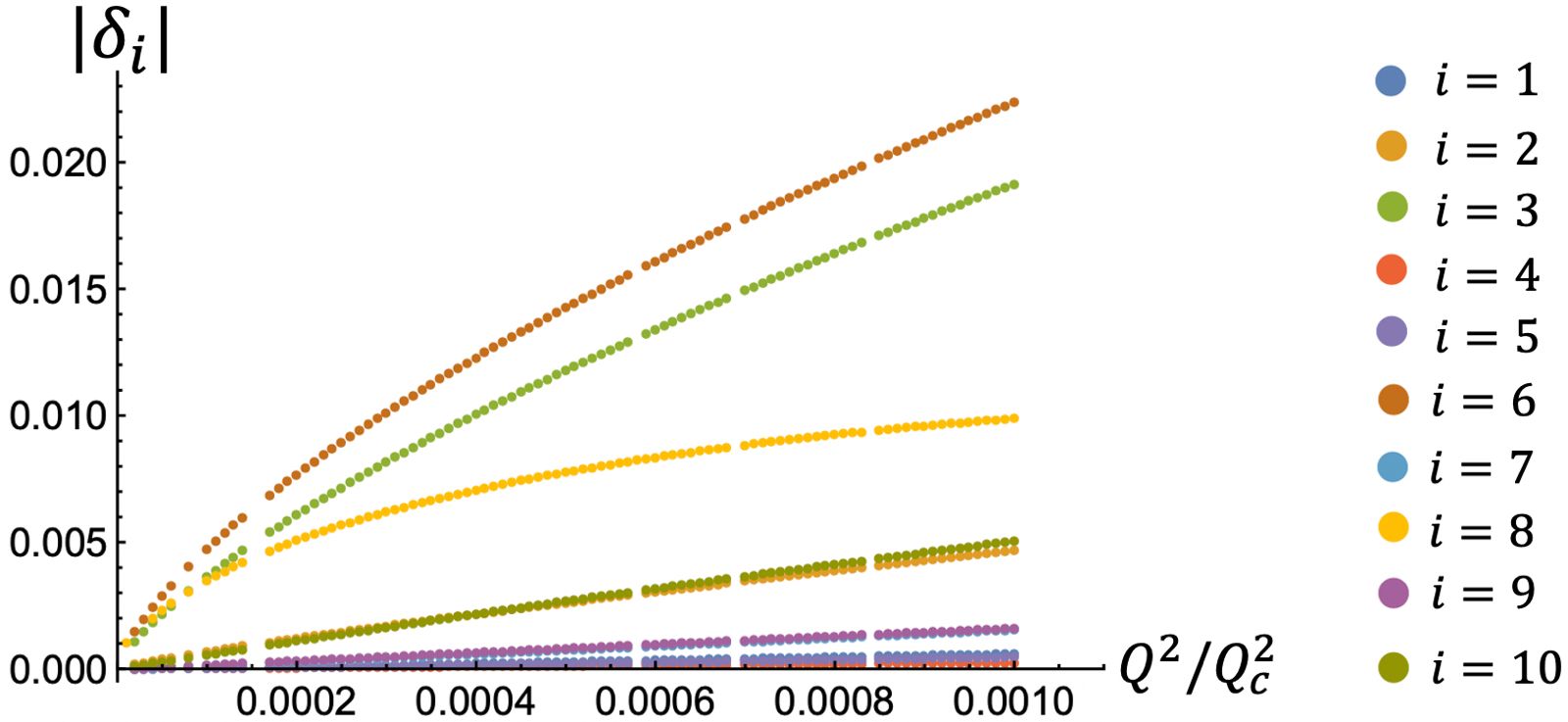}
    \caption{Eigenvalue deviation $\delta_i$ as a function of $Q$. We work in the $l=0$ scalar sector of the physical spectrum for $D=5$ and $N=1$, and include the first ten low-lying eigenvalues (indices $i=1,\ldots,10$). Different colors correspond to different $i$. \textbf{(a)} Negative light mode and \textbf{(b)} positive light mode. }
    \label{fig: deviation from sphere}
\end{figure}

Finally, we illustrate that in the limit $Q\rightarrow0$ the non-light part of the spectrum approaches the sphere spectrum, as discussed in Section~\nref{lochand}. Figure~\nref{fig: deviation from sphere} shows the deviation $\delta_i$ (defined in \eqref{eigshift}) of the first few low-lying modes in the $l=0$ scalar sector of the physical spectrum, relative to the corresponding sphere eigenvalues $\lambda_i^{(0)}$. We indeed find $\delta_i\rightarrow0$ as $Q\rightarrow0$.

\section{Axion two-point function in the sphere}
\label{axig}

In this section, we discuss the axion two-point function in the sphere, $G(x,y)$. We can find $G(x,y)$ by requiring it to be an $SO(D+1)$ invariant solution of the following equation
\begin{equation}
\label{gfct}
-\n^{2}G(x_{1},x_{2})=\frac{\delta(x_{1},x_{2})}{\sqrt{g}}-c~~\text{,with }~~ c=\frac{1}{\text{Vol}(S^{D})}
\end{equation}
with $c$ a constant, and $\text{Vol}(S^{D})$ stands for the volume of $S^{D}$. The constant $c$ is necessary because the left-hand side of \nref{gfct} is orthogonal to constant functions, since they are zero modes of $-\n^{2}$, so the right-hand side must be too.

One can solve for $G$ by imposing that $G$ is a function only of the $SO(D+1)$ invariant built of $X_{1}$ and $X_{2}$, defined in embedding space. To be more specific, we can think of $S^{D}$ as a surface in $R^{D+1}$, with coordinates $(X^{1},...,X^{D+1})$ described by
\begin{equation}
\sum_{i=1}^{D+1} X_{i}^{2}=1
\end{equation}

By parameterizing the surface via
\begin{equation}
\begin{gathered}
X_{D+1}=\cos \tau\\
X_{i}=\sin \tau\,\hat{x}_{i}~~\text{, with}~~ 1<i<D \text{, and}~~ \sum_{i=1}^{D}\hat{x}_{i}^{2}=1 
\end{gathered}
\end{equation}

We recover the usual sphere line element
\begin{equation}
ds^{2}=d\tau^{2}+\sin^{2}\tau\, d\Omega_{D-1}^{2}
\end{equation}

Note that in this reparameterization, we effectively reduce $S^{D}$ to a $\tau$ coordinate and an $S^{D-1}$ factor, which is further described via a different set of embedding coordinates $\hat{x}_{i}$. 

We can define an $SO(D+1)$ invariant function of the spherical coordinates by using the inner product in $R^{D+1}$ via
\begin{equation}
P(X_{1},X_{2})=X_{1} \cdot X_{2}=\sum_{i=1}^{D+1}(X_{1})_{i}(X_{2})_{i}=\cos \tau_{1}\cos \tau_{2}+\sin \tau_{1}\sin \tau_{2}\, \hat{x}_{1}\cdot \hat{x}_{2}
\end{equation}
with "$\cdot$" the dot product in $R^{D+1}$. Note that on the right-hand side, we defined a similar dot product, $\hat{x}_{1}\cdot \hat{x}_{2}$, for the $R^{D}$ embedding space of the $S^{D-1}$ factor. 

We have that $P=-1$ if $X_{1}$ and $X_{2}$ are antipodal points in the sphere, and $P=1$ if they are at coincident points. We can construct a $SO(D+1)$ invariant two-point function $G(x_{1},x_{2})$ by requiring that it is a function of $P$ alone. The function $G(x_{1},x_{2})$ might diverge at $P=\pm 1$. We require the function to only diverge at $P=-1$, which is the same as requiring it to be regular when $x_{1}$ and $x_{2}$ are in antipodal positions. This fixes the form of $G$, up to an additive constant, to be
\begin{equation}
G(x_{1},x_{2})=G_{\text{anti}}+c\int_{-1}^{P} \frac{dP'}{(1-P'^{2})^{\frac{D}{2}}}\int_{-1}^{P'}(1-P''^{2})^{\frac{(D-2)}{2}}dP''
\end{equation}
with $G_{\text{anti}}$ the value of the two point function when $x_{1}$ and $x_{2}$ are at antipodal positions. Note that this implies that, expanding $P$ near the antipodal configuration as $P=-1+\delta P$, we have that
\begin{equation}
G(x_{1},x_{2})\approx G_{\text{anti}}+\frac{c \delta P}{D}+O((\delta P)^{2})\approx G_{\text{anti}}+\frac{c}{2D}(\delta \tau_{1}^{2}+\delta \tau_{2}^{2}-2\delta \tau_{1}\delta \tau_{2}\, \hat{x}_{1}\cdot \hat{x}_{2})
\end{equation}

This specific combination of $\delta \tau$ and the angular coordinates comes, of course, because the function needs to be de Sitter invariant. However, we can imagine doing some gauge fixing where we fix one of the operator insertions, say the one at $\tau_{1}=0$, and move the other one by $\delta \tau_{2}$. Doing so, we see that expanding around the antipodal configuration will always increase $G(x_{1},x_{2})$, since $c>0$.

From this, we can evaluate the two-point function in \nref{goperator1} by writing
\begin{equation}
\langle e^{\frac{2 \pi i}{f_{\theta}}n \theta(x_{1})}e^{-\frac{2 \pi i}{f_{\theta}}n \theta(x_{2})} \rangle=\frac{\int D\theta\,e^{\frac{2 \pi i}{f_{\theta}}n \theta(x_{1})}e^{-\frac{2 \pi i}{f_{\theta}}n \theta(x_{2})}  e^{-\frac{1}{2}\int (\n \theta)^{2}}}{\int D\theta\,e^{-\frac{1}{2}\int (\n \theta)^{2}}}
\end{equation}

To solve for this, we simplify by writing the $\theta$ in the integral on the numerator as
\begin{equation}
\theta(x)=\theta_{cl}(x)+\delta \theta
\end{equation}
where $\theta_{cl}$ is the classical solution for $\theta$ with two-sources at $x_{1}$ and $x_{2}$. More specifically, we have that
\begin{equation}
-\n^{2}\theta_{\text{cl}}=\frac{2 \pi i \,n}{f_{\theta}}\bigg(\frac{\delta(x,x_{1})-\delta(x,x_{1})}{\sqrt{g}}\bigg)
\end{equation}
which can be solved to
\begin{equation}
\theta_{\text{cl}}(x)=\frac{2 \pi i n}{f_{\theta}}(G(x,x_{1})-G(x,x_{2}))
\end{equation}

We can then evaluate the path integral from the classical action of this configuration and find
\begin{equation}
\langle e^{\frac{2 \pi i}{f_{\theta}}n \theta(x_{1})}e^{-\frac{2 \pi i}{f_{\theta}}n \theta(x_{2})} \rangle=e^{\frac{2\pi^{2}n^{2}}{f_{\theta}}(2G(x_{1},x_{2})-G(x_{1},x_{1})-G(x_{x},x_{2}))}
\end{equation}
where we used that $G(x,y)$ is symmetric in its two entries. The $G(x_{i},x_{i})$ terms are divergent and need to be regularized properly, but since this divergence should not depend on the separation of the insertions, we neglect it.

\section{Near Einstein wormhole solution}
\label{eincl}

A feature of the classical wormhole solution that we discussed in the section \nref{prelim} of the paper is that the solution for the scale factor of $a(\tau)$ did not have a simple analytic form unless $D=3$. In fact, we do not have an analytic expression even for the period $\tau$ of the wormhole as a function of $Q$. 

However, if the charge $Q$ is near the maximum charge $Q_{c}$, we can solve for the background perturbatively in the extremal parameter. Namely, let the charge $Q$ be
\begin{equation}
Q^{2}=Q_{c}^{2}(1-\epsilon)
\end{equation}

With this parameterization, solving for the background is equivalent to solving the following differential equation
\begin{equation}
\label{fried'}
\frac{a''}{a}=\frac{\kappa^{2}Q^{2}}{a^{2(D-1)}(D-1)}-1=(1-\epsilon)\bigg(\frac{a_{c}}{a}\bigg)^{2(D-1)}-1
\end{equation}

We then note that \nref{fried'} implies that there is always a value $a=a_{e}$ in the solution where $a''=0$, more precisely $a''=0$ when
\begin{equation}
a=a_{e}=a_{c}(1-\epsilon)^{\frac{1}{2(D-1)}}
\end{equation}

Therefore, we can think of $a=a_{e}$ as the "no-force" position in the oscillatory motion of the particle $a(\tau)$. Furthermore, if the oscillations in $a$ are small, we can solve for $a$ perturbatively around this point. In other words, we can solve for $a$ perturbatively by writing $a=a_{e}(1-y)$ and solving the equation order by order in $y$. The motivation for doing so is that the amplitude of $y$ will go to zero when $\epsilon \rightarrow 0$.

The equation for $y$ is therefore
\begin{equation}
\label{yeq}
-\frac{y''}{(1-y)}=\bigg(\frac{1}{1-y}\bigg)^{2(D-1)}-1 \rightarrow y''=-2(D-1)y+O(y^{2})
\end{equation}
which we can solve as
\begin{equation}
y(\tau)=A \cos(\sqrt{2(D-1)}\tau)+O(A^{2})
\end{equation}

This solution already implies that, at leading order in $\epsilon$, the period of the near-extremal wormhole is
\begin{equation}
T \approx \frac{2\pi}{\sqrt{2(D-1)}}
\end{equation}

We can then furthermore fix the constant $A$ in terms of $\epsilon$ by using the Friedmann equation \nref{fried}, which implies that
\begin{equation}
A \approx \sqrt{\frac{\epsilon}{2(D-1)(D-2)}}
\end{equation}

This implies that the solution for $a(\tau)$ in a near-extremal solution is approximately
\begin{equation}
\label{neareina}
a(\tau)=a_{c}\bigg(1-\sqrt{\frac{\epsilon}{2(D-1)(D-2)}}\cos(\sqrt{2(D-1)}\tau)\bigg)+O(\epsilon)
\end{equation}

\section{Einstein wormhole and ghost zero modes}
\label{einghost}

In this section, we explicitly discuss whether the gauge fixing condition we chose can have ghost zero modes for the Einstein wormhole solution. For this, take the gauge fixing condition
\begin{equation}
\label{gcapp}
P_{a}=\n^{b}h_{ab}-\beta \n_{b}h
\end{equation}

If there is a pure gauge mode $h_{ab}=2\n_{(a}\xi_{b)}$ that satisfies $P_{ab}=0$, then we say the gauge fixing condition leaves residual gauge transformations. In particular, we would like to understand if there are residual gauge transformations for the gauge fixing condition used throughout the paper, namely $\beta=\frac{1}{2}$. The coordinate transformations $\xi_{a}$ are vectors, and we can therefore decompose them as
\begin{equation}
\xi_{a}=n_{a}m+\mcD_{a}s+V_{a}
\end{equation}
with $V_{a}$ a tangent transverse vector. The scalar modes will only generate scalar modes for \nref{gcapp}, and the tangent transverse vectors will only generate tangent transverse vectors. Therefore, we can study them individually.

The scalar modes will generate
\begin{equation}
P_{a}=n_{a}(2m''(1-\beta)+(1-2\beta)\mcD^{2}s'+\mcD^{2}m)+\mcD_{a}\bigg[2\mcD^{2}s(1-\beta)+2(D-1)s+s''+(1-2\beta)m\bigg]
\end{equation}

We can further decompose the scalars into modes with definite $SO(D)$ angular momentum and $U(1)$ momentum as
\begin{equation}
m=m_{n,l,+}\cos(k_{n}\tau)Y_{l}~~,~~ s=s_{n,l,-}\sin(k_{n}\tau)Y_{l}
\end{equation}
where we used the same notation and conventions of section \nref{prelimein}. That is, there is another set of modes with $+ \rightarrow -$ where we replace $\cos \rightarrow \sin$ and $\sin \rightarrow -\cos$.

We first discuss the $l=0$ modes because they are qualitatively different, since the $s$ mode does not exist for them. In this case, the gauge fixing condition will be proportional to $m''$, and therefore will be zero as long as $m$ is a constant or proportional to $\tau$. The $m$ proportional to $\tau$ solution is not invariant under $\tau \rightarrow \tau+L$ and therefore not an acceptable coordinate transformation.

Therefore, the ghost zero mode at $l=0$ is the $m=\text{const}$ mode, which corresponds to shifting the origin, $\tau=0$, in the wormhole. This mode has a finite range since we identify $\tau \sim \tau+L$.  This mode, however, is an isometry and thus generates no metric deformation. Therefore, there will be no $h_{ab}$ mode with a zero eigenvalue associated with it.

For $l>0$, expanding the modes into spherical harmonics and assuming $s_{I},m_{I} \neq 0$ we obtain that $P_{a}=0$ implies the following equation
\begin{equation}
k^{2}(1-2\beta)^{2}\lambda_{l}=(\lambda_{l}+2k^{2}(1-\beta))(2\lambda_{l}(1-\beta)+k^{2}-2(D-1))
\end{equation}

We see that at $\beta=\frac{1}{2}$, the equation above is the same as $-\n^{2}=\lambda_{l}+k^{2}=2(D-1)$. This can only have a solution for $l=1$ since at $l=2$ the left-hand side is already bigger than the right-hand side with $k=0$. 

Therefore, the gauge fixing condition has zero modes when $L$ has special values, or more specifically, when
\begin{equation}
\frac{4n^{2}\pi^{2}}{L^{2}}=\frac{(D-3)(D-1)}{(D-2)}
\end{equation}
which are precisely the possible zero modes we found in section \nref{einspec} for $l=1$ scalar fluctuations. Therefore, these zero modes are artifacts of the gauge fixing method we chose. 

\section{Wormhole solution at large $D$}
\label{app: large D}
In this section, we present an analytic solution of the Friedmann equation \eqref{fried} for general $Q$ in the large-$D$ limit. For finite $D>3$ and generic $Q$, $a(\tau)$ can only be obtained numerically.

\begin{figure}[H]
    \centering
    \includegraphics[width=1\linewidth]{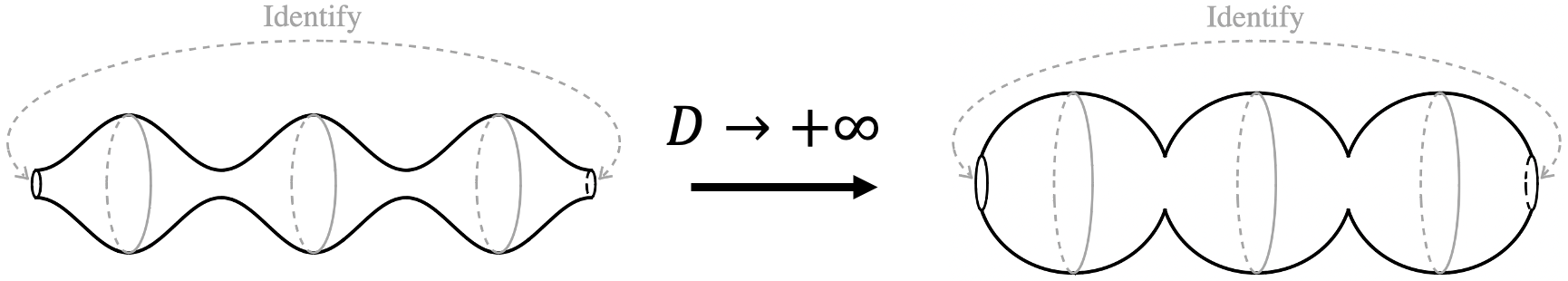}
    \caption{Illustration of the wormhole solution in the large-$D$ limit.}
    \label{fig: larege D wormhole}
\end{figure}

We first rewrite equation~\eqref{fried} as the energy-conservation equation for a one-dimensional classical particle moving in the potential $V(a):=a^2+\frac{\kappa^2Q^2}{(D-1)(D-2)}a^{-2(D-2)}$, namely $a'^2+V(a)=1$. A non-trivial large-$D$ limit is obtained by letting $Q$ scale with $D$ as
\begin{equation}
\frac{\kappa^2Q^2}{(D-1)(D-2)}\stackrel{\text{large }D}{\longrightarrow}a_0^{2(D-2)}
\label{eq: large D scaling of Q}
\end{equation}
with $a_0\in(0,1)$ a fixed parameter. We must verify that this scaling is consistent with $Q<Q_c$. Indeed, \eqref{eq: Qc} implies $\kappa^2Q_c^2= e^{-1}D+O(D^{-1})$, which grows linearly with $D$. For $a_0<1$, the scaling in \eqref{eq: large D scaling of Q} makes $Q$ exponentially small in $D$. In this limit, the potential becomes
\begin{equation}
V(a)=a^2+\left(\frac{a_0}{a}\right)^{2(D-2)}\stackrel{D\rightarrow+\infty}{\longrightarrow}\begin{cases}
+\infty ,\ &a<a_0\\
a^2, \ &a>a_0
\end{cases}
\end{equation}
Thus, for $a>a_0$ the potential is harmonic, while $a=a_0$ acts as a hard wall that bounces back the particle elastically. We further parametrize $a_0\equiv\cos(T_0/2)$, where $T_0\in(0,\frac\pi4)$ is the period of a single cycle ($N=1$) wormhole. The corresponding single cycle solution is
\begin{equation}
a(\tau)=\begin{cases}
\cos\tau,\ &0<\tau\leq T_0/2,\\
\cos(T_0-\tau),\ &T_0/2<\tau<T_0
\end{cases}
\end{equation}
The $N$-cycle solutions are obtained by repeating this segment; geometrically, they resemble $N$ round spheres that are cut and then glued together, as shown in figure~\nref{fig: larege D wormhole}.

\section{A check of the rule for small handles}
\label{checklochand}

In section \nref{lochand}, we proposed a rule to compute the ratio of functional determinants in a spacetime $\mathcal{M}$ and a spacetime $\mathcal{M}_{0}$, where $\mathcal{M}$ is $\mathcal{M}_{0}$ with a small handle. Here, we give more evidence for this rule by showing that we need it in order for the dependence on the handle size, $r_{0}$, of two quantities related by electromagnetic duality to match in odd $D$. We also assume that $D>3$ for simplicity. 

To be more specific, at odd D electromagnetic duality states that the path integral over $D-2$ forms with coupling $f_{\theta}$ in a generic manifold should match that of a massless scalar with coupling $\frac{2\pi}{f_{\theta}}$.\footnote{In even $D$ there is an relative anomaly between the two quantities} After relating some instanton contributions, this is equivalent to the following statement\footnote{Here we ignored details involving discrete torsion subgroups, which will be unimportant to our point. See \cite{Donnelly:2016mlc} for more details}
\begin{equation}
\begin{gathered}
\label{elemdual}
\frac{\text{det}^{\frac{1}{2}}\bigg(\frac{2\pi}{f_{\theta}^{2}}\Gamma_{D-2}\bigg)}{\text{Det}'\Delta_{D-2}^{\frac{1}{2}}}\times\frac{\text{Det}'\Delta_{D-3}^{1}}{\text{det}^{\frac{1}{2}}\bigg(\frac{2\pi}{f_{\theta}^{2}}\Gamma_{D-3}\bigg)}\times\frac{\text{det}^{\frac{1}{2}}\bigg(\frac{2\pi}{f_{\theta}^{2}}\Gamma_{D-4}\bigg)}{\text{Det}'\Delta_{D-4}^{\frac{3}{2}}}...\frac{\text{det}^{\frac{1}{2}}\bigg(\frac{2\pi}{f_{\theta}^{2}}\Gamma_{1}\bigg)}{\text{Det}'\Delta_{1}^{\frac{(D-2)}{2}}}\times \frac{\text{Det}'\Delta_{0}^{\frac{(D-1)}{2}}}{\text{det}^{\frac{1}{2}}\bigg(\frac{2\pi}{f_{\theta}^{2}}\Gamma_{0}\bigg)}\\
=\frac{\text{det}^{\frac{1}{2}}\bigg(\frac{f_{\theta}^{2}}{2\pi}\Gamma_{0}\bigg)}{\text{Det}'\Delta_{0}^{\frac{1}{2}}}\text{det}^{-\frac{1}{2}}\bigg(\frac{f_{\theta}^{2}}{2\pi}\Gamma_{1}\bigg)
\end{gathered}
\end{equation}
with $\Delta_{p}$ the Hodge Laplacians of $p$-forms, and where we used that $D$ is odd to relate the relative sign of $\Gamma_{D-2}$ and $\Gamma_{1}$. The $\Gamma_{p}$ are inner product matrices of a basis $w_{I}$, for the space of harmonic $p$-forms of the manifold that have integer flux over the non-contractible $p$-surfaces. 

What is relevant for us is the following: The dimension of the kernel of $\Delta_{p}$ is given by the Betti numbers of the manifold. Therefore, adding a handle to the manifold will explicitly change the spectrum of some Hodge Laplacians. We want to compare the ratio of both sides of \nref{elemdual}, between a wormhole spacetime and the sphere. We will take the wormhole spacetime to be the wormhole of the main body of the paper, in the limit that it becomes a sphere with a small handle. 

In the sphere, there are only harmonic $0$-forms and $D$-forms, one of each. In the wormhole, there is one harmonic $0$ form, one $1$-form, one $D-1$ form, and one $D$ form. Therefore, the $0$-form is a common contribution on the sphere and wormhole. Furthermore, $\Gamma_{0}$ is non-trivial in the sphere and wormhole, and $\Gamma_{1}$ is relevant only for the wormhole. 

Since the rest of the Hodge Laplacians on both sides of \nref{elemdual} do not develop new zeros, we do not expect them to develop new modes either. Therefore, we assume they approximately cancel in the ratio between the wormhole and sphere answers, or at least that their ratio does not involve powers of the handle size, $r_{0}$. Also, the associated contribution from the harmonic zero forms in $\Gamma_{0}$ is about the same in both spacetimes (see \nref{harm0form}).

However, $\Delta_{1}$ will change non-trivially between the wormhole and sphere, since it has a new zero mode in the wormhole spacetime, the harmonic $1$-form that runs in the $\Gamma_{1}$ contribution. Taking the ratio of \nref{elemdual} between the sphere with a small handle and the sphere, we obtain
\begin{equation}
\begin{gathered}
\label{elemdual2}
\frac{(\text{Det}'\Delta_{1}^{\frac{(D-2)}{2}})_{\mathcal{M}_{0}}}{(\text{Det}'\Delta_{1}^{\frac{(D-2)}{2}})_{\mathcal{M}}}\times \text{det}^{\frac{1}{2}}\bigg(\frac{2\pi}{f_{\theta}^{2}}\Gamma_{1}\bigg)_{\mathcal{M}}
 \sim \text{det}^{-\frac{1}{2}}\bigg(\frac{f_{\theta}^{2}}{2\pi}\Gamma_{1}\bigg)_{\mathcal{M}}
\end{gathered}
\end{equation}
where we used that $D>3$, so that $\Gamma_{D-2}$ was trivial in both wormhole and sphere, and $\sim$ to mean both sides have the same power dependence in the scale $r_{0}$. Note that the factors of $\frac{2\pi}{f_{\theta}^{2}}$ cancel between the two sides of \nref{elemdual2}. Moreover, from our formula for $\Gamma_{1}$ in \nref{gamma1cont}, we know that $\Gamma_{1}$ can be approximated from its contribution from the handle region in \nref{smwhreg}, such that $\det \Gamma_{1} \sim r_{0}^{-(D-2)}$. This, therefore, implies that 
\begin{equation}
\begin{gathered}
\label{elemdual3}
(\text{Det}'\Delta_{1}^{-\frac{1}{2}})_{\mathcal{M}}\times \text{det}^{\frac{1}{2}}\bigg(\frac{2\pi}{f_{\theta}^{2}}\Gamma_{1}\bigg)_{\mathcal{M}} \sim (\text{Det}'\Delta_{1}^{-\frac{1}{2}})_{\mathcal{M}_{0}}\times \text{det}^{\frac{1}{2}}\bigg(\frac{2\pi}{f_{\theta}^{2}}\Gamma_{1}\bigg)_{\mathcal{M}}\times r_{0}^{-1}
\end{gathered}
\end{equation}

Note that  we can think of the left-hand side of \nref{elemdual3} as a gauge-fixed path integral over $1$-forms with $\Delta_{1}$ as a fluctuation operator, as $Z_{1-\text{form}}=\int DA_{\mu} e^{-\frac{1}{2}(A,\Delta_{1}A)}$. Equation \nref{elemdual} is therefore telling us that, at the level of scaling, the path integral in the sphere with a small handle is given by the pure sphere answer, times a contribution from new modes in the handle spacetime, and an extra factor of $r_{0}^{-1}$. This factor of $r_{0}^{-1}$ reproduces precisely the rule we argued for in section \nref{lochand}, since $\Delta_{1}$ has one new low-lying mode in the spacetime with the handle, the harmonic 1-form. This example, therefore, provides further evidence for the rule.

\bibliographystyle{apsrev4-1long}
\bibliography{main.bib}
\end{document}